\documentclass[a4paper,11pt]{article}
\pdfoutput=1
\usepackage{jheppub}
\usepackage{amsmath,amssymb,latexsym}
\usepackage{braket}
\usepackage{bm}
\usepackage{hepunits}
\usepackage[svgnames]{xcolor}

\usepackage{xspace}
\newcommand{\air}{\texttt{AIR}\xspace}
\newcommand{\fire}{\texttt{FIRE}\xspace}
\newcommand{\kira}{\texttt{Kira}\xspace}
\newcommand{\reduze}{\texttt{Reduze}\xspace}
\newcommand{\litered}{\texttt{LiteRed}\xspace}

\allowdisplaybreaks

\DeclareUnicodeCharacter{2212}{-}

\title{Baikov representations, intersection theory, and canonical Feynman integrals}

\author[a]{Jiaqi Chen,}
\author[b]{Xuhang Jiang,}
\author[b]{Chichuan Ma,}
\author[c]{Xiaofeng Xu,}
\author[d]{Li Lin Yang}

\affiliation[a]{Institute of High Energy Physics, Chinese Academy of Sciences, Beijing 100049, China}
\affiliation[b]{School of Physics and State Key Laboratory of Nuclear Physics and Technology,\\
Peking University, Beijing 100871, China}
\affiliation[c]{Institut f\"ur Theoretische Physik, Universit\"at Bern, Sidlerstrasse 5, CH-3012 Bern, Switzerland}
\affiliation[d]{Zhejiang Institute of Modern Physics, Department of Physics, Zhejiang University, Hangzhou 310027, China}

\emailAdd{chenjq@ihep.ac.cn}
\emailAdd{xhjiang@pku.edu.cn}
\emailAdd{chichuanma@pku.edu.cn}
\emailAdd{pkuxxf@gmail.com}
\emailAdd{yanglilin@zju.edu.cn}

\abstract{The method of canonical differential equations is an important tool in the calculation of Feynman integrals in quantum field theories. It has been realized that the canonical bases are closely related to $d$-dimensional $d\log$-form integrands. In this work, we explore the generalized loop-by-loop Baikov representation, and clarify its relation and difference with Feynman integrals using the language of intersection theory. We then utilize the generalized Baikov representation to construct $d$-dimensional $d\log$-form integrands, and discuss how to convert them to Feynman integrals. We describe the technical details of our method, in particular how to deal with the difficulties encountered in the construction procedure. Our method provides a constructive approach to the problem of finding canonical bases of Feynman integrals, and we demonstrate its applicability to complicated scattering amplitudes involving multiple physical scales.}

\begin{document}

\maketitle

\clearpage

\section{Introduction}

Feynman integrals are central objects in perturbative quantum field theories (QFTs). They are the basic ingredients of correlation functions and scattering amplitudes, which are the essential bridges between fundamental theories and experimental observations. The analytic, algebraic and geometric properties of these integrals provide many new insights on QFTs themselves. In textbooks, Feynman integrals are usually represented as integrals over loop momenta or integrals over Feynman parameters. Techniques based on these representations have been greatly advanced in the past decades (see, e.g., \cite{Smirnov:2012gma, Henn:2014yza, Weinzierl:2022eaz} and references therein), leading to a proliferation of new results which cannot be obtained using traditional methods.

An important toolset in the calculation of Feynman integrals is the integration-by-parts (IBP) identities \cite{Tkachov:1981wb, Chetyrkin:1981qh} combined with the method of differential equations \cite{Kotikov:1990kg, Kotikov:1991hm, Kotikov:1991pm, Remiddi:1997ny, Gehrmann:1999as}. The IBP identities are used to reduce all scalar Feynman integrals appearing in a scattering process to a finite set of master integrals (MIs). Such a reduction can be systematically performed with the Laporta algorithm \cite{Laporta:2001dd} implemented in various program packages such as \air \cite{Anastasiou:2004vj}, \fire \cite{Smirnov:2008iw, Smirnov:2019qkx}, \litered \cite{Lee:2012cn, Lee:2013mka}, \reduze \cite{Studerus:2009ye, vonManteuffel:2012np} and \kira \cite{Maierhofer:2017gsa, Klappert:2020nbg}. The MIs satisfy a closed system of linear differential equations. If these equations can be solved, one obtains the results for the MIs and hence for all integrals under consideration.

In certain cases, the differential equations can be organized into a nice form called the $\epsilon$-form \cite{Henn:2013pwa, Henn:2014qga, Henn:2014yza}:
\begin{equation}
d\vec{f}(\vec{x},\epsilon) = \epsilon \, d\bm{A}(\vec{x}) \, \vec{f}(\vec{x},\epsilon) \, ,
\label{eq:epsform}
\end{equation}
where $\epsilon=(4-d)/2$ is the dimensional regulator with spacetime dimension $d$, $\vec{x}=\{x_i\}$ is the list of kinematic variables, $\vec{f}=\{f_i\}$ is the list of linear combinations of master integrals, and $d\bm{A}$ is a matrix of the $d\log$ form independent of $\epsilon$. Once written in the $\epsilon$-form, the solutions to the differential equations can be formally written as Chen iterated integrals \cite{Chen:1977oja}. The results can often be written in terms of generalized polylogarithms (GPLs) \cite{Goncharov:1998kja, Goncharov:2001iea} order-by-order in $\epsilon$, which allow efficient numeric evaluation~\cite{Vollinga:2004sn, Naterop:2019xaf, Wang:2021imw}. When an analytic solution is not available, it is straightforward to evaluate them numerically either by numerical integration or by a series expansion~\cite{Moriello:2019yhu, Hidding:2020ytt, Liu:2022chg}.

The list of master integrals $\vec{f}$ satisfying Eq.~\eqref{eq:epsform} is called a canonical basis. These integrals have the property of \emph{uniform transcendentality} (UT) \cite{Henn:2013pwa}. Namely, they (with suitable normalization) can be expressed as 
\begin{equation}
    f_i(\vec{x},\epsilon) = \sum_{n=0}^{\infty}\epsilon^{n} \, f_i^{(n)}(\vec{x}) \, ,
\end{equation}
where $f_i^{(n)}(\vec{x})$ is a function with transcendental weight $n$. It is conventional to assign weight $-1$ to $\epsilon$, such that the whole function $f_i(\vec{x},\epsilon)$ has weight $0$. In a practical problem, it is crucial to find such a canonical basis of UT integrals. This can be done by starting from an arbitrary set of MIs, and performing linear transformations to reduce the differential equations to the $\epsilon$-form. Algorithms for finding such kind of transformations exist \cite{Muller-Stach:2012tgj, Argeri:2014qva, Gehrmann:2014bfa, Lee:2014ioa, Meyer:2016slj, Adams:2017tga, Lee:2017oca, Dlapa:2020cwj}, and some of which have been implemented as program packages \cite{Gituliar:2017vzm, Prausa:2017ltv, Meyer:2017joq, Lee:2020zfb}. These algorithms are particularly useful when only rational transformations are needed.

An alternative way to find a canonical basis is to construct UT integrals directly without studying the differential equations. It has been realized that UT integrals are closely related to $d\log$-form integrands in $d=4$ dimensions \cite{ArkaniHamed:2010gh, Drummond:2013nda, Arkani-Hamed:2013jha, Arkani-Hamed:2014via, Bern:2014kca, Arkani-Hamed:2016byb}, i.e., the integrands can be written (usually in the momentum representation or in certain dual representations) in the form
\begin{equation}
c \, d\log\alpha_1 \wedge d\log\alpha_2 \wedge \cdots \wedge d\log\alpha_n \, ,
\label{eq:4ddlog}
\end{equation}
where $\alpha_i$ are functions of the integration variables and $c$ is constant. Integrals with such integrands are also dubbed as having constant leading singularities. However, these 4-dimensional $d\log$ integrands are not guaranteed to give rise to UT integrals in $d=4-2\epsilon$ dimensions. Further manipulation is therefore required to arrive at a canonical basis. Construction methods based on the 4-dimensional $d\log$ integrands have been considered in \cite{Henn:2014qga, Chicherin:2018old, Wasser:2018qvj, Herrmann:2019upk, Henn:2020lye, Henn:2021aco}.

Motivated by the 4-dimensional $d\log$ integrands, it was suggested \cite{Chicherin:2018old, Herrmann:2019upk, Chen:2020uyk} that one may consider $d$-dimensional $d\log$ integrands in a suitable representation (where the dimensional regulator $\epsilon$ appears as a parameter in the integrand) such as the Baikov representation \cite{Baikov:1996iu, Lee:2010wea, Bosma:2017ens, Harley:2017qut, Bosma:2017hrk, Frellesvig:2017aai}. These $d\log$-forms can be written as
\begin{equation}
c \left[ \alpha_0(\bm{z}) \right]^\epsilon d\log\alpha_1(\bm{z}) \wedge d\log\alpha_2(\bm{z}) \wedge \cdots \wedge d\log\alpha_n(\bm{z}) \, ,
\label{eq:dddlog}
\end{equation}
where $\bm{z}$ denotes the collection of integration variables (which correspond to coordinates in the base manifold for the differential $n$-forms). Such $d$-dimensional $d\log$-forms automatically give rise to UT integrals without further manipulation. This then gives strong hints on the construction of a canonical basis for a given integral family. However, finding a complete set of $d$-dimensional $d\log$-form integrands is often not a trivial task. In that case one may employ weaker constraints such as looking for integrands having constant leading singularities under certain cuts (which reduce the number of integration variables) \cite{Dlapa:2021qsl}. Integrands satisfying such weaker constraints can then be further manipulated to arrive at UT integrals.

In this paper, we develop in more detail the studies of \cite{Chen:2020uyk}, on the construction of $d$-dimensional $d\log$-form integrands in the Baikov representation as candidates for UT Feynman integrals. We first review the standard and loop-by-loop Baikov representations, and explore the \emph{generalized} loop-by-loop Baikov representation with additional polynomials in the denominators. As will be clear later (and as was mentioned in \cite{Dlapa:2021qsl}), the generalized Baikov integrals do not all correspond to Feynman integrals. We introduce the concept of FI-subspace spanned by Feynman integrals within the vector space of generalized Baikov integrals. These vector spaces are studied using the language of intersection theory \cite{Mizera:2017rqa, Mastrolia:2018uzb, Frellesvig:2019kgj, Mizera:2019gea, Frellesvig:2019uqt, Mizera:2019vvs, Mizera:2019blq, Mizera:2020wdt, Weinzierl:2020xyy, Frellesvig:2020qot}. We demonstrate how to find linear combinations of generalized Baikov integrals that belong to the FI-subspace, and how to convert them to Feynman integrals. We then elaborate on our method of constructing $d\log$-form Baikov integrands and subsequently obtaining the complete canonical basis for a given integral family. We describe how we deal with the technical difficulties encountered in this procedure. We show that our approach can be well applied to complicated problems involving multiple physical scales.

The paper is organized as follows. In section~\ref{sec:Baikov}, we review the standard and the loop-by-loop Baikov representations, and introduce the concept of generalized loop-by-loop Baikov representation. In section~\ref{sec:intersection}, we briefly review the concept of intersection theory in the context of Feynman and Baikov integrals. Special focus is put on the correspondence between the dimension of twisted cohomology groups and the number of Baikov integrals. 
In section~\ref{sec:construction}, we introduce the method for the construction of UT Baikov integrals and for the conversion to canonical Feynman integrals. In section~\ref{sec:intmassivedb} and \ref{sec:extmassivedb}, we demonstrate our method using two non-trivial examples, while technique details and further examples are presented in the appendices. We summarize in section~\ref{sec:summary}.

\section{The Baikov representation of Feynman integrals}
\label{sec:Baikov}

The Baikov representation was first proposed in \cite{Baikov:1996iu}, and since then were further developed and used to study Feynman integrals \cite{Lee:2010wea, Bosma:2017ens, Harley:2017qut, Bosma:2017hrk, Frellesvig:2017aai}. In this section, we recap the derivation of the Baikov representation both in the standard and the loop-by-loop approaches. We also propose a generalization of the loop-by-loop representation, that will be useful in our construction of $d\log$-form integrands.

\subsection{The standard Baikov representation}

We consider $L$-loop Feynman integrals with $E+1$ external legs in spacetime dimension $d=4-2\epsilon$. The loop momenta are labelled by $k_i$ ($i=1,\ldots,L$) and the independent external momenta are $p_i$ ($i=1,\ldots,E$). For later convenience we collectively refer to them as $q_i$ ($i=1,\ldots,M$), where $M \equiv L+E$, $q_i \equiv k_i$ ($i=1,\ldots,L$), and $q_{L+i} \equiv p_i$ ($i=1,\ldots,E$).
Out of these momenta, one can construct $N \equiv L(L+1)/2 + LE$ independent scalar products involving at least one of the $k_i$. An integral family is then defined by a given set of $N$ independent propagator denominators $z_i$ ($i=1,\ldots,N$), which are linear functions of the aforementioned scalar products. A generic integral in such a family is given by
\begin{equation}
F_{a_1,\ldots,a_N} = e^{\epsilon \gamma_E L} \int \bigg[ \prod_{i=1}^{L}  \frac{d^d k_i}{i \pi^{d/2}} \bigg] \frac{1}{z_1^{a_1} \, z_2^{a_2} \cdots z_N^{a_N}} \, ,
\label{eq:loop-int}
\end{equation}
where $a_i \in \mathbb{Z}$. A specific topology in the integral family is defined by a chosen subset of the powers $\{a_i\}$ whose values are positive, while the other powers are either zero or negative.

The Baikov representation of the above integral amounts to a change of integration variables from the set $\{k_i^\mu\}$ to the set $\{z_n\}$. For that purpose, we write
\begin{equation}
z_n = \sum_{i=1}^L \sum_{j=i}^M A^{ij}_n \, s_{ij} + f_n \, , \quad (n=1,\ldots,N) \, ,
\end{equation}
where $s_{ij} \equiv q_i \cdot q_j$, $A^{ij}_n$ are integer constants, and $f_n$ are functions of external momenta and internal masses. Note that the number of the ordered pairs $(ij)$ is $N$. Therefore $A^{ij}_n$ can be regarded as the elements of an $N \times N$ matrix representing the linear transformation from $\{s_{ij}\}$ to $\{z_n - f_n\}$. We denote this matrix as
\begin{equation}
\bm{A} = \bm{A}(z_1,\ldots,z_N; k_1,\ldots,k_L; p_1,\ldots,p_E) \, , \quad A^{ij}_n = \bm{A}_{n,(ij)} \, .
\end{equation}
With a slight abuse of notation, we denote the elements of the inverse of the matrix $\bm{A}$ as $A^n_{ij}$, namely,
\begin{equation}
A^n_{ij} = \big(\bm{A}^{-1}\big)_{(ij),n} \, , \quad \sum_{ij} A^m_{ij} A^{ij}_n = \delta_{mn} \, , \quad \sum_n A^{ij}_n A^n_{kl} = \delta_{(ij),(kl)} \, .
\end{equation}
Therefore we have
\begin{equation}
s_{ij} = \sum_{n=1}^N A^n_{ij} (z_n - f_n) \, , \quad (i=1,\ldots,L; \, j=i,\ldots,M) \, .
\label{eq:sij_Dn}
\end{equation}

To proceed, we decompose each loop momentum $k_i$ into two parts, $k_i^\mu = k_{i\parallel}^\mu + k_{i\perp}^\mu$, where the parallel components $k_{i\parallel}^\mu$ live in the $(M-i)$-dimensional subspace spanned by $q_j$ ($j=i+1,\ldots,M$), and the perpendicular components $k_{i\perp}^\mu$ live in the $(d-M+i)$-dimensional orthogonal subspace.\footnote{There is some subtlety in this decomposition with the Minkowski signature. We will assume that the parallel subspace contains space-like vectors (i.e., we work in the so-called ``Euclidean'' kinematic region), such that vectors in the perpendicular subspace are time-like. Results for physical kinematics can be obtained via analytic continuation.}
The integration measure over the parallel components of $k_i$ is given by
\begin{equation}
d^{M-i}k_{i\parallel} = \big| G(q_{i+1},\ldots,q_M) \big|^{-1/2} \prod_{j=i+1}^M ds_{ij} \, ,
\end{equation}
where $G(q_1,\ldots,q_n)$ is the Gram determinant defined as
\begin{equation}
G(q_1,\ldots,q_n) \equiv \det (q_i \cdot q_j) \equiv \det
\begin{pmatrix}
q_1 \cdot q_1 & q_1 \cdot q_2 & \cdots & q_1 \cdot q_n
\\
q_2 \cdot q_1 & q_2 \cdot q_2 & & \vdots
\\
\vdots & & \ddots & \vdots
\\
q_n \cdot q_1 & \cdots & \cdots & q_n \cdot q_n
\end{pmatrix}
\, .
\end{equation}
Note also that $|G(q_1,\ldots,q_n)|^{1/2}$ is the volume of the parallelogram formed by $q_1,\ldots,q_n$ (in the Euclidean sense).

For the perpendicular components $k_{i\perp}^\mu$, only the norm-squared $k_{i\perp}^2$ enters the integrand since $s_{ii} = k_i^2 = k_{i\perp}^2 + k_{i\parallel}^2$. We perform a Wick rotation for the integration contour of $k_{i\perp}^0$ from the real axis to the imaginary axis (during which the value of $k_{i\perp}^2$ is deformed into the complex plane, and in the end gets back to the real axis but with $k_{i\perp}^2 \leq 0$). We then change variable to the Euclidean vector $k_{iT}^\mu$ as usual with $k_{iT}^2 = -k_{i\perp}^2$. The norm-squared can be expressed in terms of $\{ q_i \cdot q_j \}$ through
\begin{equation}
k_{iT}^2 = -\frac{G(q_i,\ldots,q_M)}{G(q_{i+1},\ldots,q_M)} = \frac{|G(q_i,\ldots,q_M)|}{|G(q_{i+1},\ldots,q_M)|} \geq 0 \, .
\label{eq:kiT2}
\end{equation}
The integration measure for the perpendicular components can then be written as
\begin{equation}
d^{d-M+i}k_{i\perp} = \frac{i \, \pi^{(d-M+i)/2}}{\Gamma\big((d-M+i)/2\big)} \, \left| \frac{G(q_i,\ldots,q_M)}{G(q_{i+1},\ldots,q_M)} \right|^{(d-M+i-2)/2} \, ds_{ii} \, .
\end{equation}
Using the above, we are able to change the integration variables from $\{k_i^\mu\}$ to $\{s_{ij}\}$. We can further change variable to the Baikov variables $\{z_n\}$ using Eq.~\eqref{eq:sij_Dn} and
\begin{equation}
\prod_{i=1}^L \prod_{j=i}^M ds_{ij} = \big| \det(\bm{A}^{-1}) \big| \prod_{n=1}^N dz_n \, .
\end{equation}
Finally, we arrive at
\begin{equation}
F_{a_1,\ldots,a_N} = \frac{C_{L,E} \, \big| \det(\bm{A}^{-1}) \big|}{\big| G(p_1,\ldots,p_E) \big|^{(d-E-1)/2}} \int \prod_{n=1}^N dz_n \frac{u_{\text{std}}(z_1,\ldots,z_N)}{z_1^{a_1} \cdots z_N^{a_N}} \, ,
\label{eq:baikov_gen}
\end{equation}
where the prefactor is
\begin{equation}
C_{L,E} = \frac{e^{\epsilon \gamma_E L} \, \pi^{-L(L-1)/4-LE/2}}{\prod_{i=1}^L \Gamma\big((d-M+i)/2\big)} \, ,
\end{equation}
and the $u_{\text{std}}$ function takes the form
\begin{equation}
u_{\text{std}}(z_1,\ldots,z_N) \equiv \big| P_{L,E}(z_1-f_1,\ldots,z_N-f_N) \big|^{(d-M-1)/2} \, ,
\end{equation}
with the Baikov polynomial ($x_n \equiv z_n - f_n$)
\begin{equation}
P_{L,E}(x_1,\ldots,x_N) = G(q_1,\ldots,q_M)  \bigg|_{s_{ij}=A^n_{ij}x_n} \, .
\end{equation}

The integration domain for the Baikov variables can be deduce from Eq.~\eqref{eq:kiT2}. We need to require $G(q_i,\ldots,q_M)/G(q_{i+1},\ldots,q_M) \leq 0$ for each $i=1,\ldots,L$. The signs of individual Gram determinants can then be fixed according to the sign of the Gram determinant of the external momenta. These impose restrictions on the values of the Baikov variables. It is possible that the space of loop momenta is covered more than once when the variables are varied within this domain. In this case an extra normalization factor is required, which is however irrelevant to the purposes of this work.
Later on we will regard the variables as complex, and the integration in the real domain can be deformed into the complex space. To do that we need to firstly rewrite the absolute value of the Gram determinants as $\pm G(q_i,\ldots,q_M)$ according to their signs. We will often suppress these $\pm$'s when they are not important, but they should be kept in mind when considering the integration domain.

\subsection{An explicit example}

Usually one would not directly use the Baikov representation to calculate Feynman integrals, since other parameterizations are often more convenient in this respect. In this subsection we use a simple example to explicitly demonstrate how the Baikov representation works and how to deal with the integration domain which will prove to be important later. The example is the one-loop bubble integral given by
\begin{align}
    I(Q^2,\epsilon) &= e^{\epsilon\gamma_E} \int\frac{d^{d}k}{i\pi^{d/2}} \, \frac{1}{k^2}\frac{1}{(k+p)^2} = e^{\epsilon\gamma_E} \Gamma(\epsilon) \int_{0}^{1} dx \left[ Q^2 x(1-x) \right]^{-\epsilon} \nonumber
    \\
    &= e^{\epsilon\gamma_E} \left( Q^2 - i0 \right)^{-\epsilon} \frac{\Gamma^2(1-\epsilon) \, \Gamma(\epsilon)}{\Gamma(2-2\epsilon)} \, .
    \label{eq:FeynmanParam}
\end{align}
where $d=4-2\epsilon$ and $Q^2 \equiv -p^2 > 0$. We have suppressed the Feynman $+i0$ prescription until the last expression, which is important in the analytic continuation to the region $p^2 > 0$.

Now we follow the approach in the previous subsection to get the Baikov representation for the above integral. The Baikov variables are $z_1 = k^2$ and $z_2 = (k+p)^2$. The relevant Gram determinants are
\begin{align}
G(p) = p^2 = -Q^2 \, , \quad G(k, p) = k^2 \, p^2 - \left( k \cdot p \right)^2 = -\frac{1}{4} \left[ \left(z_1-z_2-Q^2\right)^2 + 4Q^2 z_1 \right] .
\end{align}
Since $G(p) < 0$, the integration domain is determined by $G(k,p) \geq 0$. We change variable to $u = (z_1 - z_2)/Q^2$, $v = z_1/Q^2$, and define the polynomial
\begin{equation}
P(u,v) = \frac{4G(k,p)}{Q^4} = -(u-1)^2 - 4v \geq 0 \, .
\end{equation}
The integration domain for $u$ and $v$ is then
\begin{equation}
u \in (-\infty,+\infty) \, , \quad v \in \left( -\infty, -\frac{(u-1)^2}{4} \right) .
\end{equation}

The Baikov representation can be written in the form
\begin{equation}
I(Q^2,\epsilon) = \mathcal{N}_\epsilon(Q^2) \, f(\epsilon) \, ,
\label{eq:baikov_explicit}
\end{equation}
where
\begin{align}
\mathcal{N}_\epsilon(Q^2) &= \frac{e^{\epsilon\gamma_E} \, \Gamma(1-\epsilon)}{2\pi \Gamma(2-2\epsilon)} \left(Q^2\right)^{-\epsilon} \, , \nonumber
\\
f(\epsilon) &= \int_{P \geq 0} du \, dv \, \frac{\left[ P(u,v) \right]^{1/2-\epsilon}}{v \, (v-u)} \, .
\end{align}
The integration over $v$ can be carried out using partial fraction, and we arrive at
\begin{equation}
f(\epsilon) = \frac{\pi}{\cos(\pi\epsilon)} \int_{-\infty}^{+\infty} \frac{du}{u} \left[ \left( (u+1)^2 \right)^{1/2-\epsilon} - \left( (u-1)^2 \right)^{1/2-\epsilon} \right] .
\end{equation}
Note that the integrand is not singular at $u = 0$ due to the cancellation between the two terms. However in practice, it is more convenient to perform the integration for each term separately, which then requires some extra regularization. We employ the analytic regulator $u^{-1} \to (u^2)^\delta u^{-1}$, and take the limit $\delta \to 0$ in the end. This gives
\begin{equation}
\int_{-\infty}^{+\infty} \frac{du}{u} \left( (u+1)^2 \right)^{1/2-\epsilon} = - \int_{-\infty}^{+\infty} \frac{du}{u} \left( (u-1)^2 \right)^{1/2-\epsilon} = \cos(\pi\epsilon) \, \Gamma(\epsilon) \, \Gamma(1-\epsilon) \, .
\end{equation}
Hence we have
\begin{equation}
f(\epsilon) = 2\pi \, \Gamma(\epsilon) \, \Gamma(1-\epsilon) \, .
\end{equation}
Plugging the above back to Eq.~\eqref{eq:baikov_explicit}, we find a result in agreement with that from Feynman parameterization \eqref{eq:FeynmanParam}.

\subsection{The (generalized) loop-by-loop Baikov representation}
\label{sec:genlbl}

The standard Baikov representation \eqref{eq:baikov_gen} works generically for multi-loop integrals. On the other hand, for $L > 1$ the number of positive $a_i$'s in a given integral is often smaller than $N$. Those $z_i$'s with zero or negative powers are called irreducible scalar products (ISPs). They may not directly appear in the corresponding Feynman integrals (or they may appear as numerators), but is necessary for a unique definition of the integral family, and is also necessary for the construction of the standard Baikov representation. We start with one of the ISPs, which, without loss of generality, is taken to be $z_N$. That is, we assume $a_N \leq 0$ in the following discussion. Starting from the standard representation, it is possible to integrate out $z_N$ to arrive at a different, but equivalent representation of the same integral.

If $a_N = 0$, $z_N$ only appears in the polynomial $P_{L,E}$, and hence it is often straightforward to integrate over it. The same practice may be carried out for other ISPs as well. The resulting representation is equivalent to Eq.~\eqref{eq:baikov_gen}, but with fewer integration variables. This representation is the same as the so-called loop-by-loop (LBL) Baikov representation if the same set of Baikov variables are chosen in the latter. In the loop-by-loop approach, one performs the change of variables for a single loop momentum at a time, treating the others as external. 

\begin{figure}[t!]
    \centering
    \includegraphics[width=0.6\textwidth]{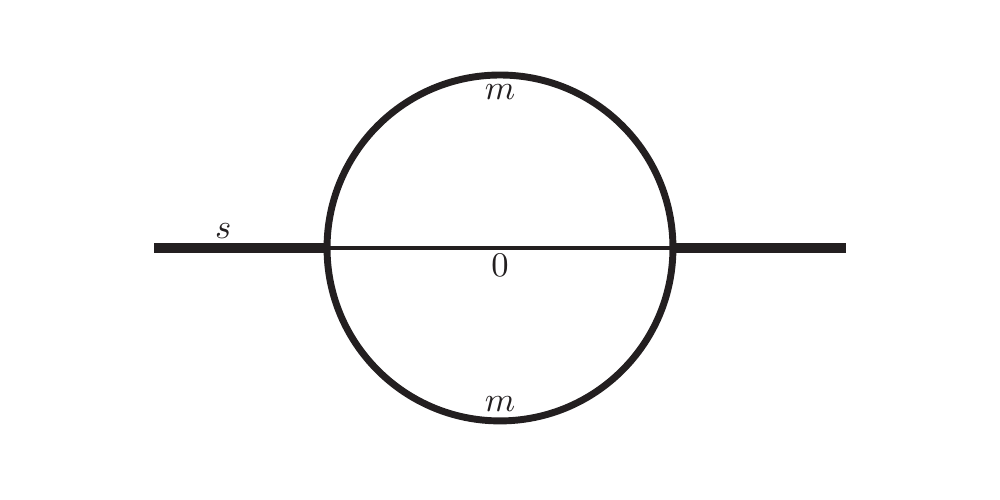}
    \caption{\label{fig:equalmasssr}The sunrise diagram with two equal-mass propagators and one massless propagator.}
\end{figure}

We take the sunrise integral family as an example. The diagram is shown in Fig.~\ref{fig:equalmasssr}. The integral family is defined by the propagator denominators
\begin{equation}
\{z_1 = k_1^2-m^2, \, z_2 = (k_1-k_2)^2, \, z_3 = (k_2-p)^2-m^2, \, z_4 = k_2^2-m^2, \, z_5 = (k_1-p)^2-m^2\} \, ,
\label{eq:sunrise_propagators}
\end{equation}
where $p^2 = s \neq 0$. Suppose that we are interested in integrals where only the first three propagators appear, namely, $F_{a_1,a_2,a_3,0,0}$. In the standard Baikov representation, we still need to include the last two denominators as ISPs. On the other hand, in the loop-by-loop approach, as the first step we perform the change of variables from $k_1^\mu$ to $z_1$ and $z_2$, treating $k_2$ as an external momentum. In the second step we perform the variable change from $k_2^\mu$ to $z_3$ and $z_4$. Here, the variable $z_5$ does not appear in the representation, and only one ISP, $z_4$, is needed. The resulting representation reads
\begin{align}
F_{a_1,a_2,a_3,0,0} &\propto \int dz_1 dz_2 \int \frac{d^d k_2}{i \pi^{d/2}} \frac{[G(k_1,k_2)]^{(d-3)/2}}{[G(k_2)]^{(d-2)/2}} \, \frac{1}{z_1^{a_1} \, z_2^{a_2} \, z_3^{a_3}} \nonumber
\\
&\propto \int dz_1 dz_2 dz_3 dz_4 \, u_{\text{LBL}}(z_1,z_2,z_3,z_4) \, \frac{1}{z_1^{a_1} \, z_2^{a_2} \, z_3^{a_3}} \, ,
\label{eq:lbl}
\end{align}
where we have omitted some constant prefactors, and the function
\begin{equation}
u_{\text{LBL}}(z_1,z_2,z_3,z_4)= [G(k_2)]^{-1+\epsilon} \, [G(k_1,k_2)]^{1/2-\epsilon} \, [G(k_2,p)]^{1/2-\epsilon} \, .
\label{eq:u_LBL_sunrise}
\end{equation}

Apparently, the LBL representation \eqref{eq:lbl} can be straightforwardly applied to integrals with a non-zero $a_4$. On the other hand, it fails to capture those integrals with a non-zero $a_5$ (even if $z_5$ appears only in the numerator of the integrand, i.e., $a_5 < 0$).\footnote{It should be noted that had we started from $k_2$ in the first step, we would end up with an alternative loop-by-loop representation in terms of the Baikov variables $z_1$, $z_2$, $z_3$ and $z_5$. This can be used to represent integrals with a non-zero $a_5$, but not those with a non-zero $a_4$. In any case, the conventional LBL approach cannot reduce the number of integration variables if both $a_4$ and $a_5$ are non-zero.} 
The problem is that when we change variables from $k_1^\mu$ to the Baikov variables, we have to include $z_5$ since the integrand depends on $k_1 \cdot p$. As a result, we will end up with the standard Baikov representation following this approach. In this case, it is then useful to consider the LBL representation as the result of performing the integration over $z_5$ in the standard Baikov representation. From this viewpoint, it is possible to start from the standard Baikov representation with $z_5$ in the numerator, integrate out $z_5$, and arrive at a new representation without $z_5$. More generically, we consider a Feynman integral where $z_N$ only appears in the numerator (i.e., $a_N \leq 0$). We construct its standard Baikov representation with the Baikov polynomial $P(z_1,\ldots,z_N)$. We consider $P(z_1,\ldots,z_N)$ as a quadratic polynomial of $z_N$ while treating other variables as constants: $P(z_N) = -A_N z_N^2 + B_N z_N - C_N$, with $A_N$, $B_N$ and $C_N$ being polynomials of $\bm{z} \equiv \{z_1,\ldots,z_{N-1}\}$. The two roots of the polynomial are given by:
\begin{equation}
r_{\pm} = \frac{B_N \pm \sqrt{B_N^2 - 4A_N C_N}}{2A_N} \, .
\end{equation}
The integration over $z_N$ then gives
\begin{align}
F_{a_1,\ldots,a_N} &\propto \int d^{N-1}\bm{z} \, z_1^{-a_1} \cdots z_{N-1}^{-a_{N-1}} \int_{r_-}^{r_+} dz_N \, z_N^{-a_N} \left[ P(z_N) \right]^\gamma \nonumber
\\
&\propto \int d^{N-1}\bm{z} \, z_1^{-a_1} \cdots z_{N-1}^{-a_{N-1}} \, A_N^{-1-\gamma} \left( B_N^2 - 4A_N C_N \right)^{1/2+\gamma} \nonumber
\\
&\hspace{8em} \times \left( r_- \right)^{-a_N} \, {}_2F_1 \left( a_N, 1+\gamma, 2+2\gamma, 1-\frac{r_+}{r_-} \right) ,
\label{eq:gen_lbl}
\end{align}
where $\gamma$ is a parameter depending on $\epsilon$.
Since $a_N \leq 0$, the hypergeometric function in the above is in fact a polynomial of its argument:
\begin{equation}
{}_2F_1 \left( a_N, 1+\gamma, 2+2\gamma, 1-\frac{r_+}{r_-} \right) = \sum_{n=0}^{-a_N} (-1)^n \binom{-a_N}{n} \frac{\Gamma(1+\gamma+n) \, \Gamma(2+2\gamma)}{\Gamma(1+\gamma) \, \Gamma(2+2\gamma+n)} \left( 1 - \frac{r_+}{r_-} \right)^n .
\end{equation}

If $a_N = 0$, Eq.~\eqref{eq:gen_lbl} simply reduces to the conventional loop-by-loop representation.
The more interesting cases are those with $a_N < 0$. They describe integrals with $z_N$ in the numerator, albeit $z_N$ does not appear in the final integrand. For illustration purposes, we consider again the sunrise family with $a_5=-1$. The standard Baikov polynomial is $P(z_5) = G(k_1,k_2,p)$. The coefficient of $-z_5^2$ in $P(z_5)$ can be easily seen to be $A_5 = G(k_2)/4$, while the discriminant of $P(z_5)$ can be shown to be\footnote{In this simple case, these relations can be easily deduced by brute-force expansion of the Gram determinants. We will give more generalized relations of this kind in later sections.}
\begin{equation*}
B_5^2 - 4 A_5 C_5 = G(k_1,k_2) \, G(k_2,p) \, .
\end{equation*}
Eq.~\eqref{eq:gen_lbl} in this case then gives
\begin{align*}
&\int \frac{dz_1 dz_2 dz_3 dz_4}{z_1^{a_1} z_2^{a_2} z_3^{a_3} z_{4}^{a_{4}}} \int_{r_-}^{r_+} dz_5 \, z_5 \left[ P(z_5) \right]^{-\epsilon}
\\
&\propto \int \frac{dz_1 dz_2 dz_3 dz_4}{z_1^{a_1} z_2^{a_2} z_3^{a_3} z_{4}^{a_{4}}} \, A_5^{-1+\epsilon} \left( B_5^2 - 4A_5 C_5 \right)^{1/2-\epsilon} \frac{B_{5}}{A_{5}}
\\
&\propto \int \frac{dz_1 dz_2 dz_3 dz_4}{z_1^{a_1} z_2^{a_2} z_3^{a_3} z_{4}^{a_{4}}} 
\, u_{\text{LBL}}(z_1,z_2,z_3,z_4) \, \frac{1}{G(k_2)} \frac{\partial G(k_1,k_2,p)}{\partial z_5} \bigg|_{z_5=0} \, ,
\end{align*}
where we have used the fact that $B_5$ is the coefficient of $z_5$ in $G(k_1,k_2,p)$, and $u_{\text{LBL}}(z_1,z_2,z_3,z_4)$ is the same as Eq.~\eqref{eq:u_LBL_sunrise}.
For example, the integral $F_{1,1,1,0,-1}$ in the sunrise family can be represented by
\begin{equation}
F_{1,1,1,0,-1} \propto \int dz_1 dz_2 dz_3 dz_4 \, u_{\text{LBL}}(z_1,z_2,z_3,z_4) \frac{1}{z_1 z_2 z_3 \, G(k_2)} \, \frac{\partial G(k_1,k_2,p)}{\partial z_5} \bigg|_{z_5=0} \, .
\end{equation}

An important fact about the above representation is that certain polynomials of $\{z_i\}$ (e.g., $G(k_2)=z_4+m^2$ in the above example) can appear in the denominators of the integrands. In generic situations where more than one ISPs are integrated out, more than one polynomials may appear in the denominators. These polynomials are factors of the $u_{\text{LBL}}$ function. From the loop-by-loop approach described below Eq.~\eqref{eq:sunrise_propagators}, one can see that at $L$ loops there are $m=2L-1$ such polynomial factors. We denote them as $P_1,\ldots,P_m$. We will then refer to integrals of the form
\begin{equation}
\int \prod_{i=1}^n dz_i \frac{u_{\text{LBL}}(z_1,\ldots,z_n)}{z_1^{a_1} \cdots z_n^{a_n} \, P_1^{b_1} \cdots P_m^{b_m}}  \, ,
\label{eq:gen_lbl_Pm}
\end{equation}
as \emph{generalized} loop-by-loop Baikov integrals, where the variables $z_1,\ldots,z_n$ are those not integrated out. A Feynman integral in this \emph{generalized} loop-by-loop representation is written as a linear combination of integrals with the above form.

As we will see later, the introduction of polynomials in the denominators greatly broadens the possible forms of the integrands among which we will search for $d\log$ ones. That said, it is also clear that the polynomial denominators cannot appear arbitrarily, but must be accompanied by a suitable numerator. Otherwise it is possible that the expression does not correspond to (a combination of) Feynman integrals.\footnote{This fact has also been observed in \cite{Dlapa:2021qsl}, where suitable combinations of generalized LBL integrals are treated as Feynman integrals in shifted spacetime dimensions.} We will come back to this point later from the viewpoint of the intersection theory.

\subsection{Cuts of integrals in the Baikov representation}
\label{sec:baikov_cut}

It is often useful to consider cuts of integrals in the Baikov representation \cite{Frellesvig:2017aai}. Cutting a propagator variable $z_i$ amounts to localize its integration contour around the point $z_i = 0$. For example, consider cutting the variables $z_1,\ldots,z_r$ in a (standard or generalized loop-by-loop) Baikov representation. The result is given by:
\begin{equation}
\int \prod_{j=r+1}^n dz_j \prod_{i=1}^r \oint_{z_i=0} dz_i \, \frac{u(z_1,\ldots,z_n)}{z_1^{a_1} \cdots z_n^{a_n} \, P_1^{b_1} \cdots P_m^{b_m}} \, .
\end{equation}
Apparently, cutting a variable $z_i$ is equivalent to taking the residue of the integrand at $z_i = 0$.

Cut Baikov integrals are useful due to the fact that they satisfy the same IBP relations and the same differential equations as the uncut ones. Let's take again the sunrise integral family as an example. For simplicity we consider cases with $a_4=a_5=0$, and omit them from the subscripts. Any integral in this family $F_{a_1,a_2,a_3}$ can be expressed as a linear combination of three master integrals, chosen as $F_{1,1,1}$, $F_{1,1,2}$ and $F_{1,0,1}$:
\begin{equation}
    F_{a_1,a_2,a_3} = c_1 F_{1,1,1} + c_2 F_{1,1,2} + c_3 F_{1,0,1} \, .
\end{equation}
We can now take the maximal cut (i.e., cutting $z_1$, $z_2$ and $z_3$) on both sides of the above equality. Note that cutting $z_2$ on $F_{1,0,1}$ leads to a vanishing result. Therefore, we have the relation
\begin{equation}
    F_{a_1,a_2,a_3} \big|_{\text{$3$-cut}} = c_1 F_{1,1,1} \big|_{\text{$3$-cut}} + c_2 F_{1,1,2} \big|_{\text{$3$-cut}} \, .
\end{equation}
Determining the coefficients $c_1$ and $c_2$ from the cut-version of the IBP relations is simpler than solving the full IBP relations. The same is true when using the intersection theory to calculate the coefficients. The simplification is much more pronounced in more complicated situations. Note however, after taking the cuts, we lose the information about $c_3$ completely, which can be recovered in the next step by loosing the cuts.

From the definition of the cut, it is clear that if the power $a_i$ is non-positive, cutting $z_i$ will lead to a vanishing result. On the other hand, if $a_i > 0$, the $z_i$-cut integral is usually non-zero. This property is often used to select integrals belonging to a particular sector. However, one should be careful with some exceptions to the above rule, especially when cutting multiple variables. It is possible that when taking several variables to zero, the function $u(z_1,\ldots,z_n)$ vanishes. Since the $u$ function consists of polynomials raised to non-integer powers, this means that all its derivatives also vanish in this limit. In this case, even if all the $a_i$'s are positive, the cut integral still vanishes. This does not necessarily mean that this sector is reducible, but is just an accidental fact of this particular representation.
There exist other exceptional cases where a cut on variables in the denominator could lead to a vanishing result. It is possible that localizing the variables to zero may force the integration over the remaining variables to be scaleless, or the integrand may become a total derivative. In all the above situations, if one still wants to study this particular cut, an alternative representation has to be used. We will see examples in later sections.

\section{The intersection theory of Baikov and Feynman integrals}
\label{sec:intersection}

From the discussions in the previous section, it is clear that we sometimes need to consider integrals in the generalized LBL representation, where polynomials of Baikov variables may appear in the denominator of the integrand. We will need to convert them to linear combinations of Feynman integrals appearing in scattering amplitudes. This can be achieved via generalized IBP relations \cite{Dlapa:2021qsl} or via the method of intersection theory \cite{Frellesvig:2019kgj, Mizera:2019gea, Frellesvig:2019uqt, Mizera:2019vvs, Weinzierl:2020xyy, Frellesvig:2020qot}. In this section, we briefly introduce the concept of intersection theory in the context of Baikov and Feynman integrals. For a more detailed explanation, we refer the readers to the original literature.

We will be dealing with Aomoto-Gelfand general hypergeometric functions \cite{aomoto2011theory} which can be defined via integrals of the form
\begin{equation}
\label{eq:hypergeometricfunction}
I[\varphi] = \int_{\mathcal{C}} u(\bm{z}) \varphi(\bm{z}) \, ,
\end{equation}
where $\varphi(\bm{z})$ is a single-valued differential $n$-form on an $n$-dimensional manifold, and $u(\bm{z})$ is a multi-valued function which vanishes on the boundary $\partial\mathcal{C}$ of the integration domain $\mathcal{C}$. It is required that $\varphi(\bm{z})$ can only be singular on the boundary $\partial\mathcal{C}$, where the singularity is regularized by the vanishing $u(\bm{z})$.
We will often work with a particular coordinate system. In that case the point $\bm{z}$ is parametrized by $n$ variables $\{z_1,z_2,...,z_n\}$, and the $n$-form can be written as $\varphi(\bm{z})=\hat{\varphi}(\bm{z}) d^{n}\bm{z}$, where $\hat{\varphi}(\bm{z})$ is a single-valued function and $d^n\bm{z} = dz_1 \wedge \cdots \wedge dz_n$.

We are interested in the relations among integrals with a given $u(\bm{z})$ and a given $\mathcal{C}$. It is clear that different $\varphi$'s may give rise to the same integral due to the IBP identity:
\begin{equation}
0 = \int_{\mathcal{C}} d \left( u(\bm{z}) \xi(\bm{z}) \right) = \int_{\mathcal{C}} u(\bm{z}) \left( d + \omega \wedge \right) \xi(\bm{z}) \equiv \int_{\mathcal{C}} u(\bm{z}) \nabla_{\omega} \xi(\bm{z}) \, ,
\end{equation}
where $\xi(\bm{z})$ is a differential $(n-1)$-form, $\omega \equiv d\log u(\bm{z})$ is a $1$-form, and $\nabla_\omega \equiv d + \omega \wedge$ is a covariant derivative with $\omega$ as the connection. It follows that for a given $\varphi$ and an arbitrary $\xi$, the following relation holds:
\begin{equation}
I[\varphi] = I[\varphi+\nabla_{\omega}\xi] \, .
\end{equation}
The above identity can be understood as an equivalence relation between the two $n$-forms:
\begin{equation}
\varphi \sim \varphi+\nabla_{\omega}\xi \, .
\end{equation}
We collect all $n$-forms equivalent to $\varphi$ into an equivalence class denoted as a bra $\bra{\varphi}$, which is also called a twisted cocycle. The set of all twisted cocycles forms a vector space called the $n$th twisted cohomology group $H^n_\omega$ with respect to the connection $\omega$.

It is easy to see that the generalized LBL Baikov representation introduced in the last section is a special case of general hypergeometric functions. The $u(\bm{z})$ function corresponds to the $u_{\text{LBL}}$ function consisting of Gram determinants raised to non-integer powers:
\begin{equation}
\label{eq:u_as_P}
u(\bm{z}) = \left[ P_1(\bm{z}) \right]^{\gamma_1} \cdots \left[ P_m(\bm{z}) \right]^{\gamma_m} \, .
\end{equation}
The $n$-forms $\varphi(\bm{z})$ are linear combinations of the building blocks
\begin{equation}
\frac{dz_1 \wedge \cdots \wedge dz_n}{z_1^{a_1} \cdots z_n^{a_n} \, P_1^{b_1} \cdots P_m^{b_m}} \, ,
\end{equation}
The non-integer power $\gamma_i$ serves as a regulator for the possible singularity of $\varphi(\bm{z})$ when $P_i \to 0$. On the other hand, the singularity at $z_i \to 0$ is not regularized by $u(\bm{z})$. Therefore, it is necessary to multiply $u(\bm{z})$ by an extra factor $z_i^{\rho_i}$ for each $a_i > 0$ in order to satisfy the requirement of general hypergeometric functions. One takes the limit $\rho_i \to 0$ at the end of calculations.

The dimension $\nu$ of the twisted cohomology group $H^n_\omega$ counts the number of independent integrals of the form \eqref{eq:hypergeometricfunction}. It can be computed by counting the number of proper critical points \cite{Lee:2013hzt, Bitoun:2018afx, Frellesvig:2019kgj, Frellesvig:2020qot, Mizera:2020wdt}.\footnote{We assume that all critical points are non-degenerate and isolated.} A critical point is a solution to the set of equations\footnote{The powers $\{\gamma_i\}$ in the $u(\bm{z})$ function are assumed to be generic non-integers, e.g., containing the dimensional regulator $\epsilon$. Otherwise the number of solutions could be smaller than the actual number of independent integrals. In this case, it is necessary to add an extra regulator for these $\gamma_i$'s, and take the regulators to zero in the last step.}
\begin{equation}
\omega_{i} \equiv \partial_{z_i} \log u(\bm{z}) = 0 \, , \quad (i = 1,\ldots,n) \, .
\end{equation}
Given the form of the $u(\bm{z})$ function in Eq.~\eqref{eq:u_as_P}, the equations can be recasted to
\begin{align}
\beta_i(\bm{z}) &= 0 \, , \quad (i=1,\ldots,n) \, , \nonumber
\\
P_j(\bm{z}) &\neq 0 \, , \quad (j=1,\ldots,m) \, ,
\end{align}
where
\begin{align}
\beta_i(\bm{z}) \equiv \sum_{j=1}^{m} \partial_{z_i}P_{j}(\bm{z})\prod_{k\neq j}P_{k}(\bm{z}) \, .
\end{align}
We introduce an additional variable $z_0$ and define the polynomial
\begin{equation}
\beta_{n+1}(z_0,\bm{z}) \equiv z_0 \prod_{j=1}^{m}P_{j}(\bm{z}) - 1 \, .
\end{equation}
The conditions $P_j(\bm{z}) \neq 0$ can then be imposed by asking for a solution of $z_0$ to the equation $\beta_{n+1}(z_0,\bm{z}) = 0$. The number of solutions to the set of equations $\beta_i = 0, (i=1,\ldots,n+1)$ is equal to the dimension of the quotient ring
\begin{equation}
\mathbb{C}[z_0,z_1,\ldots,z_n] / \mathcal{I} \, ,
\end{equation}
where $\mathcal{I}$ is the ideal generated by the polynomials $\{ \beta_i \}$, i.e.,
\begin{equation}
\mathcal{I} = \Braket{ \beta_{1},\ldots,\beta_{n},\beta_{n+1}} .
\end{equation}
The dimension of the quotient ring can be obtained using methods from computational algebraic geometry.

When working with generalized LBL representations, it is often the case where the dimension $\nu$ is different from the number of independent Feynman integrals found by reduction programs. The dimension $\nu$ can be larger than the number of independent integrals if there exist certain symmetry relations among the integrals which are not captured by the IBP relations (but are considered by reduction programs). This is apparently harmless since these symmetries can be easily incorporated later. 
After taking into account the symmetry relations, it is still possible that $\nu$ is larger than the number of independent Feynman integrals. This leads us to conclude that, certain integrals of the form \eqref{eq:hypergeometricfunction} actually do not correspond to Feynman integrals, as we have already mentioned in the previous section. Therefore, the space of Feynman integrals can be regarded as a subspace of the vector space $H^n_\omega$. We will refer to this subspace as the FI-subspace. It is our quest to identify the FI-subspace, and look for $d\log$-form integrands inside it.

Before considering the subspace, we briefly discuss how to work with $H^n_\omega$ using the intersection theory. Since $H^n_\omega$ is a vector space of dimension $\nu$, one may choose a basis of it consisting of vectors  $\bra{e_{1}}, \bra{e_{2}}, \ldots, \bra{e_{\nu}}$, such that any vector $\bra{\varphi} \in H^n_\omega$ can be expressed as a linear combination of the basis vectors:
\begin{equation}
\bra{\varphi} = c_1 \bra{e_{1}} + c_2 \bra{e_{2}} + \cdots + c_\nu \bra{e_{\nu}} \, .
\end{equation}
In the context of Feynman integrals, this gives the reduction of an integral as a linear combination of MIs. To calculate the coefficients $c_i$, one introduces the dual space of $H^n_\omega$, denoted as $(H^n_\omega)^*$. It turns out that $(H^n_\omega)^*$ is isomorphic to $H^n_{-\omega}$, i.e., the twisted cohomology group with respect to the connection $-\omega$. We denote a vector in $(H^n_\omega)^*$ as a ket $\ket{\varphi}$, which is the equivalence class
\begin{equation}
\ket{\varphi}: \varphi \sim \varphi - \nabla_{\omega}\xi \, .
\end{equation}
Between a bra $\bra{\varphi_L}$ and a ket $\ket{\varphi_R}$ one can define a bilinear pairing $\braket{\varphi_L | \varphi_R}$ called an \emph{intersection number} \cite{kita1994intersection, cho1995intersection, aomoto2011theory, yoshida2013hypergeometric, eisenbud20163264}.
This serves as an inner product between the vector space $H^n_\omega$ and its dual. With this, it is straightforward to compute the coefficients $c_i$ by first choosing a basis $\{\ket{h_{1}}, \ket{h_{2}}, \ldots, \ket{h_{\nu}}\}$ of the dual space $(H^n_\omega)^*$, and then use
\begin{equation}
c_i = \sum_{j=1}^\nu \braket{\varphi | h_j} \left( \bm{C}^{-1} \right)_{ji} \, ,
\end{equation}
where $\bm{C}$ is a $\nu \times \nu$ matrix with elements $\bm{C}_{ij} \equiv \braket{e_i | h_j}$. We will not discuss the computation of the intersection numbers in detail, but refer the interested readers to the original articles. It suffices to mention that, if both $e_i(\bm{z})$ and $h_j(\bm{z})$ are $d\log$-forms (which have only simple poles), the computation of $\braket{e_i | h_j}$ is greatly simplified. Therefore, having a $d\log$ basis not only simplifies the differential equations, but also helps the integral reduction using the intersection theory.

We now come back to the possible cases where \emph{not} all linear combinations of $\{\bra{e_i}\}$ correspond to Feynman integrals. In this case the dimension $\nu$ of $H^n_\omega$ is larger than the number $\nu_f$ of independent Feynman integrals. Equipped with the intersection theory, it is straightforward to identify the FI-subspace: one chooses a set of $\nu_f$ master Feynman integrals\footnote{This task can be accomplished using any suitable reduction method, e.g., momentum-space IBP, Baikov IBP, or intersection theory.}, and projects them onto the basis $\{\bra{e_i}\}$ using intersection theory. These $\nu_f$ linear combinations of $\{\bra{e_i}\}$ span a $\nu_f$-dimensional subspace, and we will look for $d\log$-form integrals inside this subspace.

Let's look at an example in the sunrise family introduced in the previous section. For simplicity, we consider cutting the two variables $z_1$ and $z_3$ (i.e., the two massive propagators) in the generalized LBL representation with $z_4$ as the ISP. We do not introduce the regulator for $z_2$, which means that $a_2$ can only be nonpositive. The integrals then take the form
\begin{equation}
\int u(\bm{z}) \varphi(\bm{z}) = \int u_{\text{cut}}(z_2,z_4) \, \frac{z_2^{-a_2} z_4^{-a_4} \, dz_2 \wedge dz_4}{\left[P_1(z_4)\right]^{b_1}  \left[P_2(z_2,z_4)\right]^{b_2}  \left[P_3(z_4)\right]^{b_3}} \, ,
\label{eq:sunrise_cut13}
\end{equation}
where
\begin{align}
u_{\text{cut}}(z_2,z_4) &= \left[P_1(z_4)\right]^{-1+\epsilon}  \left[P_2(z_2,z_4)\right]^{1/2-\epsilon}  \left[P_3(z_4)\right]^{1/2-\epsilon} \, , \nonumber
\\
P_1(z_4) &= G(k_2) \bigg|_{z_1=z_3=0} = z_4+m^2 \, , \nonumber
\\
P_2(z_2,z_4) &= G(k_1,k_2) \bigg|_{z_1=z_3=0} = z_2m^2 - \frac{1}{4} (z_2-z_4)^2 \, , \nonumber
\\
P_3(z_4) &= G(k_2,p) \bigg|_{z_1=z_3=0} = sm^2 - \frac{1}{4} (s-z_4)^2  \, .
\end{align}
It is easy to determine the dimension of the corresponding twisted cohomology group by computing the connection $\omega=d\log u_\text{cut}$ and counting the number of critical points which are solutions of $\omega = 0$. The result is $\nu = 2$, which means that there are two independent integrals of the form \eqref{eq:sunrise_cut13}. Indeed, we can choose a basis $\{\bra{e_1}, \bra{e_2}\}$ where $e_i = \hat{e}_i(z_2,z_4) \, dz_2 \wedge dz_4$ with
\begin{equation}
\hat{e}_1(z_2,z_4) = \frac{m^2}{P_2 P_3} \, , \quad \hat{e}_2(z_2,z_4) = \frac{z_4}{P_2 P_3} \, ,
\end{equation}
which can be shown to be independent.\footnote{The basis is of course not unique. We have made this choice for the sake of simplicity (both in the computation of intersection numbers and in the final expression \eqref{eq:sunrise_phi_e}).} On the other hand, the topology under consideration is just the product of two massive tadpoles. It is easy to see that there is only one independent Feynman integral in this sector, i.e., $\nu_f = 1$. We can arbitrarily choose a Feynman integral, e.g., $F_{1,0,1,0,0}$, whose corresponding $2$-form is simply $\varphi = dz_2 \wedge dz_4$. Computing the intersection numbers, we get
\begin{equation}
\bra{\varphi} = \frac{(1-2\epsilon)^2 s m^4}{4(1-\epsilon)^2} \left( \bra{e_1} + \bra{e_2} \right) .
\label{eq:sunrise_phi_e}
\end{equation}
Therefore, we conclude that Feynman integrals live in the $1$-dimensional subspace spanned by $\bra{e_1} + \bra{e_2}$.

\section{Constructing \texorpdfstring{$d\log$}{dlogform}-form integrals}
\label{sec:construction}

We now come to the construction of UT Feynman integrals satisfying canonical differential equations in a given integral family. The idea \cite{Chen:2020uyk} is very simple: we conjecture that each UT Feynman integral should admit a representation of a generalized $d\log$-form, i.e., can be written as
\begin{equation}
I[\varphi] = \mathcal{N}_\epsilon \int_{\mathcal{C}} u(\bm{z}) \varphi(\bm{z}) = \tilde{\mathcal{N}}_\epsilon \int_{\mathcal{C}} \left[ G(\bm{z}) \right]^\epsilon \bigwedge_{j=1}^n d\log f_j(\bm{z}) \, ,
\label{eq:Idlog}
\end{equation}
where $G(\bm{z})$ is a rational function and $f_j(\bm{z})$'s are algebraic functions of the Baikov variables. $\mathcal{N}_\epsilon$ is the prefactor arising from the Baikov representation, while $\tilde{\mathcal{N}}_\epsilon$ is a UT factor depending on $\epsilon$ and external variables (i.e., masses and scalar products of external momenta). In this section, we will ignore the factor $\tilde{\mathcal{N}}_\epsilon / \mathcal{N}_\epsilon$ which needs to be built into $\varphi(\bm{z})$. For applications in later sections, this factor can be easily deduced from the Gamma functions appearing in $\mathcal{N}_\epsilon$. With a slight abuse of notation, we will call $\varphi(\bm{z})$ a $d\log$ $n$-form, although it needs to be combined with some factors in $u(\bm{z})$ to be written as a $d\log$ integrand.

It should be noted that a UT integral can have many different representations, some of which are of the $d\log$-form while others are not. For example, a UT integral might be $d\log$ in the loop-by-loop Baikov representation, while the same is not true in the standard representation. For our purpose, it is sufficient to construct \emph{one} $d\log$ representation for each candidate of a UT integral. To do that, it is sometimes necessary to try different representations until an appropriate one is found.  

The $n$-form $\varphi(\bm{z})$ is a linear combination of the building blocks
\begin{equation}
\frac{dz_1 \wedge \cdots \wedge dz_n}{z_1^{a_1} \cdots z_n^{a_n} \, P_1^{b_1} \cdots P_m^{b_m}} \, ,
\label{eq:varphi_gen}
\end{equation}
where $P_1,\ldots,P_m$ are irreducible polynomial factors of $G(\bm{z})$ (and hence of $u(\bm{z})$). It should be emphasized that $\varphi(\bm{z})$ must be a single-valued differential $n$-form, whose denominator can only contain Baikov variables and the polynomial factors of $u(\bm{z})$. The $d\log$-form of Eq.~\eqref{eq:Idlog} puts further constraints on the properties of $\varphi(\bm{z})$. Most importantly, $u(\bm{z}) \varphi(\bm{z})$ can only have simple poles in all the variables. This requirement puts upper and lowers bounds on the powers $\{a_i\}$ and $\{b_i\}$ (note the poles at infinity). In the following, we first show a systematic way to construct such $d\log$ $n$-forms both in the univariate and the multivariate cases, and then discuss how to convert them to UT Feynman integrals.

\subsection{The univariate case}

We start with the cases where only one Baikov variable is involved in the integrals. This can happen when we consider the maximal cuts of many integrals. We refer to this variable simply as $z$, and the $u(z)$ function can always be factorized into the form\footnote{In this expression, we have dropped some possible minus signs for the $(z-c_i)$ factors. In this section we'll not be worried about these signs and the integration domain. They will be recovered for the examples in later sections.}.
\begin{equation}
u(z) = \frac{\mathcal{K}_1^\epsilon}{\mathcal{K}_0} \prod_{i=0}^{\nu} \left( z-c_i \right)^{-\gamma_i-\beta_i\epsilon} \, ,
\label{eq:uz_fac}
\end{equation}
where $\beta_i$ are integers and $\gamma_i$ can be either integers or half-integers; $\mathcal{K}_0$ is an algebraic function and $\mathcal{K}_1$ is a rational function of external variables, respectively. In the above expression, we assume that the roots $c_i$ are all distinct. The dimensionality of the cohomology group is then given by $\nu$.

If the number of half-integer $\gamma_i$'s in $u(z)$ is larger than two, this sector involves elliptic integrals or more complicated functional structures. In such case it is still possible to construct several $d\log$-form integrals, but one does not expect to have a complete canonical basis. If none of the $\gamma_i$'s is a half-integer, we can choose
\begin{equation}
\varphi_i(z) = \frac{\mathcal{K}_0 \, dz}{z-c_i} \prod_{j=0}^\nu (z-c_j)^{\gamma_j} \, , \quad (i = 0,\ldots,\nu) \, .
\label{eq:phi_zero_sqrt}
\end{equation}
This gives
\begin{equation}
u(z) \varphi_i(z) = \left( \mathcal{K}_1 \prod_{j=0}^\nu (z-c_j)^{-\beta_j} \right)^\epsilon \, d\log(z-c_i) \, ,
\label{eq:uphi_zero_sqrt}
\end{equation}
which takes the desired $d\log$-form \eqref{eq:Idlog}. Note that there are $\nu+1$ $1$-forms in the above, but only $\nu$ of them are independent.

If there's one half-integer $\gamma_i$, without loss of generality, we take it to be $\gamma_0$. We can perform the construction using the identity
\begin{equation}
d\log \frac{1 + \sqrt{\frac{c_0-z}{c_0-c}}}{1 - \sqrt{\frac{c_0-z}{c_0-c}}} = -\frac{\sqrt{c-c_0} \, dz}{(z-c) \sqrt{z-c_0}} \, ,
\end{equation}
for arbitrary $c \neq c_0$. Evidently we can choose
\begin{equation}
\varphi_i(z) = -\mathcal{K}_0 \, dz \, \frac{\sqrt{c_i-c_0}}{z-c_i} \, (z-c_0)^{\gamma_0-1/2} \prod_{j=1}^\nu (z-c_j)^{\gamma_j} \, , \quad (i=1,\ldots,\nu) \, ,
\label{eq:phi_one_sqrt}
\end{equation}
such that $u(z) \varphi_i(z)$ takes the $d\log$-form:
\begin{equation}
u(z) \varphi_i(z) = \left( \mathcal{K}_1 \prod_{j=0}^\nu (z-c_j)^{-\beta_j} \right)^\epsilon \, d\log \frac{1 + \sqrt{\frac{c_0-z}{c_0-c_i}}}{1 - \sqrt{\frac{c_0-z}{c_0-i}}} \, .
\label{eq:uphi_one_sqrt}
\end{equation}

Things are quite similar in the case of two half-integer $\gamma_i$'s. We take them to be $\gamma_0$ and $\gamma_1$. Here we employ the identities
\begin{align}\label{eq:dlogatomexpr}
d\log \frac{1+\sqrt{\frac{(c_0-z)}{(c_1-z)}}}{1-\sqrt{\frac{(c_0-z)}{(c_1-z)}}} &= \frac{dz}{\sqrt{(z-c_0)(z-c_1)}} \, , \nonumber
\\
d\log \frac{1+\sqrt{\frac{(c_1-c)(c_0-z)}{(c_0-c)(c_1-z)}}}{1-\sqrt{\frac{(c_1-c)(c_0-z)}{(c_0-c)(c_1-z)}}} &= -\frac{\sqrt{(c_0-c)(c_1-c)} \, dz}{(z-c)\sqrt{(z-c_0)(z-c_1)}} \, .
\end{align} 
We can then construct the following $1$-forms:
\begin{align}
\varphi_1(z) &= \mathcal{K}_0 \, dz \, (z-c_0)^{\gamma_0-1/2} (z-c_1)^{\gamma_1-1/2} \prod_{j=2}^\nu (z-c_j)^{\gamma_j} \, , \nonumber
\\
\varphi_i(z) &= \mathcal{K}_0 \, dz \, \frac{\sqrt{(c_0-c_i)(c_1-c_i)}}{z-c_i} \, (z-c_0)^{\gamma_0-1/2} (z-c_1)^{\gamma_1-1/2} \prod_{j=2}^\nu (z-c_j)^{\gamma_j} \, ,
\label{eq:phi_two_sqrt}
\end{align}
where $i=2,\ldots,\nu$. The corresponding $u(z) \varphi_i(z)$ again have forms similar to Eqs.~\eqref{eq:uphi_zero_sqrt} and \eqref{eq:uphi_one_sqrt}.

\subsection{The multivariate cases}

We now want generalize the above procedure to multivariate cases. Our approach is to perform the construction one-by-one for each variable. In the first step, we select a variable which allows us to apply the methodology of the previous subsection, while treating the other variables as ``external'' at the moment. We call this variable $z_1$.\footnote{The order of variables is sometimes important, and one may need to try different orders to arrive at a successful construction.} Using the univariate constructions, we can construct functions $\hat{\varphi}_i^{(1)}(\bm{z})$ such that
\begin{equation}
u(\bm{z}) \, \hat{\varphi}_i^{(1)}(\bm{z}) = \left[ G(\bm{z}) \right]^\epsilon \frac{\partial}{\partial z_1} \log f_i^{(1)}(\bm{z}) \, .
\end{equation}
We call the combination $u(\bm{z}) \hat{\varphi}_i^{(1)}(\bm{z}) dz_1$ as a partial-$d\log$-form integrand in $z_1$.
Here it should be noted that $z_1$ could be a propagator denominator instead of an ISP. In that case there is a regularization factor $z_1^\rho$ in $u(\bm{z})$, and $z_1$ itself should be regarded as one of the ``polynomial factors'' of $u(\bm{z})$. This means that one of the $c_i$'s in Eq.~\eqref{eq:uz_fac} is zero. Note that $\hat{\varphi}_i^{(1)}(\bm{z})$ is in general not a rational function, which is a problem to be dealt with later.

Given the above partial results, the next step is to pick a variable $z_2$ and repeat the procedure. Namely, we try to construct functions $\hat{\varphi}_{i,j}^{(2)}(\bm{z}')/\Lambda_i(\bm{z}')$ such that $u(\bm{z}) \hat{\varphi}_i^{(1)}(\bm{z}) dz_1 \wedge \hat{\varphi}_{i,j}^{(2)}(\bm{z}') dz_2 / \Lambda_i(\bm{z}') $ are partial-$d\log$-form integrands in the two variables $z_1$ and $z_2$, where $\bm{z}' = \{z_2,\ldots,z_n\}$. The algebraic functions $\Lambda_i(\bm{z}')$ are meant to cancel certain factors in $\hat{\varphi}_i^{(1)}(\bm{z})$, such that $\hat{\varphi}_i^{(1)}(\bm{z}) / \Lambda_i(\bm{z}')$ become rational functions. Such a recursive procedure, if succeeded, leads to $d\log$-form integrals we want, with the full $\hat{\varphi}(\bm{z})$ a rational function. There is, however, a few complications in the second step (and further steps). We will address them in the following.

\subsubsection{Square roots from the previous step}
\label{sec:square_roots}

First of all, in Eqs.~\eqref{eq:phi_zero_sqrt}, \eqref{eq:phi_one_sqrt} and \eqref{eq:phi_two_sqrt}, denominators of the form $(z-c_i)$ appear. These are allowed in the univariate case according to the generic form \eqref{eq:varphi_gen}, since $(z-c_i)$ \emph{is} a polynomial factor of $u(z)$.\footnote{Here ``polynomial'' regards the Baikov variables only. The coefficients can be algebraic functions of external variables.} However, this is problematic in the multivariate case, since $c_i$ is in general an algebraic function of the remaining Baikov variables $\bm{z}'$. Hence $(z_1-c_i(\bm{z}))$ may \emph{not} be a polynomial factor of the full $u(\bm{z})$, and cannot appear in the denominator alone. In this case it is necessary to make a linear combination of several terms, such that their common denominator becomes one of the polynomial factors $P_i(\bm{z})$ of $u(\bm{z})$. Fortunately, such linear combinations can be worked out rather generically, which depend on which of the formulas \eqref{eq:phi_zero_sqrt}, \eqref{eq:phi_one_sqrt} and \eqref{eq:phi_two_sqrt} was used in the previous step.

The simplest case is Eq.~\eqref{eq:phi_zero_sqrt}, where no square roots are involved. One can simply make linear combinations of the form
\begin{equation}
u(\bm{z}) \, \hat{\varphi}^{(1)}(\bm{z}) = \left[ G(\bm{z}) \right]^\epsilon \sum_{i} \frac{r_i}{z_1-c_i(\bm{z}')} \, ,
\label{eq:uphi_comb}
\end{equation}
with rational coefficients $r_i$. The sum in the above expression is over a subset of $\{0,\ldots,n\}$ such that $\mathcal{K}(\bm{z}') \prod_{i} (z_1-c_i(\bm{z}'))$ is an irreducible polynomial factor (say, $P(\bm{z})$) of $u(\bm{z})$, where $\mathcal{K}(\bm{z}')$ is a polynomial in $\bm{z}'$ (which is the coefficient of the highest power of $z_1$ in $P(\bm{z})$). The coefficients $r_i$ need to be chosen such that the numerator (after combining the denominators into $P(\bm{z})$) is either a rational function of $\bm{z}$, or the square root of a rational function of $\bm{z}'$. In the former case $\hat{\varphi}^{(1)}(\bm{z})$ is already single-valued with the correct denominator, and one can continue the construction for the remaining variables. In the latter case one needs to incorporate the square root in the next step to make the whole $\varphi(\bm{z})$ a single-valued differential form. In practice we most often encounter cases where $P(\bm{z})$ is quadratic in $z_1$, and it is straightforward to choose $r_i = \pm 1$. The two candidates are then simply given by the symmetric and anti-symmetric combinations:
\begin{align}
u(\bm{z}) \, \varphi_{+}(\bm{z}) &= \left[ G(\bm{z}) \right]^\epsilon \frac{1}{P(\bm{z})} \frac{\partial P(\bm{z})}{\partial z_1} dz_1 \wedge \hat{\varphi}'(\bm{z}') d^{n-1}\bm{z}' \, , \nonumber
\\
u(\bm{z}) \, \varphi_{-}(\bm{z}) &= \left[ G(\bm{z}) \right]^\epsilon \frac{\mathcal{K}(\bm{z}') \left[ c_i(\bm{z}') - c_j(\bm{z}') \right] dz_1}{P(\bm{z})} \wedge \frac{\hat{\varphi}'(\bm{z}') d^{n-1}\bm{z}'}{c_i(\bm{z}') - c_j(\bm{z}')} \, ,
\end{align}
where $c_i$ and $c_j$ are the two roots of $P(\bm{z})$ in $z_1$, and $\hat{\varphi}'(\bm{z}')$ is a rational function of $\bm{z}'$ which remains to be constructed.

We now turn to the case where the second line of Eq.~\eqref{eq:phi_two_sqrt} is used in the previous step of construction (if the first line is used, the situation is very simple). Here we have
\begin{equation}
u(\bm{z}) \, \hat{\varphi}_i^{(1)}(\bm{z}) = \left[ G(\bm{z}) \right]^\epsilon \frac{\sqrt{\left( c_0(\bm{z}')-c_i(\bm{z}') \right) \left( c_1(\bm{z}')-c_i(\bm{z}') \right)}}{\left( z_1-c_i(\bm{z}') \right) \sqrt{\left( z_1-c_0(\bm{z}') \right) \left( z_1-c_1(\bm{z}') \right)}} \, ,
\end{equation}
where $i=2,\ldots,\nu$. We again need to make linear combinations of the above to proceed with the next variable. We first note that $c_0$ and $c_1$ are the two roots of a quadratic polynomial $Q(z_1,\bm{z}')$ with respect to $z_1$, and therefore the square roots in the numerator and the denominator can be rescaled to $Q(c_i,\bm{z}')$ and $Q(z_1,\bm{z}')$, respectively. As before, we also assume that $c_i$ is a root of the irreducible polynomial factor $P(z_1,\bm{z}')$. If $c_i$ itself is a polynomial of $\bm{z}'$ (including the case where $c_i$ is a constant), $P(z_1,\bm{z}')$ is simply $z_1-c_i$. Hence we can readily write down the candidate
\begin{equation}
u(\bm{z}) \, \varphi(\bm{z}) = \left[ G(\bm{z}) \right]^\epsilon \frac{dz_1 \sqrt{Q(c_i,\bm{z}')}}{(z_1-c_i) \sqrt{Q(z_1,\bm{z}')}} \wedge \frac{\hat{\varphi}'(\bm{z}') d^{n-1}\bm{z}'}{\sqrt{Q(c_i,\bm{z}')}} \, .
\end{equation}
The construction can then be continued recursively. 

On the other hand, more generally $c_i$ is an algebraic function of $\bm{z}'$,  and $P(z_1,\bm{z}')$ is a non-linear polynomial of $z_1$. There seems to be no way to continue the construction with $\sqrt{Q(c_i,\bm{z}')}$ (square root inside a square root) in the denominator. Fortunately, very often it can be expressed in a simpler form due to relations among Gram determinants.\footnote{These relations have also been presented in the appendix of \cite{Dlapa:2021qsl}.}
To understand that, we first define a generalized Gram determinant with two sets of momenta:
\begin{equation}
G(\{q_1,\ldots,q_n\},\{l_1,\ldots,l_n\}) \equiv \det (q_i \cdot l_j)
	\equiv \det
	\begin{pmatrix}
		q_1 \cdot l_1 & q_1 \cdot l_2 & \cdots & q_1 \cdot l_n
		\\
		q_2 \cdot l_1 & q_2 \cdot l_2 & & \vdots
		\\
		\vdots & & \ddots & \vdots
		\\
		q_n \cdot l_1 & \cdots & \cdots & q_n \cdot l_n
	\end{pmatrix}
	\, .
\end{equation}
It then follows from Sylvester's determinant identity that
\begin{multline}
\big[ G(\{q_1,\ldots,q_n,k\},\{q_1,\ldots,q_n,q_{n+1}\}) \big]^2 = 
\\
G(k,q_1,\ldots,q_{n}) \, G(q_1,\ldots,q_{n+1}) - G(k,q_1,\ldots,q_{n+1}) \, G(q_1,\ldots,q_{n}) \, .
\label{eq:Sylvester}
\end{multline}

The above identity can be used in various ways. As an example (which often appears in practice), consider $Q(z_1,\bm{z}') = G(k,q_1,\ldots,q_{n+1})$ and $P(z_1,\bm{z}') = G(k,q_1,\ldots,q_n)$, where $z_1$ is one of the Baikov variables associated with the loop momentum $k$ (in the loop-by-loop sense). We introduce a short-hand notation for the polynomial
\begin{equation}
R(z_1,\bm{z}') \equiv G(\{q_1,\ldots,q_n,k\},\{q_1,\ldots,q_n,q_{n+1}\}) = - \frac{1}{2} \frac{\partial}{\partial (k \cdot q_{n+1})} G(k,q_1,\ldots,q_{n+1}) \, ,
\end{equation}
where the last equal sign follows from Jacobi's formula. 
Because $z_1 = c_i$ is a zero point of $P(z_1,\bm{z}')$, we immediately find that
\begin{equation}\label{eq:generalsqrtQ}
\sqrt{Q(c_i,\bm{z}')} = \frac{\pm R(c_i,\bm{z}')}{\sqrt{-G(q_1,\ldots,q_n)}} \, .
\end{equation}
Since $G(q_1,\ldots,q_n)$ is independent of $z_1$, the above expression is actually a polynomial of $c_i$. It is then possible to build linear combinations of the form
\begin{equation}
u(\bm{z}) \, \hat{\varphi}^{(1)}(\bm{z}) = \left[ G(\bm{z}) \right]^\epsilon \frac{1}{\sqrt{-G(q_1,\ldots,q_n)} \, \sqrt{Q(z_1,\bm{z}')} } \sum_{i} \frac{r_i \, R(c_i,\bm{z}')}{(z_1-c_i)} \, ,
\end{equation}
where the rational coefficients $r_i$ are chosen to satisfy conditions similar to the discussions below Eq.~\eqref{eq:uphi_comb}.
To see how such linear combinations can be found generically in the above situation, it is enough to consider $P(z_1,\bm{z})$ to be quadratic in $z_1$ (since $Q$ is quadratic and the degree of $P$ cannot exceed $Q$ in terms of $z_1$). We can then use the fact that $R(z_1,\bm{z}')$ is a linear function of $z_1$ to write
\begin{equation}\label{eq:generalR}
R(c_i,\bm{z}') = \left( 1 - (z_1-c_i) \frac{\partial}{\partial z_1} \right) R(z_1,\bm{z}') \, .
\end{equation}
Taking $r_i = \pm 1$, the two linear combinations are then given by
\begin{align}
u(\bm{z}) \, \hat{\varphi}_{+}^{(1)}(\bm{z}) &= \left[ G(\bm{z}) \right]^\epsilon \frac{1}{\sqrt{-G(q_1,\ldots,q_n)} \, \sqrt{Q(z_1,\bm{z}')} } \left( \frac{R(z_1,\bm{z}')}{P(\bm{z})} \frac{\partial P(\bm{z})}{\partial z_1} - 2\frac{\partial R(z_1,\bm{z}')}{\partial z_1} \right) , \nonumber
\\
u(\bm{z}) \, \hat{\varphi}_{-}^{(1)}(\bm{z}) &= \left[ G(\bm{z}) \right]^\epsilon \frac{1}{\sqrt{-G(q_1,\ldots,q_n)} \, \sqrt{Q(z_1,\bm{z}')} } \frac{\mathcal{K}(\bm{z}') \left[ c_i(\bm{z}') - c_j(\bm{z}') \right]}{P(\bm{z})} \, R(z_1,\bm{z}') \, ,
\end{align}
where the meaning of $\mathcal{K}$, $c_i$ and $c_j$ are the same as in the discussions below Eq.~\eqref{eq:uphi_comb}. Similar constructions can be applied for the case $P = G(k,q_1,\ldots,q_{n+1})$ and $Q = G(k,q_1,\ldots,q_n)$, which also appears in practice quite often. The considerations outlined above is also valid if Eq.~\eqref{eq:phi_one_sqrt} is used for the construction of $z_1$, and we will not go into details about that.

\subsubsection{Higher-degree polynomial with a half-integer power}

Unlike the univariate case, the appearance of higher-degree polynomials with half-integer powers does not necessarily mean that the result is elliptic. It is possible that a canonical basis is known to exist (from the maximal cuts of various sectors), but at some point in the construction procedure one encounters square root of a polynomial with degree higher than two for each remaining variable. Hence the construction cannot proceed straightforwardly. There are a couple of ways to circumvent this issue, and we will briefly discuss them in the following.

An apparent possibility is to perform a variable transformation such that the polynomial becomes quadratic in one of the new variables (which is similar in spirit to \cite{Henn:2020lye}). This can be illustrated by a simple example:
\begin{equation}
\int \left[ G(x,y) \right]^\epsilon \frac{\hat{\varphi}(x,y) \, dx \wedge dy}{\sqrt{(x^2+y^2-6xy)(x^2+y^2+2xy+2x+2y-2)}} \, .
\end{equation}
It is not easy to see how to construct the function $\hat{\varphi}(x,y)$ such that the above integrand becomes $d\log$. However, it is easy to find a change of variables $u=x+y$ and $v=x-y$, and the square root becomes
\begin{equation}
\frac{du \wedge dv}{2\sqrt{(2v^2-u^2)((u+1)^2-3)}} \, .
\end{equation}
It is now straightforward to perform the construction first in $v$ and then in $u$.

A different type of variable transformation is to ``rationalize'' part of the square root. Suppose that we have a square root of the polynomial $P(\bm{z})$. It is possible to find a rational change of variable $\bm{z} \to \tilde{\bm{z}}$, such that
\begin{equation}
P(\bm{z}) = R^2(\tilde{\bm{z}}) \, Q(\tilde{\bm{z}}) \, ,
\end{equation}
where $R$ is a rational function of the new variables $\tilde{\bm{z}}$, and $Q$ is a polynomial which is quadratic in some of the new variables. This possibility has also been used in \cite{Dlapa:2021qsl}. Note that for this single square-root, the rationalizing transformation can be found algorithmically when it exists \cite{Besier:2018jen, Besier:2019kco}.

Another possibility to avoid higher-degree polynomials is to perform the construction in reducible super-sectors. Given a sector with some propagator denominators and several ISPs, a super-sector is a sector where some of the ISPs are allowed to appear in the denominator. Sometimes a super-sector can be reducible. This means that all integrals in that super-sector can be expressed as linear combinations of integrals in lower sectors. We find that a canonical integral may have a very complicated form in its own sector (involving variable transformations as mentioned above), but is much simpler when expressed in the Baikov representation of a reducible super-sector. For this reason we usually look into the super-sectors first before attempting variable transformations.

In the beginning of this Section, we have emphasized that a UT integral can have many equivalent but different representations, and it is enough for us to find one representation that is $d\log$. In the above, we have searched only in the generalized LBL representations. It is sometimes useful to extend these representations by introducing an additional fold of integration over an extra variable in the intermediate steps of the construction. We will call them ``extended Baikov representations''. As will be demonstrated in a practical example later, this extra variable is not randomly chosen, but is often motivated by (but not the same as) the variables in the standard Baikov representations. In this way it is easy to show that such an extended integral is indeed equivalent to integrals (we will refer to their integrands as ``equivalently-$d\log$ integrands'') in the original representation. A benefit of such an extension is that higher-degree polynomials might disappear, leading to a successful construction. We will see an example of this method later in the top sector of outer-massive double box family.

\subsection{From $d\log$-forms to canonical Feynman integrals}

The generic procedure outlined in the previous subsections allows us to construct $d\log$ Baikov integrals for a given $u(\bm{z})$. On the other hand, the main goal of this Section is to construct UT Feynman integrals satisfying a canonical set of differential equations. In this subsection we show how to convert between these two in a systematic way.

As extensively discussed in Section~\ref{sec:Baikov} and \ref{sec:intersection}, Feynman integrals live in a subspace of the space of generalized Baikov integrals. It is hence easy to understand that the $d\log$ Baikov integrals are not necessarily expressible as linear combinations of Feynman integrals. As a result, we usually need to construct more $d\log$ Baikov integrals than the number of independent Feynman integrals. We call these extra ones \emph{auxiliary $d\log$ forms}. With them, we can make linear combinations belonging to the FI-subspace using the method outlined at the end of Section~\ref{sec:intersection}. We note that these combinations, after being put into a common denominator, generically take the form
\begin{equation}
\frac{N(\bm{z}) \, dz_1 \wedge \cdots \wedge dz_n}{z_1 \cdots z_n \, P_1^{b_1} \cdots P_m^{b_m}} \, ,
\end{equation}
where $b_i$ is either $0$ or $1$, and $N(\bm{z})$ is a polynomial in the numerator. This provides hint on which auxiliary $d\log$ forms can be combined together, and is very helpful in many cases.

To summarize this section, we list below the procedure for constructing UT Feynman integrals for a given integral family:
\begin{enumerate}

\item Starting from a top sector, use IBP programs such as \fire, \litered, \reduze or \kira to find all irreducible sectors containing master Feynman integrals (unique sectors in the notation of \litered), as well as the number of master Feynman integrals in each sector. Also compute the number of critical points to identify the dimension of the cohomology group for each sector, which corresponds to the number of independent Baikov integrals (here we also consider reducible sectors if necessary, which may provide simpler super-sector constructions as well as auxiliary $d\log$ forms).

\item Write down the generalized LBL Baikov representation for each sector, and apply the construction method to find enough $d\log$ Baikov integrals.

\item Identify linear combinations of $d\log$ Baikov integrals that belong to the FI-subspace, and transform them to Feynman integrals using either Baikov IBP, dimensional recurrence relations, or intersection theory.

\end{enumerate}
In the following sections we demonstrate this procedure in several non-trivial examples. More examples can be found in the appendices.

\section{Inner-massive double box}
\label{sec:intmassivedb}

\begin{figure}[t!]
    \centering
    \includegraphics[width=0.6\textwidth]{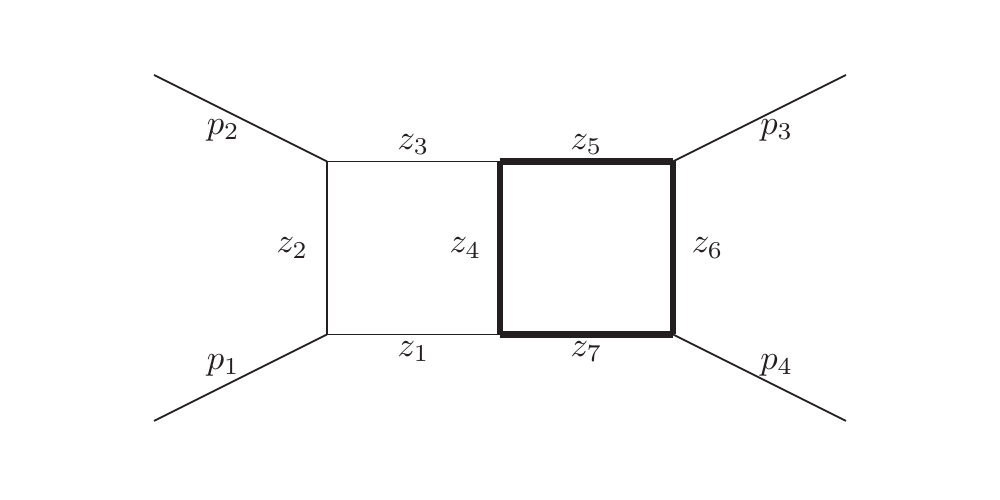}
    \caption{\label{fig:intmassivedb}Inner-massive double box family. Thick lines represent propagators with a mass $m$, while thin lines represent massless propagators.}
\end{figure}

As the first example, we consider the double box integral family where the propagators in an ``inner'' loop have the same mass $m$, while the other propagators as well as external legs are massless. We take all external momenta to be incoming. The propagator denominators and the relevant scalar products are given by
\begin{align}
\label{eq:intmassivedbplist}
&\{k_1^2,\,(k_1-p_1)^2,\,(k_1-p_1-p_2)^2,\,(k_1-k_2)^2-m^2,\,(k_2-p_1-p_2)^2-m^2, \nonumber
\\
&\;(k_2-p_1-p_2-p_3)^2-m^2,\,k_2^2-m^2,\,(k_2-p_1)^2-m^2,\,(k_1-p_1-p_2-p_3)^2\} \, , \nonumber
\\
&p_{i}^2=0 \, , \quad (p_1+p_2)^2=s \, , \quad (p_2+p_3)^2=t \, ,
\end{align}
where the last two propagator denominators appear as ISPs. The corresponding diagram is depicted in Fig.~\ref{fig:intmassivedb}.

This integral family has been considered in \cite{Becchetti:2017abb}. There are 20 unique sectors with 32 master integrals in total. The construction of $d\log$ forms in most sectors is straightforward. In the following, we discuss three representative sectors where special treatments are required. The complete results will be given in Appendix~\ref{app:imdballsector}.

\subsection{Sector \{1,1,0,1,1,1,0,0,0\}}

We construct the LBL Baikov representation for this sector with $z_7$ and $z_8$ as ISPs. The relevant ingredients are given by:
\begin{align}
    \mathcal{N}_{\epsilon}&=-\frac{e^{2\epsilon\gamma_{E}}}{4\pi^3\Gamma(1-2\epsilon)} \left[-st(s+t)\right]^{\epsilon} \, , \nonumber
    \\
    u(\bm{z})&=P_1^{-1/2+\epsilon} \, P_{2}^{-\epsilon} \, P_{3}^{-1/2-\epsilon} \, ,
    \label{eq:u110111}
\end{align}
where the three polynomials are
\begin{align}
    P_{1}(z_7,z_8) &= 4 \, G(k_2,p_1) = -(z_7-z_8)^2 \, , \nonumber
    \\
    P_{2}(z_1,z_2,z_4,z_7,z_8) &= 4 \, G(k_1,k_2,p_1) \, , \nonumber
    \\
    P_{3}(z_5,z_6,z_7,z_8) &= 16 \, G(k_2,p_1,p_2,p_3) \, .
    \label{eq:u110111_poly}
\end{align}
It is attempting to apply maximal cut to this representation, to count the dimension of the corresponding cohomology group. However, in this case the maximal cut does not work for this specific LBL representation. The reason is that $P_2$ equals to 0 when $z_1$ and $z_2$ are set to zero. Since $P_2$ comes with a power of $-\epsilon$, the cut integral is identically zero in dimensional regularization. This situation has been discussed in Section~\ref{sec:baikov_cut}, and we need to search for other representations to perform the counting.

We can try a different LBL Baikov representation with $z_3$, $z_7$ and $z_8$ as ISPs. Applying maximal cut, we have
\begin{equation}
    u_{\text{cut}}(\bm{z}) = P_{1,\text{cut}}^{\epsilon} \, P_{2,\text{cut}}^{-1/2-\epsilon} \, P_{3,\text{cut}}^{-1/2-\epsilon} \, ,
\end{equation}
where
\begin{align}
P_{1,\text{cut}}(z_7,z_8) &= z_8 (s-z_7+z_8) + s m^2 \, , \nonumber
\\
P_{2,\text{cut}}(z_3,z_7,z_8) &= \left[s z_8+z_3 \left(z_7-z_8\right)\right]^2 \, , \nonumber
\\
P_{3,\text{cut}}(z_7,z_8) &= 4m^2 s t(s + t) - (s z_8+t z_7-s t)^2 \, .
\end{align}
Note that the integration domain is determined by $P_{3,\text{cut}} \geq 0$ and $P_{2,\text{cut}}/P_{1,\text{cut}} \geq 0$. These conditions do not constrain $z_3$, which means that the integration range of $z_3$ is $(-\infty,+\infty)$. In dimensional regularization such kind of integrals vanish. Hence we still cannot study the maximal cut using this representation.

In the end, we find that we need to employ the LBL Baikov representation which keeps $z_3$, $z_7$ and $z_9$ as ISPs in order to apply the maximal cut. The result is
\begin{equation}
u_{\text{cut}}(\bm{z})=P_{1,\text{cut}}^{\epsilon} \, P_{2,\text{cut}}^{-1/2-\epsilon} \, P_{3,\text{cut}}^{-1/2-\epsilon} \, ,
\end{equation}
where
\begin{align}
P_{1,\text{cut}}(z_3,z_9)&=z_9 (s-z_3+z_9) \, , \nonumber
\\
P_{2,\text{cut}}(z_3,z_7,z_9)&=(z_7z_9-z_3z_7-sz_9)^2-4m^2 s z_9(s - z_3 + z_9) \, , \nonumber \\
P_{3,\text{cut}}(z_3,z_9)&=\left[ s (z_9-t)+t z_3 \right]^2 \, .
\end{align}
Counting the number of critical points we get the result $\nu = 2$, which coincides with the number of master integrals in this sector found by \kira and \reduze.

Now we need to construct two $d\log$-forms corresponding to Feynman integrals in this sector. This can be done in any representation that is convenient for the purpose, and we choose to work in the representation \eqref{eq:u110111}. Since this is the first practical example in this paper, we will demonstrate the construction procedure step-by-step. We first note that, for this sector we require all of $z_1$, $z_2$, $z_4$, $z_5$ and $z_6$ to appear in the denominator, otherwise the integrand likely belongs to a sub-sector. We also need to add regulators for these five variables in the context of intersection theory, such that the $u(\bm{z})$ function now becomes
\begin{equation}
u(\bm{z})= z_1^\rho z_2^\rho z_4^\rho z_5^\rho z_6^\rho \, P_1^{-1/2+\epsilon} \, P_{2}^{-\epsilon} \, P_{3}^{-1/2-\epsilon} \, ,
\end{equation}
where the polynomials $P_1$, $P_2$ and $P_3$ are given in Eq.~\eqref{eq:u110111_poly}, and the regulator $\rho$ will be taken to zero in the end.

Observing that $z_1$, $z_2$ and $z_4$ do not appear in the polynomials $P_1$ and $P_3$,\footnote{The polynomial $P_2$ is irrelevant here since it comes with a power of $-\epsilon$ in the $u(\bm{z})$ function.} the construction for them is straightforward according to Eq.~\eqref{eq:phi_zero_sqrt}. And we immediately know that the desired 7-form takes the form
\begin{equation}
    \frac{dz_1}{z_1} \wedge \frac{dz_2}{z_2} \wedge \frac{dz_4}{z_4} \wedge \varphi(z_5,z_6,z_7,z_8) \, .
\end{equation}
We now need to find a 4-form $\varphi(z_5,z_6,z_7,z_8)$ such that $\varphi(z_5,z_6,z_7,z_8) / \sqrt{P_1 P_3}$ is a $d\log$ form. Since $z_5$ and $z_6$ only appear in $P_3$ but not $P_1$, we choose to work with them first. For the readers' convenience, we recall that
\begin{align}
P_{1} &= -(z_7-z_8)^2 \, , \nonumber
\\
P_{3} &= (st-sz_8-tz_7+sz_6+tz_5)^2 - 4st \left[ tz_5+sz_6+(z_5-z_6)(z_7-z_8)+m^2(s+t) \right] .
\end{align}
$P_3$ is a quadratic polynomial of $z_5$, and we can apply the second equation in \eqref{eq:dlogatomexpr} where $c=0$. This give rise to the $d\log$ factor
\begin{equation}
\frac{\varphi(z_5,z_6,z_7,z_8)}{\sqrt{P_1} \sqrt{P_3}} = \frac{\sqrt{P_{3}(z_5=0)}}{z_5\sqrt{P_{3}}} \, dz_5 \wedge \cdots \, ,
\end{equation}
where the ellipsis denotes the part yet to be constructed.
We now need to take care of a factor of $1/\sqrt{P_{3}(z_5=0)}$ in the construction for the remaining variables. Note that $P_{3}(z_5=0)$ is again a quadratic polynomial of $z_6$, and hence we can apply the second equation in \eqref{eq:dlogatomexpr} here. This leads to
\begin{equation}
\frac{\varphi(z_5,z_6,z_7,z_8)}{\sqrt{P_1} \sqrt{P_3}} = \frac{\sqrt{P_{3}(z_5=0)}}{z_5\sqrt{P_{3}}} \, dz_5 \wedge \frac{\sqrt{P_{3}(z_5=0,z_6=0)}}{z_6\sqrt{P_{3}(z_5=0)}} \, dz_6 \wedge \cdots \, .
\end{equation}

We can now continue with the construction for $z_7$ and $z_8$. $P_{3}(z_5=0,z_6=0)$ can now be regarded as a quadratic polynomial of $z_7$, and $\sqrt{P_1}$ can be written as a constant factor multiplying $(z_7-z_8)$.\footnote{A constant factor does not affect the construction of $d\log$ integrands, and we will simply drop it.} We apply again the second equation of \eqref{eq:dlogatomexpr} with $c=z_8$, and obtain
\begin{multline}
\frac{\varphi(z_5,z_6,z_7,z_8)}{\sqrt{P_1} \sqrt{P_3}} = \frac{\sqrt{P_{3}(z_5=0)}}{z_5\sqrt{P_{3}}} \, dz_5 \wedge \frac{\sqrt{P_{3}(z_5=0,z_6=0)}}{z_6\sqrt{P_{3}(z_5=0)}} \, dz_6
\\
\wedge \frac{\sqrt{P_{3}(z_5=0,z_6=0,z_7=z_8)}}{(z_7-z_8)\sqrt{P_{3}(z_5=0,z_6=0)}} \, dz_7 \wedge \cdots \, .
\end{multline}
We are left with the last variable $z_8$, with
\begin{equation}
    P_{3}(z_5=0,z_6=0,z_7=z_8) = \left[ st-(s+t)z_8 \right]^2 - 4m^2st(s+t) \, .
\end{equation}
We now apply the first equation in \eqref{eq:dlogatomexpr} to arrive at
\begin{align}
\frac{\varphi(z_5,z_6,z_7,z_8)}{\sqrt{P_1} \sqrt{P_3}} = \cdots \wedge
\frac{s+t}{\sqrt{P_{3}(z_5=0,z_6=0,z_7=z_8)}} \, dz_8 \, ,
\end{align}
which is the final answer for a $d\log$-form integrand in this sector. The corresponding $\hat{\varphi}$ function is simply given by
\begin{equation}
    \hat{\varphi}_{9} = \frac{s+t}{z_1z_2z_4z_5z_6} \, ,
\end{equation}
where the numbering follows the list in Appendix~\ref{app:imdballsector}.

It is possible to construct the second $d\log$-form within this sector, albeit a bit tricky. It is much simpler to employ the reducible super-sector $\{1,1,0,1,1,1,1,0,0\}$. Namely, we allow $z_7$ to appear in the denominator of the generalized LBL Baikov representation, and add the necessary $z_7^\rho$ factor in the $u(\bm{z})$ function. All integrals in this super-sector can be reduced to the sector under consideration and its sub-sectors. It is now straightforward to construct the second $d\log$-form:
\begin{equation}
    \hat{\varphi}_{10}=\frac{\sqrt{st(st-4m^2(s+t))}}{z_1z_2z_4z_5z_6z_7} \, .
\end{equation}
It is clear that both $d\log$-forms correspond to Feynman integrals, and hence the construction for this sector completes.

As a final remark here, we note that the prefactor $\mathcal{N}_\epsilon$ in Eq.~\eqref{eq:u110111} is already a UT function with weight $-3$. Hence we can directly take $\tilde{\mathcal{N}}_\epsilon=\mathcal{N}_\epsilon$ in Eq.~\eqref{eq:Idlog}. We also note that the 7-fold integrations over $d\log$-forms leads to UT functions with weight $+7$, and the final results $\bra{\varphi_9}$ and $\bra{\varphi_{10}}$ are weight $+4$ functions. This should be kept in mind when constructing other sectors, since we would like to have a canonical basis with the same transcendental weight.

\subsection{Sector \{0,1,0,1,1,1,1,0,0\}}

We construct the LBL Baikov representation for this sector with $z_8$ as the ISP. To do that we first perform the momenta shifts $k_1\rightarrow k_1+p_1$ and $k_2\rightarrow k_2+p_1$ in the propagator denominators in Eq.~\eqref{eq:intmassivedbplist}. The resulting ingredients are:
\begin{align}
    \mathcal{N}_{\epsilon}&=\frac{1}{1-2\epsilon}\, \frac{e^{2\epsilon\gamma_{E}}\Gamma^2(-\epsilon)}{16\pi^3\Gamma^{2}(-2\epsilon)} \left[st(s+t)\right]^{\epsilon} \, , \nonumber \\
    u(\bm{z})&=(z_8+m^2)^{-1+\epsilon}P_{2}^{1/2-\epsilon}P_{3}^{-1/2-\epsilon} \, ,
\end{align}
where the two polynomials are
\begin{align}
    P_{2}(z_2,z_4,z_8)&=4 \, G(k_1,k_2) \, , \nonumber \\
    P_{3}(z_5,z_6,z_7,z_8)&=16 \, G(k_2,p_1,p_2,p_3) \, .
\end{align}
Under maximal cut there are three critical points, corresponding to three MIs in this sector. Here $\mathcal{N}_\epsilon$ is not a UT function due to the factor of $(1-2\epsilon)$. Taking that into account and performing the construction, we arrive at three $d\log$-forms:
\begin{align}
    &\hat{\varphi}'_{14}= \frac{(1-2 \epsilon ) \left(m^2+z_8\right) \sqrt{s t \left(s t-4 m^2 (s+t)\right)}}{\epsilon z_2 z_4 z_5 z_6 z_7 P_{2}} \, , \nonumber \\
    &\hat{\varphi}'_{15}= \frac{(1-2 \epsilon ) z_8\sqrt{s^2 \left(t-m^2\right)^2-4 m^2 s t^2}}{\epsilon z_2 z_4 z_5 z_6 z_7 P_{2}} \, , \nonumber \\
    &\hat{\varphi}'_{16}= \frac{sz_8 (1-2 \epsilon ) \left(m^2+z_8\right)}{\epsilon z_2 z_4 z_5 z_6 z_7 P_{2}} \, .
\end{align}
Note that the polynomial $P_2$ appears in the denominators, and hence it is not clear at first sight whether the above three correspond to Feynman integrals. 

Since we require that the constructed $d\log$-forms are Feynman integrals without any cuts, it is necessary to consider the equivalence classes of the full 6-forms without any cuts
\begin{equation}
\frac{z_8^{-a_8}}{z_2^{a_2} z_4^{a_4} z_5^{a_5} z_6^{a_6} z_7^{a_7} (z_8+m^2)^{b_1} P_2^{b_2} P_3^{b_3}} \, .
\label{eq:IMDB_010111100}
\end{equation}
Computing the number of critical points we get $\nu=20$ (and after taking into account a symmetry between $z_5$ and $z_7$, there are 18 independent integrals), while from an IBP reduction we know that there are only $\nu_f=12$ independent Feynman integrals in this sector including sub-sectors. It is therefore not surprising that some 6-forms like Eq.~\eqref{eq:IMDB_010111100} do not correspond to Feynman integrals.
Following the strategy outlined in Sec.~\ref{sec:intersection}, we could use the intersection theory to find the FI-subspace of the twisted cohomology group. In practice, the 6-fold intersection numbers are computationally heavy. However, observing that $P_2$ only depends on $z_2$, $z_4$ and $z_8$, we find it sufficient to consider integrals with cut on $z_5$, $z_6$ and $z_7$, such that only 3-fold intersection numbers are involved. The number of critical points is $\nu=5$ in this situation, and the number of master Feynman integrals is $\nu_f = 4$ (three in this sector and one in a sub-sector).

We choose the following basis for this 5-dimensional cohomology group:
\begin{align}
    \hat{e}_1 = \frac{1}{z_2z_4} \, , \quad \hat{e}_2 = \frac{1}{z_2z_4^2} \, , \quad  \hat{e}_3 = \frac{1}{z_2z_4^3} \, , \quad \hat{e}_4 = \frac{1}{z_4^3} \, , \quad \hat{e}_5 = \frac{1}{z_2 z_4 P_{2,\text{cut}}} \, .
\end{align}
It is clear that the first four vectors correspond to Feynman integrals (under the cut), while the last one does not (since it is linearly independent from the first four). To compute the intersection numbers, we need to multiply $u(\bm{z})$ by $z_2^\rho z_4^\rho$ as a regulator, and take the limit $\rho \to 0$ by the end of the calculation. Performing the decomposition, we have
\begin{align}
    \bra{\varphi'_{14,\text{cut}}} &=  \frac{\sqrt{s t \left(s t-4 m^2 (s+t)\right)}}{\epsilon} \left( \bra{e_2} + \frac{m^2}{\epsilon} \bra{e_3} \right) , \nonumber
    \\
    \bra{\varphi'_{15,\text{cut}}} &=  \frac{\sqrt{s^2 \left(t-m^2\right)^2-4 m^2 s t^2}}{\epsilon} \left( \bra{e_2} + \frac{m^2}{\epsilon} \bra{e_3} - (1-2\epsilon)m^2 \bra{e_5} \right) , \nonumber
    \\
    \bra{\varphi'_{16,\text{cut}}} &= -\frac{s}{\epsilon} \left( (1-2\epsilon) \bra{e_1} - m^2 \bra{e_2} + \frac{m^2}{\epsilon} \bra{e_4} \right) .
\end{align}
We can see that $\bra{\varphi'_{14,\text{cut}}}$ and $\bra{\varphi'_{16,\text{cut}}}$ have no components in $\bra{e_5}$, and hence they are candidates for canonical Feynman integrals. On the other hand, $\bra{\varphi'_{15,\text{cut}}}$ is not a Feynman integral since the coefficient in front of $\bra{e_5}$ is non-zero. We discuss how to transform it into a canonical Feynman integral in the following.

According to the discussion in Section~\ref{sec:intersection}, the quest is to find a $d\log$-form $\tilde{\varphi}_{15}$ such that projection from $\bra{\varphi'_{15}} + \bra{\tilde{\varphi}_{15}}$ to $\bra{e_5}$ vanishes. We call the $d\log$-forms such as $\tilde{\varphi}_{15}$ as \emph{auxiliary} $d\log$-forms. We know that $\bra{\tilde{\varphi}_{15}}$'s decomposition coefficient in front of $\bra{e_5}$ must be negative to that of $\bra{\varphi'_{15}}$. This requirement greatly constrains the possible forms of $\tilde{\varphi}_{15}$, and it's easy to construct two candidates:
\begin{align}
    \hat{\phi}_1 &= \frac{(1-2 \epsilon ) \sqrt{s^2 \left(t-m^2\right)^2-4 m^2 s t^2}}{\epsilon z_4 z_5 z_6 z_7 P_{2}} \, , \nonumber
    \\
    \hat{\phi}_2 &= \frac{(1-2 \epsilon ) \sqrt{s^2 \left(t-m^2\right)^2-4 m^2 s t^2}}{\epsilon z_2 z_5 z_6 z_7 P_{2}} \, .
    \label{eq:inmdbaux}
\end{align}
Cutting on $z_5$, $z_6$, $z_7$, and projecting them onto our basis we find
\begin{align}
    \bra{\phi_{1,\text{cut}}} &= \frac{\sqrt{s^2 \left(t-m^2\right)^2-4 m^2 s t^2}}{\epsilon} \left( -\frac{m^2}{\epsilon}\bra{e_3} + (1-2\epsilon)m^2 \frac{\epsilon-\rho}{\epsilon-2\rho} \bra{e_5} \right) , \nonumber
    \\
    \bra{\phi_{2,\text{cut}}} &= \frac{\sqrt{s^2 \left(t-m^2\right)^2-4 m^2 s t^2}}{\epsilon} \left( (1-2\epsilon)m^2 \frac{\rho}{\epsilon-2\rho} \bra{e_5} \right) ,
\end{align}
where we have taken the limit $\rho \to 0$ except for the coefficient in front of $\bra{e_5}$. Note that $\bra{\phi_{2,\text{cut}}}$ actually vanishes in that limit, which means that it is zero to begin with. In fact, the uncut version $\bra{\phi_{2}}$ is also zero. This is something similar to the usual rule that scaleless integrals in dimensional regularization is zero. While it is not necessary, we will keep it in order to subtract exactly the $\bra{e_5}$ term from $\bra{\varphi'_{15}}$.

From the above results, one can identify the auxiliary $d\log$-form as $\tilde{\varphi}_{15} = \phi_1 - \phi_2$. It turns out that this is enough even in the uncut case, and one doesn't need to introduce more auxiliary $d\log$-forms in the sub-sectors. Hence the correct $d\log$-form and its relation to Feynman integrals are given by
\begin{equation}
\bra{\varphi_{15}} = \bra{\varphi'_{15}} + \bra{\phi_1} - \bra{\phi_2} = \frac{\sqrt{s^2 \left(t-m^2\right)^2-4 m^2 s t^2}}{\epsilon} \bra{F_{010211100}} \, .
\label{eq:imdbvarphi15}
\end{equation}
Note that the above combination actually takes the form
\begin{align}
        \hat{\varphi}_{15} &= -\frac{(1-2 \epsilon ) \sqrt{s^2 \left(t-m^2\right)^2-4 m^2 s t^2}(z_4-z_2-z_8)}{\epsilon z_2 z_4 z_5 z_6 z_7 P_{2}} \nonumber
        \\
        &= -\frac{(1-2 \epsilon ) \sqrt{s^2 \left(t-m^2\right)^2-4 m^2 s t^2}}{\epsilon z_2 z_4 z_5 z_6 z_7} \frac{1}{2P_2} \frac{\partial P_{2}}{\partial z_4} \, ,
\end{align}
which is manifestly a Feynman integral after performing an IBP with respect to $z_4$. This motivates another way to look for linear combinations of auxiliary $d\log$-forms, similar to the idea of syzygies \cite{Georgoudis:2016wff}.

\subsection{Top-sector \{1,1,1,1,1,1,1,0,0\}}\label{subsec:intmtop}

We now turn to the top sector, and construct the LBL Baikov representation with $z_8$ as the ISP. The relevant ingredients are:
\begin{align}
    \mathcal{N}_{\epsilon}&=\frac{e^{2\epsilon\gamma_{E}}\Gamma^{2}(-\epsilon)}{16\pi^4\Gamma^{2}(-2\epsilon)} \left[s^2t(s+t)\right]^{\epsilon} \, , \nonumber \\
    u(\bm{z})&=P_{1}^{\epsilon} \, P_{2}^{-1/2-\epsilon}P_{3}^{-1/2-\epsilon} \, ,
\end{align}
where the three polynomials are
\begin{align}
    P_{1}(z_5,z_7,z_8)&= -\frac{4}{s}\, G(k_2,p_1,p_2)\, , \nonumber \\
    P_{2}(z_5,z_6,z_7,z_8)&= 16 \, G(k_2,p_1,p_2,p_3)\, , \nonumber \\
    P_{3}(z_1,z_2,z_3,z_4,z_5,z_7,z_8)&= 16 \, G(k_1,k_2,p_1,p_2) \, .
    \label{eq:imdbtop}
\end{align}
There are four master integrals in this sector. Performing the construction, we arrive at the following four $d\log$-forms:
\begin{align}
    &\hat{\varphi}_1= \frac{s \sqrt{s t \left(s t-4 m^2 (s+t)\right)}}{z_1 z_2 z_3 z_4 z_5 z_6 z_7} \, , \nonumber
    \\
    &\hat{\varphi}_2= \frac{s^2 z_8}{z_1 z_2 z_3 z_4 z_5 z_6 z_7} \, , \nonumber
    \\
    &\hat{\varphi}'_3= \frac{z_8}{z_1 z_2 z_3 z_4 z_5 z_6 z_7} \left( \frac{\partial^2 P_{2}}{\partial z_6 \partial z_8} - \frac{1}{2P_1} \frac{\partial P_{2}}{\partial z_6}\frac{\partial P_{1}}{\partial z_8} \right) , \nonumber
    \\
    &\hat{\varphi}'_4= \frac{\sqrt{s(s-4m^2)} \, z_8}{z_1 z_2 z_3 z_4 z_5 z_6 z_7} \frac{1}{2P_1} \frac{\partial P_2}{\partial z_6} \, .
    \label{eq:imdbtopdlog}
\end{align}
While $\hat{\varphi}_{1}$ and $\hat{\varphi}_{2}$ can be straightforwardly identified as Feynman integrals, $\hat{\varphi}'_3$ and $\hat{\varphi}'_4$ are not. We again need to add linear combinations of auxiliary $d\log$-forms to bring them into the FI-subspace. The suitable auxiliary $d\log$-forms can be constructed systematically, and we leave the details to Appendix~\ref{app:imdb}. The final combinations are given by
\begin{align}
    \hat{\varphi}_{3} &= \frac{1}{z_1z_2z_3z_4z_5z_6z_7}\left[-2s^2t+2s^2(z_8+z_9)+4sz_8z_9 \right. \nonumber \\
                            &\hspace{8em}\left. +2tz_1(s+2z_5)+2tz_3(s+2z_7)-2sz_2(s-z_5+2z_6-z_7)-4stz_4 \right] \, , \nonumber \\
    \hat{\varphi}_{4} &= \frac{-2st+2sz_9+2t(z_1+z_3)\sqrt{s(s-4m^2)}}{z_1z_2z_3z_4z_5z_6z_7} \, .
    \label{eq:imdb_phi3}
\end{align}

\subsection{The complete canonical basis as Feynman integrals}

We now list the complete canonical basis of the inner-massive double box family, written as linear combinations of Feynman integrals:
\begin{align}
    \bra{\varphi_1}&=s\sqrt{st(st-4m^2(s+t))}\bra{F_{111111100}} \, , \nonumber \\
    \bra{\varphi_2}&=s^2\bra{F_{1111111-10}} \, , \nonumber \\
    \bra{\varphi_3}&=-2s^2t\bra{F_{111111100}}+2s^2\bra{F_{1111111-10}}+2s^2\bra{F_{11111110-1}}+4s\bra{F_{1111111-1-1}} \nonumber\\
    &-2s^2\bra{F_{101111100}}+4st\bra{F_{110111100}}-4s\bra{F_{101110100}}+4s\bra{F_{101111000}}\nonumber \\
    &-\frac{4t(1-2\epsilon)}{\epsilon}\bra{F_{101011100}}+8t\bra{F_{110111000}} \, , \nonumber \\
    \bra{\varphi_4}&=\sqrt{s(s-4m^2)}\left(-2st\bra{F_{111111100}}+2s\bra{F_{11111110-1}}+4t\bra{F_{110111100}} \right) \, , \nonumber \\
    \bra{\varphi_5}&=st\bra{F_{111111000}} \, , \nonumber \\
    \bra{\varphi_6}&=s\sqrt{s(s-4m^2)}\bra{F_{101111100}} \, , \nonumber \\
    \bra{\varphi_7}&=\frac{s\sqrt{t(t-4m^2)}}{\epsilon}\bra{F_{111102000}} \, , \nonumber \\
    \bra{\varphi_8}&=\frac{1-2\epsilon}{\epsilon^2}\bra{F_{101200000}}-\frac{s(1-2\epsilon)}{\epsilon}\bra{F_{111101000}}-\frac{s(t-4m^2)}{\epsilon}\bra{F_{111102000}} \, , \nonumber \\
    \bra{\varphi_9}&=(s+t)\bra{F_{110111000}} \, , \nonumber \\
    \bra{\varphi_{10}}&=\sqrt{st(st-4m^2(s+t))}\bra{F_{110111100}} \, , \nonumber \\
    \bra{\varphi_{11}}&=s\bra{F_{101111000}} \, , \nonumber \\
    \bra{\varphi_{12}}&=\frac{s(1-2\epsilon)}{\epsilon}\bra{F_{101110100}} \, , \nonumber \\
    \bra{\varphi_{13}}&=\frac{s(1-2\epsilon)}{\epsilon}\bra{F_{101011100}} \, , \nonumber \\
    \bra{\varphi_{14}}&=\frac{\sqrt{st(st-4m^2(s+t))}}{\epsilon}\bra{F_{010211100}} + \frac{m^2\sqrt{st(st-4m^2(s+t))}}{\epsilon^2}\bra{F_{010311100}} \, , \nonumber \\
    \bra{\varphi_{15}}&=\frac{\sqrt{s(s(t-m^2)^2-4t^2m^2)}}{\epsilon}\bra{F_{010211100}} \, , \nonumber \\
    \bra{\varphi_{16}}&=-\frac{sm^2}{\epsilon^2}\bra{F_{000311100}}-\frac{s(1-2\epsilon)}{\epsilon}\bra{F_{010111100}}+\frac{sm^2}{\epsilon}\bra{F_{010211100}} \, , \nonumber \\
    \bra{\varphi_{17}}&=\frac{s}{\epsilon}\bra{F_{101102000}} \, , \nonumber \\
    \bra{\varphi_{18}}&=\frac{s\sqrt{s(s+4m^2)}}{\epsilon^2}\bra{F_{102102000}}-\frac{\sqrt{s(s+4m^2)}}{2m^2\epsilon(1+2\epsilon)}\bra{F_{000220000}} \, , \nonumber \\
    \bra{\varphi_{19}}&=-\frac{\sqrt{s(s-4m^2)}(1-2\epsilon)}{\epsilon^2}\bra{F_{101010200}} \, , \nonumber \\
    \bra{\varphi_{20}}&=\frac{s}{2\epsilon}\bra{F_{100211000}}-\frac{s}{4\epsilon^2}\bra{F_{100220000}} \, , \nonumber \\
    \bra{\varphi_{21}}&=-\frac{t}{2\epsilon}\bra{F_{010211000}}+\frac{t}{4\epsilon^2}\bra{F_{010202000}} \, , \nonumber \\
    \bra{\varphi_{22}}&=-\frac{s}{\epsilon}\bra{F_{010210100}}-\frac{sm^2}{\epsilon^2}\bra{F_{010310100}} \, , \nonumber \\
    \bra{\varphi_{23}}&=-\frac{\sqrt{s(s-4m^2)}(1-2\epsilon)}{\epsilon^2}\bra{F_{010110200}}-\frac{\sqrt{s(s-4m^2)}}{\epsilon}\bra{F_{010210100}} \, , \nonumber \\
    \bra{\varphi_{24}}&=\frac{s}{\epsilon}\bra{F_{010210100}}+\frac{2sm^2}{\epsilon^2}\bra{F_{010310100}} \, , \nonumber \\
    \bra{\varphi_{25}}&=\frac{sm^2}{\epsilon^2}\bra{F_{000311100}} \, , \nonumber \\
    \bra{\varphi_{26}}&=\frac{1-2\epsilon}{\epsilon}\bra{F_{101200000}} \, , \nonumber \\
    \bra{\varphi_{27}}&=\frac{\sqrt{s(s-4m^2)}}{4\epsilon^2}\bra{F_{100220000}}+\frac{\sqrt{s(s-4m^2)}}{2\epsilon^2}\bra{F_{200120000}} \, , \nonumber \\
    \bra{\varphi_{28}}&=\frac{1}{4\epsilon^2}\bra{F_{000220000}}-\frac{s}{2\epsilon^2}\bra{F_{100220000}} \, , \nonumber \\
    \bra{\varphi_{29}}&=\frac{\sqrt{t(t-4m^2)}}{4\epsilon^2}\bra{F_{010202000}}+\frac{\sqrt{t(t-4m^2)}}{2\epsilon^2}\bra{F_{020102000}} \, , \nonumber \\
    \bra{\varphi_{30}}&=\frac{1}{4\epsilon^2}\bra{F_{000220000}}-\frac{t}{2\epsilon^2}\bra{F_{010202000}} \, , \nonumber \\
    \bra{\varphi_{31}}&=\frac{\sqrt{s(s-4m^2)}}{2\epsilon^2}\bra{F_{000210200}} \, , \nonumber \\
    \bra{\varphi_{32}}&=\frac{1}{\epsilon^2}\bra{F_{000220000}} \, .
    \label{eq:imdb_allbasis}
\end{align}
We have worked out the differential equations of the above basis, and verified that they indeed take the $\epsilon$-form.

\section{Outer-massive double box}
\label{sec:extmassivedb}

\begin{figure}[t!]
    \centering
    \includegraphics[width=0.6\textwidth]{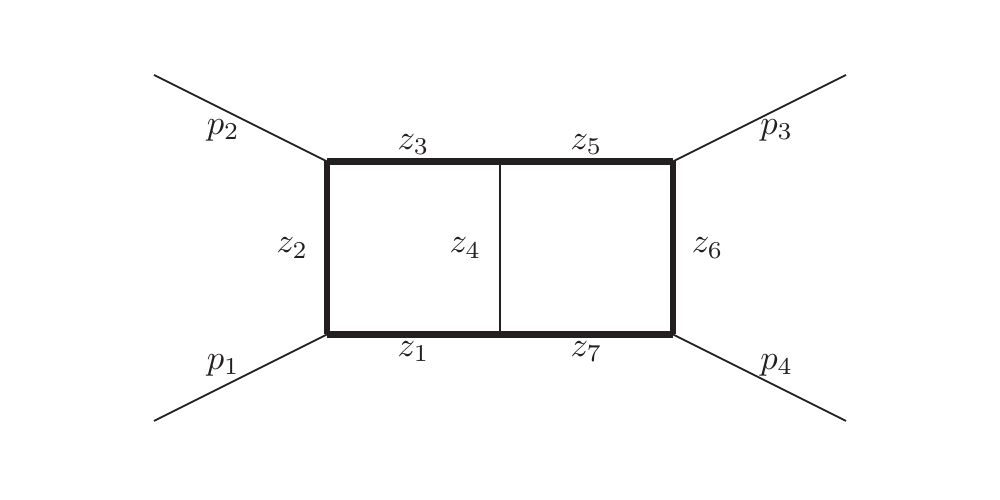}
    \caption{\label{fig:bigmassivedb}The outer-massive double-box integral family. Thick lines represent propagators with mass $m$, while thin lines represent massless propagators.}
\end{figure}

Now we tackle a different double-box family with one mass, where the propagators in the ``outer'' loop are taken to have the mass $m$. The propagator denominators and kinematic invariants are
\begin{align}
&\{k_1^2-m^2,\,(k_1-p_1)^2-m^2,\,(k_1-p_1-p_2)^2-m^2,\,(k_1-k_2)^2,\,(k_2-p_1-p_2)^2-m^2, \nonumber
\\
&(k_2-p_1-p_2-p_3)^2-m^2,\,k_2^2-m^2,\,(k_2-p_1)^2-m^2,\,(k_1-p_1-p_2-p_3)^2-m^2 \} \, , \nonumber
\\
&p_{i}^2=0 \, , \quad (p_1+p_2)^2=s \, , \quad (p_2+p_3)^2=t \, .
\end{align}
The integral family with $z_8$ and $z_9$ as ISPs corresponds to the diagram in Fig.~\ref{fig:bigmassivedb}.

This integral family has already been considered in \cite{Xu:2018eos}. There are 17 unique sectors with 29 master integrals found by \litered~\cite{Lee:2013mka} and \kira~\cite{Klappert:2020nbg}. The construction of canonical integrals in most sectors is straightforward following the procedure in the last section. The only non-trivial sector is the top sector with 7 propagators.

\subsection{Top sector \{1,1,1,1,1,1,1,0,0\}}

We first construct the loop-by-loop Baikov representation with $z_8$ as ISP. The results are:
\begin{align}
    \mathcal{N}_{\epsilon}&= \frac{e^{2\epsilon\gamma_{E}}\Gamma^2(-\epsilon)}{16\pi^4\Gamma^2(-2\epsilon)}[s^2t(s+t)]^{\epsilon} \, , \nonumber
    \\
    u(\bm{z})&=P_{1}^{\epsilon}P_{2}^{-\epsilon-1/2}P_{3}^{-\epsilon-1/2} \, ,
\end{align}
where
\begin{align}
    P_{1}(z_5,z_7,z_8)&= -\frac{4}{s} \, G(k_2,p_1,p_2) \, , \nonumber
    \\
    P_{2}(z_1,z_2,z_3,z_4,z_5,z_7,z_8)&= 16 \, G(k_1,k_2,p_1,p_2) \, , \nonumber
    \\
    P_{3}(z_6,z_5,z_7,z_8)&= 16 \, G(k_2,p_1,p_2,p_3) \, .
    \label{eq:emdbtopsector}
\end{align}

The construction can be performed easily for the variables $z_1$, $z_2$, $z_3$, $z_4$ and $z_6$. However, after that we are left with the expression
\begin{equation}
u_\epsilon(\bm{z}) \, d\log f_1 \wedge d\log f_2 \wedge d\log f_3 \wedge d\log f_4 \wedge d\log f_6 \wedge \frac{\hat{\varphi}(z_5,z_7,z_8) dz_5 \wedge dz_7 \wedge dz_8}{\sqrt{\bar{P}_2 \bar{P}_3}} \, ,
\label{eq:phi578}
\end{equation}
where $\hat{\varphi}(z_5,z_7,z_8)$ is the remaining function we need to construct, and
\begin{align}
u_\epsilon(\bm{z})&=P_{1}^{\epsilon}P_{2}^{-\epsilon}P_{3}^{-\epsilon} \, , \nonumber
\\
\bar{P}_2(z_5,z_7,z_8) &= P_2(0,0,0,0,z_5,z_7,z_8) = s \left(4 m^2 \left(z_7-z_8\right) \left(z_8-z_5\right)+s z_8^2\right) , \nonumber
\\
\bar{P}_3(z_5,z_7,z_8) &= P_3(0,z_5,z_7,z_8) = (sz_8-st+t(z_5+z_7))^2-4t(s+t)(z_5z_7+sm^2) \, .
\end{align}
Since $\bar{P}_2 \bar{P}_3$ is a quartic polynomial of $z_8$ and is also cubic in $z_5$ and $z_7$, the usual construction strategy breaks down here. On the other hand, from the maximal cut we know that there are no elliptic integrals involved in this sector. Hence it is expected that a $d\log$ representation must exist somehow.

We may continue the construction by recalling that the LBL Baikov representation can be obtained by integrating out $z_9$ from the standard Baikov representation, as shown in Section~\ref{sec:genlbl}. In particular, the polynomials $P_2$ and $P_3$ are related to the two roots of the polynomial $P_0(\bm{z},z_9) = 16 G(k_1,k_2,p_1,p_2,p_3)/s$ with respect to $z_9$ ($P_2 P_3 \propto B_N^2-4A_NC_N$ in Eq.~\eqref{eq:gen_lbl}). Let $r_{\pm}(\bm{z})$ denote the two roots, we can write
\begin{equation}
P_0^{-\delta} \, \frac{d^8\bm{z} dz_9}{P_0(\bm{z},z_9)} = P_0^{-\delta} \, \frac{d^8\bm{z}}{\sqrt{P_2(\bm{z}) P_3(\bm{z})}} \, d\log\frac{z_9-r_+}{z_9-r_-} \, ,
\end{equation}
where $\delta$ serves as a regulator.
Note that the above relation still holds if we set $z_1$, $z_2$, $z_3$, $z_4$ and $z_6$ to zero in all of $P_i(\bm{z})$ and $r_{\pm}(\bm{z})$. Hence we have
\begin{equation}
\bar{P}_0^{-\delta} \, \frac{f(z_5,z_7,z_8,z_9) dz_8 dz_9}{\bar{P}_0(z_5,z_7,z_8,z_9)} = \bar{P}_0^{-\delta} \, \frac{f(z_5,z_7,z_8,z_9) dz_8}{\sqrt{\bar{P}_2 \bar{P}_3}} \, d\log\frac{z_9-\bar{r}_+}{z_9-\bar{r}_-} \, ,
\end{equation}
where the notations $\bar{P}_0$ and $\bar{r}_{\pm}$ should be clear to the readers, and we have suppressed the other factors in Eq.~\eqref{eq:phi578}. Here, $f(z_5,z_7,z_8,z_9)$ is an arbitrary rational function whose singularities are properly regularized. The left-hand side of the above equation can also be written as
\begin{equation}
\bar{P}_0^{-\delta} \frac{f(z_5,z_7,z_8,z_9) dz_8 dz_9}{\bar{P}_0(z_5,z_7,z_8,z_9)} = d\log\frac{z_8-\bar{t}_+}{z_8-\bar{t}_-} \, \bar{P}_0^{-\delta} \frac{f(z_5,z_7,z_8,z_9) dz_9}{\sqrt{\bar{Q}_2(z_9) \bar{Q}_3(z_5,z_7,z_9)}} \, ,
\label{eq:f5789}
\end{equation}
where $\bar{t}_{\pm}$ are the two roots of $\bar{P}_0$ with respect to $z_8$, and the two polynomials $\bar{Q}_2$ and $\bar{Q}_3$ are given by
\begin{align}
\bar{Q}_{2}(z_9) &= 16 \, G(k_1,p_1,p_1+p_2,p_3)\Big|_{z_1,z_2,z_3 \to 0} \nonumber
\\ 
&= s \left(4 m^2 s t+4 m^2 t^2-s t^2+2 s t z_9-s z_9^2\right) , \nonumber
\\
\bar{Q}_{3}(z_5,z_7,z_9) &= 16 \, G(k_2,k_1,p_1+p_2,p_3)\Big|_{z_1,z_3,z_4,z_6 \to 0} \nonumber
\\ 
&= 4 m^2 s z_9^2+4 m^2 s z_5 z_7+4 m^2 s z_5 z_9+4 m^2 s z_7 z_9-s^2 z_9^2+2 s z_5 z_9^2 \nonumber
\\
&\quad +2 s z_7z_9^2+4 s z_5 z_7 z_9-z_5^2 z_9^2-z_7^2 z_9^2+2 z_5 z_7 z_9^2 \, .
\end{align}

Now comes the crucial observation: the function $f(z_5,z_7,z_8,z_9)$ can be seen as a representative of an equivalence class under IBP relations. It is possible that the following equivalence relation holds:
\begin{equation}
f(z_5,z_7,z_8,z_9) \sim \hat{\varphi}(z_5,z_7,z_8) \sim \hat{\phi}(z_5,z_7,z_9) \, ,
\label{eq:phi579}
\end{equation}
where $\hat{\varphi}$ does not depend on $z_9$, and $\hat{\phi}$ does not depend on $z_8$. In this case, we can construct the function $\hat{\phi}(z_5,z_7,z_9)$ such that the right-hand size of Eq.~\eqref{eq:f5789} becomes a $d\log$-form (with $\hat{\phi}$ in place of $f$). Then by inserting the corresponding $\hat{\varphi}(z_5,z_7,z_8)$ into Eq.~\eqref{eq:phi578}, we find a candidate for canonical integrals in the original LBL Baikov representation. A dictionary of equivalent $\hat{\varphi}$'s and $\hat{\phi}$'s can be generated by putting different forms of $f(z_5,z_7,z_8,z_9)$ into Eq.~\eqref{eq:f5789}, and integrating out $z_9$ (which gives a $\hat{\varphi}$) or $z_8$ (which gives a $\hat{\phi}$). Since we are interested in integrands with simple poles, it is enough to consider four possible kinds of the function $f$: $f=c$, $f=c z_8$, $f = c z_9$ and $f=c z_8 z_9$, where $c$ denotes a ``constant'' polynomial independent of $z_8$ and $z_9$. The integration over $z_8$ or $z_9$ can be performed using Eq.~\eqref{eq:gen_lbl} and the relation
\begin{equation}
\bar{P}_0(z_5,z_7,z_8,z_9) = P_{1}(z_5,z_7,z_8) \, (r_+-z_9)(z_9-r_-) = Q_{1}(z_9) \, (t_+-z_8)(z_8-t_-) \, ,
\label{eq:P0bar}
\end{equation}
where $P_1$ was given in Eq.~\eqref{eq:emdbtopsector}, and
\begin{equation}
Q_{1}(z_9) = -\frac{4}{s} \, G(k_1,p_1+p_2,p_3)\Big|_{z_1,z_3 \to 0} = z_9^2 + s z_9 + s m^2 \, .
\end{equation}
We then find the following correspondences between $\hat{\varphi}(z_5,z_7,z_8)$ and $\hat{\phi}(z_5,z_7,z_9)$:
\begin{align}
1 &\longleftrightarrow 1 \, , \nonumber
\\
-\frac{R_1(z_5,z_7,z_8)}{2s \, P_1(z_5,z_7,z_8)} &\longleftrightarrow z_9 \, , \nonumber
\\
z_8 &\longleftrightarrow -\frac{R_2(z_5,z_7,z_9)}{2s \, Q_1(z_9)} \, , \nonumber
\\
\frac{z_8 \, R_1(z_5,z_7,z_8)}{P_1(z_5,z_7,z_8)} &\longleftrightarrow \frac{z_9 \, R_2(z_5,z_7,z_9)}{Q_1(z_9)} \, ,
\label{eq:emdb_dic}
\end{align}
with the polynomials
\begin{align}
R_1 &= -2 s t \left[ m^2 (z_5+z_7) + z_5 z_7 \right] + s z_8 \left[ 2 m^2 (s+2t) - s t +t(z_5+z_7) \right] + s^2 z_8^2 \, , \nonumber
\\
R_2 &= z_9 N_1 + (z_5+z_7) z_9 N_2 + 2t(z_5+z_7)Q_1 \, , \nonumber
\\
N_1 &= s(2m^2(s+2t) - st + sz_9) \, , \nonumber
\\
N_2 &= -st - (s+2t) z_9 \, .
\label{eq:imdb_R2}
\end{align}

Now, since $\bar{Q}_2$ doesn't depend on $z_5$ and $z_7$ and $\bar{Q}_3$ is quadratic in all variables, it is straightforward to construct the following candidates for $\hat{\phi}(z_5,z_7,z_9)$ such that $\hat{\phi}/\sqrt{\bar{Q}_2\bar{Q}_3}$ is $d\log$:
\begin{align}
&\frac{\sqrt{s(s-4m^2)}\sqrt{st(st-4m^2(s+t))}}{z_5z_7} \, , \nonumber
\\
&\frac{\sqrt{s(s-4m^2)}}{z_5z_7} \frac{z_9N_1}{Q_1} \, , \quad \frac{s(s-4m^2)}{z_5z_7}\frac{z_9N_2}{Q_1} \, ,\nonumber
\\
&\frac{\sqrt{s(s-4m^2)}(z_5+z_7)}{z_5z_7}\frac{z_9N_2}{Q_1} \, , \frac{(z_5+z_7)}{z_5 z_7}\frac{z_9N_1}{Q_1} \, , \quad \frac{s(z_5+z_7)z_9}{z_5 z_7} \, .
\end{align}
We need to identify three linear combinations of the above candidates which can be converted to $\hat{\varphi}(z_5,z_7,z_8)$ using the dictionary Eq.~\eqref{eq:emdb_dic}. These are
\begin{align}
\hat{\phi}_1 &= \frac{\sqrt{s(s-4m^2)}\sqrt{st(st-4m^2(s+t))}}{z_5z_7} \, , \nonumber
\\
\hat{\phi}_2 &= -\frac{\sqrt{s(s-4m^2)}}{z_5z_7} \frac{z_9N_1 + (z_5+z_7)z_9N_2}{2Q_1} \nonumber
\\
&= -\frac{\sqrt{s(s-4m^2)}}{z_5z_7} \left[ \frac{R_2}{2Q_1} - t(z_5+z_7) \right] , \nonumber
\\
\hat{\phi}_3 &= \frac{(z_5+z_7)}{z_5z_7}\frac{z_9N_1}{2Q_1}-\frac{s(z_5+z_7)z_9}{z_5z_7}+\frac{s(s-4m^2)}{z_5z_7}\frac{z_9N_2}{2Q_1} \nonumber
\\
&= \frac{s}{z_5z_7} \left[ \frac{z_9R_2}{sQ_1}- sz_9 + \frac{R_2}{2Q_1} - t(z_5+z_7) \right] ,
\end{align}
where in the last equation we have used the following identity
\begin{align}
    z_9R_2 = s(s-z_5-z_7)z_9Q_1 + \frac{1}{2}z_9(z_5+z_7)(N_1-sN_2) - \frac{1}{2}sz_9(N_1-(s-4m^2)N_2) \, .
    \label{eq:imdb_z9R2}
\end{align}
Their corresponding $\hat{\varphi}(z_5,z_7,z_8)$'s can be easily found. Multiplying them by the other factors from the construction of $z_1$, $z_2$, $z_3$, $z_4$ and $z_6$, we obtain
\begin{align}
    \hat{\varphi}_{1}&=\frac{\sqrt{s(s-4m^2)}\sqrt{st(st-4m^2(s+t))}}{z_1z_2z_3z_4z_5z_6z_7} \, , \nonumber \\
    \hat{\varphi}_{2}&=\frac{\sqrt{s(s-4m^2)}(s z_8+t(z_5 + z_7))}{z_1z_2z_3z_4z_5z_6z_7} \, , \nonumber \\
    \hat{\varphi}'_{3}&=\frac{(s+2z_8)R_1}{2z_1z_2z_3z_4z_5z_6z_7P_{1}}-\frac{s(s z_8+t(z_5 + z_7))}{z_1z_2z_3z_4z_5z_6z_7} \, .
    \label{eq:topsectordlogs}
\end{align}
One can see that $\varphi_1$ and $\varphi_2$ are Feynman integrals while $\varphi_3$ contains a polynomial denominator. We needs to again introduce auxiliary $d\log$s from sub-sectors to convert it to Feynman integrals. Further details can be found in Appendix~\ref{app:omdb}. The final results are:\footnote{Here we have utilized some symmetry relations such as $F_{111101100}=F_{111111000}$ and $F_{011101100}=F_{110111000}$. These relations can be automatically detected by \kira.}
\begin{align}
    \bra{\varphi_1}&=\sqrt{s(s-4m^2)}\sqrt{st(st-4m^2(s+t))} \, F_{111111100} \, , \nonumber
    \\
    \bra{\varphi_2}&=\sqrt{s(s-4m^2)} \left(sF_{1111111-10}+2tF_{111101100}\right) , \nonumber
    \\
    \bra{\varphi_3}&=-2sF_{1111111-1-1}-2s^2F_{1111111-10}-4stF_{111111000}+2stF_{111011100} \nonumber
    \\
    &\quad +2s^2F_{101111100}+2sF_{101110100}-4tF_{110111000} \, .
    \label{eq:topsectorfeynint}
\end{align}

Note that in Eqs.~\eqref{eq:imdb_R2} and \eqref{eq:imdb_z9R2}, we have used the fact that both $R_2$ and $z_9R_2$ can be written as linear combinations of $z_9N_1$, $z_9N_2$ and $Q_1$, where the coefficients of $z_9N_1$ and $z_9N_2$ are independent of $z_9$ and the coefficient of $Q_1$ is at most linear in $z_9$. This is not a coincidence as is shown in Appendix~\ref{app:omdb}.

\subsection{The canonical basis}

The $d\log$-forms in the other sectors can be constructed and converted to Feynman integrals similarly. We list the canonical basis from our construction in the following:
\begin{align}
    \bra{\varphi_1}&= \sqrt{s(s-4m^2)}\sqrt{st(st-4m^2(s+t))}\bra{F_{111111100}} \nonumber\\
    \bra{\varphi_2}&= \sqrt{s(s-4m^2)}\left(s\bra{F_{1111111-10}}+2t\bra{F_{111111000}}\right) \nonumber \\
    \bra{\varphi_3}&= -2s\bra{F_{1111111-1-1}}-2s^2\bra{F_{1111111-10}}-4st\bra{F_{111111000}}+2st\bra{F_{111011100}} \nonumber \\
    &+2s^2\bra{F_{101111100}} +2s\bra{F_{101110100}}-4t\bra{F_{110111000}} \nonumber \\
    \bra{\varphi_4}&= \sqrt{s t \left(s t-4 m^2 (s+t)\right)} \bra{F_{111111000}} \nonumber \\
    \bra{\varphi_5}&= s \bra{F_{110111000}} - s \bra{F_{11111100-1}} \nonumber \\
    \bra{\varphi_6}&= s^2 \bra{F_{111011100}} \nonumber \\
    \bra{\varphi_7}&= \frac{\sqrt{s t \left(s t - 4m^2(s+t)\right)}}{2 \epsilon }\bra{F_{111102000}}+\frac{\sqrt{s t \left(s t - 4m^2(s+t)\right)}}{2 \epsilon }\bra{F_{111201000}} \nonumber \\
    \bra{\varphi_8}&= \frac{\sqrt{s \left(s (t-m^2)^2-4 m^2 t^2\right)}}{\epsilon }\bra{F_{111102000}} \nonumber \\
    \bra{\varphi_9}&= \frac{s m^2}{\epsilon }\bra{F_{111102000}}-\frac{ s (1-2\epsilon )}{\epsilon }\bra{F_{111101000}}+\frac{ s}{2 \epsilon }\bra{F_{111020000}} \nonumber \\
    \bra{\varphi_{10}}&= \frac{ s \sqrt{s \left(s-4 m^2\right)}}{\epsilon }\bra{F_{111010200}} \nonumber \\
    \bra{\varphi_{11}}&= (s+t) \bra{F_{110111000}} \nonumber \\
    \bra{\varphi_{12}}&= \frac{ \sqrt{s t \left(s t-4 m^2 (s+t)\right)}}{\epsilon }\bra{F_{110211000}} \nonumber \\
    \bra{\varphi_{13}}&= \frac{s}{\epsilon}\bra{F_{11021100-1}}+\frac{t}{\epsilon}\bra{F_{11-1211000}} \nonumber \\
    \bra{\varphi_{14}}&= s \bra{F_{101111000}} \nonumber \\
    \bra{\varphi_{15}}&= s \sqrt{s \left(s-4 m^2\right)} \bra{F_{101111100}} \nonumber \\
    \bra{\varphi_{16}}&= \frac{ s}{\epsilon }\bra{F_{101102000}} \nonumber \\
    \bra{\varphi_{17}}&= -\frac{ s}{2 \epsilon }\bra{F_{101102000}}-\frac{ s}{2 \epsilon }\bra{F_{101201000}} \nonumber \\
    \bra{\varphi_{18}}&= \frac{ \sqrt{s \left(s-4 m^2\right)}}{\epsilon }\bra{F_{101102000}}+\frac{ (1-2 \epsilon)\sqrt{s \left(s-4 m^2\right)}}{\epsilon^2}\bra{F_{102101000}} \nonumber \\
    \bra{\varphi_{19}}&= \frac{ s}{\epsilon }\bra{F_{111020000}} \nonumber \\
    \bra{\varphi_{20}}&= t \bra{F_{011111000}} \nonumber \\
    \bra{\varphi_{21}}&= \frac{ s}{2 \epsilon }\bra{F_{110120000}}-\frac{ s}{4 \epsilon^2}\bra{F_{200120000}} \nonumber \\
    \bra{\varphi_{22}}&= \frac{ t}{2 \epsilon }\bra{F_{110102000}}-\frac{ t}{4 \epsilon ^2}\bra{F_{020102000}} \nonumber \\
    \bra{\varphi_{23}}&= \frac{ s \left(s-4 m^2\right)}{\epsilon^2}\bra{F_{102010200}} \nonumber \\
    \bra{\varphi_{24}}&= \frac{ \sqrt{s \left(s-4 m^2\right)}}{2 \epsilon ^2}\bra{F_{100220000}}+\frac{ \sqrt{s\left(s-4 m^2\right)}}{4 \epsilon^2} \bra{F_{200120000}} \nonumber \\
    \bra{\varphi_{25}}&= \frac{ s}{4 \epsilon ^2}\bra{F_{200120000}} \nonumber \\
    \bra{\varphi_{26}}&= \frac{ \sqrt{t \left(t-4 m^2\right)}}{2 \epsilon^2}\bra{F_{010202000}}+\frac{ \sqrt{t\left(t-4 m^2\right)}}{4 \epsilon^2}\bra{F_{020102000}} \nonumber \\
    \bra{\varphi_{27}}&= \frac{ t}{4 \epsilon^2}\bra{F_{020102000}} \nonumber \\
    \bra{\varphi_{28}}&= \frac{ \sqrt{s \left(s-4 m^2\right)}}{\epsilon^2}\bra{F_{102020000}} \nonumber \\
    \bra{\varphi_{29}}&= \frac{1}{\epsilon ^2}\bra{F_{200020000}}
\end{align}
We have checked that their differential equations with respect to $s$, $t$ and $m^2$ all take the $\epsilon$-form.

\section{Summary and outlook}
\label{sec:summary}

In this paper, we have explored the properties of the generalized Baikov representation, which allows additional polynomials of the Baikov variables to appear in the denominator. We have investigated its difference and relation with the usual Baikov representation of Feynman integrals using the language of intersection theory. We find that Feynman integrals span a subspace of the vector space of generalized Baikov integrals. This explains why the dimension-counting by computing the number of critical points in the loop-by-loop Baikov representation often gives a number larger than that of independent Feynman integrals. We have further discussed how to identify this so-called FI-subspace using intersection theory, optionally supplemented with IBP relations.

Utilizing the generalized Baikov integrals, we have proposed a novel method to construct canonical Feynman integrals satisfying $\epsilon$-form differential equations. The method start with constructing $d\log$-form integrands in the generalized Baikov representation. The construction is performed variable-by-variable, and we show in detail how to deal with the square roots appearing in the intermediate steps using properties of Gram determinants. The $d\log$ Baikov integrals are then converted to Feynman integrals by looking for linear combinations belonging to the FI-subspace. In this way a complete canonical basis is obtained. We emphasize that the constructed $d\log$ Baikov integrals are fully $d$-dimensional without any cuts. The resulting Feynman integrals therefore automatically have uniform transcendentality without further manipulations.

We have demonstrated our method using several examples including two kinds of one-mass double box families, and further examples are given in the appendices. In all cases we have verified the differential equations for the constructed canonical bases, which indeed take the $\epsilon$-form. Such equations allow solutions in terms of iterated integrals which satisfy nice algebraic properties and can be easily evaluated numerically.

At one-loop, the $d\log$-form of the UT integrals helps to determine the letters appearing in the differential equations, as well as the symbols of the solutions \cite{Abreu:2017ptx, Abreu:2017enx, Abreu:2017mtm, Chen:2022fyw}. It is interesting to investigate whether similar results can be obtained for higher loops using the construction in this work. It is also interesting to extend our framework to study integral families involving elliptic integrals. We leave these investigations to future works.

\vspace{3ex}

\textbf{Acknowledgment.} We'd like to thank Sebastian Mizera and Yang Zhang for useful discussions. This work was supported in part by the National Natural Science Foundation of China under Grant No. 11975030, 11635001 and 11925506. The research of X. Xu was supported in part by the Swiss National Science Foundation (SNF) under Grant No. $200020\_182038$.

\appendix

\section{Further details for the inner-massive double box family}
\label{app:imdb}

In this Appendix, we give further details for the inner-massive double box family. We first elaborate on the construction of auxiliary $d\log$-forms which are necessary to find the UT Feynman integrals in the top sector. We then list all the $d\log$-forms for sectors corresponding to the canonical basis \eqref{eq:imdb_allbasis} in the main text. We also provide an alternative way to convert $d\log$ Baikov integrals to Feynman integrals.

\subsection{The auxiliary $d\log$-forms for the top sector}
\label{app:imdbadlog}

In this subsection, we discuss the construction of auxiliary $d\log$-forms that allow us to arrive at $\hat{\varphi}_3$ and $\hat{\varphi}_4$ in Eq.~\eqref{eq:imdb_phi3}. These auxiliary $d\log$s must belong to sub-sectors, since the number of maximally-cut Baikov integrals is the same as the number of master Feynman integrals in the top sector. We therefore choose to subtract the top-sector components from $\varphi'_3$ and $\varphi'_4$ as the first step. Note that both $\varphi'_3$ and $\varphi'_4$ have the polynomial $P_1$ in the denominator. This fact leads us to choose the following four master Feynman integrals in the top sector:
\begin{equation}
    F_{111111100}\,, \; F_{1111111-10}\,, \; F_{11111110-1}\,, \; F_{1111111-1-1}\,,
\end{equation}
where the last two have $z_9$ in the numerator (which gives $P_1$ denominator in the generalized loop-by-loop representation without $z_9$ ISP).

Performing the projections using intersection theory we have
\begin{align}
\bra{\varphi'_3} &= -2s^2t \bra{F_{111111100}} + 2s^2 \bra{F_{1111111-10}} + 2s^2 \bra{F_{11111110-1}} \nonumber
\\
&+ 4s \bra{F_{1111111-1-1}} + \text{sub-sector integrals} \, , \nonumber
\\
\bra{\varphi'_4} &= \sqrt{s(s-4m^2)}\left(-2st \bra{F_{111111100}} + 2s \bra{F_{11111110-1}} + \text{sub-sector integrals}\right) \, .
\end{align}
Subtracting the top-sector components, we arrive at the remainders
\begin{align}
R_{3} &= \frac{N_{3}}{z_1z_2z_3z_4z_5z_6z_7P_{1}} \,, \nonumber
\\
R_{4} &= \frac{\sqrt{s(s-4m^2)} \, N_{4}}{z_1z_2z_3z_4z_5z_6z_7P_{1}} \, ,
\end{align}
where the numerators are given by
\begin{align}
N_3&=z_1[-(z_5+z_7)(z_5-z_8)Q_1-(z_5-z_8)Q_2+2(st+sz_5-sz_8+2tz_5)P_{1}] \nonumber
\\
&+z_3[-(z_5+z_7)(z_7-z_8)Q_1-(z_7-z_8)Q_2+2(st+sz_7-sz_8+2tz_7)P_{1}] \nonumber
\\
&+z_2[-TQ_1-(z_5+z_7)Q_2-2s(s-z_5+2z_6-z_7)P_{1}] \nonumber
\\
&+z_4[-s(z_5+z_7)Q_1+s(Q_2-2(s+2t)P_{1})] \nonumber
\\
&-sz_5z_8Q_1-sz_7z_8Q_1 \, , \nonumber
\\
N_4&=z_1[2tP_{1}-(z_5-z_8)Q_1] + z_3 [2tP_{1}-(z_7-z_8)Q_1] + z_4sQ_{1}-z_2Q_{2} \, , \nonumber
\\
T&=\lambda(s,z_5,z_7)-4sm^2 \, ,
    \label{eq:imdbT}
\end{align}
with the K\"all\'en function $\lambda(x,y,z) \equiv x^2+y^2+z^2-2xy-2xz-2yz$, while $Q_1$ and $Q_2$ are the two polynomials already present in $\varphi'_3$ and $\varphi'_4$:
\begin{align}\label{eq:imdbQ1Q2}
    Q_1&= \frac{1}{2s}\frac{\partial P_{2}}{\partial z_6} \nonumber \\
       &= -s t+s z_6-s z_8+t z_5+t z_7-2 t z_8 \, , \nonumber \\
    Q_2&= \frac{1}{2s}\left(2 P_{1} \frac{\partial^2 P_{2}}{\partial z_6 \partial z_8} - \frac{\partial P_{2}}{\partial z_6}\frac{\partial P_{1}}{\partial z_8} \right) \nonumber \\
    &= 2 m^2 s^2+4 m^2 s t-s^2 t+s^2 z_8+2 s t z_7+z_5 \left(2 z_7 (s+t)+s \left(2 t-z_8\right)-sz_6\right) \nonumber \\
    &-s z_7 z_8+s z_6 \left(s-z_7+2 z_8\right)-t z_5^2-t z_7^2 \, .
\end{align}

\begin{table}[t!]
    \centering
    \begin{tabular}{cc}
        \hline
        Sector & Auxiliary $d\log$s \\[8pt] 
        011111100 & $\frac{\sqrt{s(s-4m^2)}(z_5-z_8)Q_1}{z_2z_3z_4z_5z_6z_7P_{1}},\,\frac{(z_5-z_8)Q_2}{z_2z_3z_4z_5z_6z_7P_{1}},\,\frac{s(z_5-z_8)}{z_2z_3z_4z_5z_6z_7}$  \\[6pt]
        011101100 & $\frac{(z_5-z_8)Q_1}{z_2z_3z_4z_6z_7P_{1}}$ \\[6pt]
        011111000 & $\frac{(z_5-z_8)Q_1}{z_2z_3z_4z_5z_6P_{1}}$ \\[6pt]
        110111100 & $\frac{\sqrt{s(s-4m^2)}(z_7-z_8)Q_1}{z_1z_2z_4z_5z_6z_7P_{1}},\,\frac{\sqrt{s(s-4m^2)}(z_7-z_8)Q_2}{z_1z_2z_4z_5z_6z_7P_{1}},\,\frac{s(z_7-z_8)}{z_1z_2z_4z_5z_6z_7}$ \\[6pt]
        110101100 & $\frac{(z_7-z_8)Q_1}{z_1z_2z_4z_6z_7P_{1}}$ \\[6pt]
        110111000 & $\frac{(z_7-z_8)Q_1}{z_1z_2z_4z_5z_6P_{1}}$ \\[6pt]
        101111100 & $\frac{TQ_1}{z_1z_3z_4z_5z_6z_7P_{1}},\,\frac{\sqrt{s(s-4m^2)}Q_2}{z_1z_3z_4z_5z_6z_7P_{1}},\,\frac{\sqrt{s(s-4m^2)}s}{z_1z_3z_4z_5z_6z_7}$ \\[6pt]
        101101100 & $\frac{Q_2}{z_1z_3z_4z_6z_7P_{1}}$ \\[6pt]
        101111000 & $\frac{Q_2}{z_1z_3z_4z_5z_6P_{1}}$ \\[6pt]
        111011100 & $\frac{s\sqrt{s(s-4m^2)}Q_1}{z_1z_2z_3z_5z_6z_7P_{1}},\,\frac{sQ_2}{z_1z_2z_3z_5z_6z_7P_{1}},\,\frac{s^2}{z_1z_2z_3z_5z_6z_7}$ \\[6pt]
        111001100 & $\frac{sQ_1}{z_1z_2z_3z_6z_7P_{1}}$ \\[6pt]
        111011000 & $\frac{sQ_1}{z_1z_2z_3z_5z_6P_{1}}$ \\[6pt]
        111101100 & $\frac{sz_8Q_1}{z_1z_2z_3z_4z_6z_7P_{1}},\,\frac{st}{z_1z_2z_3z_4z_6z_7}$ \\[6pt]
        111111000 & $\frac{sz_8Q_1}{z_1z_2z_3z_4z_6z_7P_{1}},\,\frac{st}{z_1z_2z_3z_4z_6z_7}$ \\[6pt]
        \hline
    \end{tabular}
    \caption{Auxiliary $d\log$s in the sub-sectors of the inner-massive double box family.}
    \label{tab:innerauxdlog}
\end{table}

The expressions of $R_3$ and $R_4$ seem to be rather complicated. However, since we know that they are composed of $d\log$-forms in the sub-sectors, we can systematically construct these $d\log$-forms and use them to subtract all terms with $P_1$ in the denominator. After such a subtraction, the results must belong to the FI-subspace and it is then straightforward to convert them to Feynman integrals. In Table~\ref{tab:innerauxdlog} we list all relevant $d\log$s in the sub-sectors. From these it is easy to deduce the required combinations:
\begin{align}
\hat{\phi}_3 &= \frac{z_1(z_5-z_8)\left(-Q_2 - (z_5 + z_7)Q_1+ 2sP_1\right) + z_3(z_7-z_8)\left(-Q_2 - (z_5 + z_7)Q_1+ 2sP_1\right)}{z_1z_2z_3z_4z_5z_6z_7 P_1} \nonumber \\
&+\frac{- z_2\left(T Q_1+ (z_5 + z_7)Q_2\right) + z_4s\left(Q_2- (z_5+z_7)Q_1- 2sP_1\right)-z_5sz_8Q_1-z_7sz_8Q_1}{z_1z_2z_3z_4z_5z_6z_7 P_1} \, , \nonumber
\\
\hat{\phi}_4 &= \frac{\sqrt{s^2-4 m^2 s}\left(-z_1(z_5 - z_8)Q_1 -z_3(z_7 - z_8)Q_1 -z_2Q_2 + z_4 s Q_1\right)}{z_1z_2z_3z_4z_5z_6z_7 P_{1}} \, .
\end{align}
Our final results for the last two UT Feynman integrals are hence
\begin{equation}
\hat{\varphi}_3 = \hat{\varphi}'_3 - \hat{\phi}_3 \, , \quad \hat{\varphi}_4 = \hat{\varphi}'_4 - \hat{\phi}_4 \, .
\end{equation}

\subsection{List of $d\log$-forms for all sectors}
\label{app:imdballsector}

Here we list all $d\log$-forms and hints for their construction. The Gram determinants in the following should be rewritten as functions of the propagator denominators $z_i$. These denominators are given by
\begin{align}
&\{k_1^2,\,(k_1-p_1)^2,\,(k_1-p_1-p_2)^2,\,(k_1-k_2)^2-m^2,\,(k_2-p_1-p_2)^2-m^2, \nonumber
\\
&\;(k_2-p_1-p_2-p_3)^2-m^2,\,k_2^2-m^2,\,(k_2-p_1)^2-m^2,\,(k_1-p_1-p_2-p_3)^2\} \, .
\end{align}
\begin{itemize}

\item Sector \{111111100\}: $z_8$ as ISP.
\begin{align}
    \hat{\varphi}_1 &= \frac{s \sqrt{s t \left(s t-4 m^2 (s+t)\right)}}{z_1 z_2 z_3 z_4 z_5 z_6 z_7} \, , \nonumber\\
    \hat{\varphi}_2 &= \frac{s^2 z_8}{z_1 z_2 z_3 z_4 z_5 z_6 z_7} \, , \nonumber \\
    \hat{\varphi}_{3} &= \frac{1}{z_1z_2z_3z_4z_5z_6z_7}\left[-2s^2t+2s^2(z_8+z_9)+4sz_8z_9 \right. \nonumber \\
                            &\hspace{6em}\left. +2tz_1(s+2z_5)+2tz_3(s+2z_7)-2sz_2(s-z_5+2z_6-z_7)-4stz_4 \right] \, , \nonumber \\
    \hat{\varphi}_{4} &= \frac{-2st+2sz_9+2t(z_1+z_3)\sqrt{s(s-4m^2)}}{z_1z_2z_3z_4z_5z_6z_7} \, .
\end{align}

\item Sector \{111111000\}: $z_9$ as ISP.
\begin{align}
\hat{\varphi}_5 &= \frac{s t}{z_1 z_2 z_3 z_4 z_5 z_6} \, .
\end{align}

\item Sector \{101111100\}: no ISP.
\begin{align}
\hat{\varphi}_6 &= \frac{s \sqrt{s \left(s-4 m^2\right)}}{z_1 z_3 z_4 z_5 z_6 z_7} \, .
\end{align}

\item Sector \{111101000\}: $z_9$ as ISP.
\begin{align}
\hat{\varphi}_7 &= \frac{s z_9 (1-2 \epsilon ) \sqrt{t \left(t-4 m^2\right)}}{4\epsilon z_1 z_2 z_3 z_4 z_6 \, G(\tilde{k}_1,\tilde{k}_2)} \, , \nonumber\\
\hat{\varphi}_8 &= \frac{s z_9 (1-2 \epsilon ) \left(z_9-t\right)}{4\epsilon z_1 z_2 z_3 z_4 z_6 \, G(\tilde{k}_1,\tilde{k}_2)} \, ,
\end{align}
where $\tilde{k}_i = k_i-p_1-p_2-p_3$.

\item Sector \{110111000\}: $z_7$ and $z_8$ as ISP.
\begin{align}
    \hat{\varphi}_9 &= \frac{s+t}{z_1 z_2 z_4 z_5 z_6} \, , \nonumber \\
    \hat{\varphi}_{10} &= \frac{\sqrt{st(st-4m^2(s+t))}}{z_1 z_2 z_4 z_5 z_6 z_7} \, .
\end{align}

\item Sector \{101111000\}: $z_9$ as ISP.
\begin{align}
\hat{\varphi}_{11} &= \frac{s}{z_1 z_3 z_4 z_5 z_6} \, .
\end{align}

\item Sector \{101110100\}: no ISP.
\begin{align}
\hat{\varphi}_{12} &= \frac{s (1-2 \epsilon )}{\epsilon z_1 z_3 z_4 z_5 z_7 } \, .
\end{align}

\item Sector \{101011100\}: no ISP.
\begin{align}
\hat{\varphi}_{13} &= - \frac{s^3 (1-2 \epsilon )}{4\epsilon z_1 z_3 z_5 z_6 z_7 \, G(k_1,p_1+p_2)}
\end{align}

\item Sector \{010111100\}: $z_8$ as ISP.
\begin{align}
    \hat{\varphi}_{14} &= \frac{(1-2 \epsilon ) \left(m^2+z_8\right) \sqrt{s t \left(s t-4 m^2 (s+t)\right)}}{4\epsilon z_2 z_4 z_5 z_6 z_7 \, G(k_1-p_1,k_2-p_1)} \, , \nonumber \\
    \hat{\varphi}_{15} &=\frac{(1-2 \epsilon ) \sqrt{s^2 (t-m^2)^2-4 m^2 s t^2}}{2\epsilon z_2 z_4 z_5 z_6 z_7 \, G(k_1-p_1,k_2-p_1)}\frac{\partial G(k_1-p_1,k_2-p_1)}{\partial z_4} \, , \nonumber \\
    \hat{\varphi}_{16} &= \frac{sz_8 (1-2 \epsilon ) \left(m^2+z_8\right)}{4\epsilon z_2 z_4 z_5 z_6 z_7 \, G(k_1-p_1,k_2-p_1)} \, .
\end{align}

\item Sector \{101101000\}: $z_9$ as ISP.
\begin{align}
\hat{\varphi}_{17} &= \frac{(1-2\epsilon)s z_9}{4\epsilon z_1 z_3 z_4 z_6 \, G(\tilde{k}_1,\tilde{k}_2)} \, , \nonumber \\
\hat{\varphi}_{18} &= -\frac{(1-2\epsilon)s^2 z_9^2\sqrt{s(s+4m^2)}}{16\epsilon z_1 z_3 z_4 z_6 \, G(\tilde{k}_1,\tilde{k}_2) \, G(\tilde{k}_1,p_1+p_2,p_3)} \, ,
\end{align}
where $\tilde{k}_i = k_i-p_1-p_2-p_3$.

\item Sector \{101010100\}: no ISP.
\begin{align}
\hat{\varphi}_{19} &= \frac{s^3 (1-2 \epsilon )^2 \sqrt{s \left(s-4 m^2\right)}}{16\epsilon^2 z_1 z_3 z_5 z_7 \, G(k_1,p_1+p_2) \, G(k_2,p_1+p_2)} \, .
\end{align}

\item Sector \{100111000\}: $z_7$ as ISP.
\begin{align}
\hat{\varphi}_{20} &= \frac{(1-2\epsilon)s \, G(k_2)}{4\epsilon z_1 z_4 z_5 z_6 \, G(k_1,k_2)} \, .
\end{align}

\item Sector \{010111000\}: $z_8$ as ISP.
\begin{align}
\hat{\varphi}_{21} &= \frac{(1-2\epsilon)t \, G(k_2-p_1)}{\epsilon z_2 z_4 z_5 z_6 \, G(k_1-p_1,k_2-p_1)}
\end{align}

\item Sector \{010110100\}: $z_8$ as ISP.
\begin{align}
\hat{\varphi}_{22} &= -\frac{s (1-2 \epsilon ) \left(m^2+z_8\right)}{4\epsilon z_2 z_4 z_5 z_7 \, G(k_1-p_1,k_2-p_1)} \, , \nonumber \\
\hat{\varphi}_{23} &=  \frac{s^2 z_8 (1-2 \epsilon ) \sqrt{s \left(s-4 m^2\right)} \left(m^2+z_8\right)}{16\epsilon z_2 z_4 z_5 z_7 \, G(k_1-p_1,k_2-p_1) \, G(k_2-p_1,p_1,p_2)} \, , \nonumber \\
\hat{\varphi}_{24} &= -\frac{s z_8 (1-2 \epsilon ) \left(m^2+z_8\right)}{4\epsilon z_2 z_4 z_5 z_7 \, G(k_1-p_1,k_2-p_1) \, G(k_2-p_1,p_1,p_2)}\frac{\partial G(k_2-p_1,p_1,p_2)}{\partial z_8} \, .
\end{align}

\item Sector \{000111100\}: $z_1$ as ISP.
\begin{align}
\hat{\varphi}_{25} &= -\frac{(1-2 \epsilon )s \, G(k_2)}{4\epsilon z_4 z_5 z_6 z_7 \, G(k_1,k_2)} \, .
\end{align}

\item Sector \{101100000\}: no ISP.
\begin{align}
\hat{\varphi}_{26} &= \frac{s^2 (1-2 \epsilon)(1-\epsilon)}{4\epsilon^2 z_1 z_3 z_4 \, G(k_2) \, G(k_1,p_1+p_2)} \, .
\end{align}

\item Sector \{100110000\}: $z_3$ as ISP.
\begin{align}
\hat{\varphi}_{27} &= \frac{s z_3 (1-2 \epsilon )^2 \sqrt{s \left(s-4 m^2\right)}}{16\epsilon^2 z_1 z_4 z_5 \, G(\tilde{k}_1,\tilde{k}_2) \, G(\tilde{k}_1,p_1+p_2)} \, , \nonumber \\
\hat{\varphi}_{28} &= \frac{s z_3 (1-2 \epsilon )^2 \left(z_3-s\right)}{16\epsilon^2 z_1 z_4 z_5 \, G(\tilde{k}_1,\tilde{k}_2) \, G(\tilde{k}_1,p_1+p_2)} \, ,
\end{align}
where $\tilde{k}_i=k_i-p_1-p_2$.

\item Sector \{010101000\}: $z_9$ as ISP.
\begin{align}
\hat{\varphi}_{29} &= \frac{t z_9 (1-2 \epsilon )^2 \sqrt{t \left(t-4 m^2\right)}}{16\epsilon^2 z_2 z_4 z_6 \, G(\tilde{k}_1,\tilde{k}_2) \, G(\tilde{k}_1,p_2+p_3)} \, , \nonumber \\
\hat{\varphi}_{30} &= \frac{t z_9 (1-2 \epsilon )^2 \left(z_9-t\right)}{16\epsilon^2 z_2 z_4 z_6 \, G(\tilde{k}_1,\tilde{k}_2) \, G(\tilde{k}_1,p_2+p_3)} \, ,
\end{align}
where $\tilde{k}_i=k_i-p_1-p_2-p_3$.

\item Sector \{000110100\}: $z_1$ as ISP.
\begin{align}
\hat{\varphi}_{31} &= \frac{s (1-2 \epsilon )^2 \sqrt{s \left(s-4 m^2\right)} \, G(k_2)}{16\epsilon^2 z_4 z_5 z_7 \, G(k_1,k_2) \, G(k_2,p_1+p_2)} \, .
\end{align}

\item Sector \{000110000\}: no ISP.
\begin{align}
\hat{\varphi}_{32}= \frac{(1-\epsilon )^2}{\epsilon^2 z_4 z_5 \, G(k_1-k_2) \, G(k_2-p_1-p_2)} \, .
\end{align}
\end{itemize}

\section{Further details for the outer-massive double box family}
\label{app:omdb}

\subsection{Auxiliary $d\log$s for the top sector}

\begin{table}[t!]
    \centering
    \begin{tabular}{cc}
        \hline
        Sector & Auxiliary $d\log$s \\[8pt]
        011111100 & $\frac{s(z_5-z_8)}{z_2z_3z_4z_5z_6z_7},\, \frac{(z_5-z_8)}{z_2z_3z_4z_5z_6z_7}\frac{E_1}{P_1},\, \frac{(z_5-z_8)(z_5+z_7)}{z_2z_3z_4z_5z_6z_7}\frac{E_2}{P_1}$ \\[6pt]
        110111100 & $\frac{s(z_7-z_8)}{z_1z_2z_4z_5z_6z_7},\, \frac{(z_7-z_8)}{z_1z_2z_4z_5z_6z_7}\frac{E_1}{P_1},\, \frac{(z_7-z_8)(z_5+z_7)}{z_1z_2z_4z_5z_6z_7}\frac{E_2}{P_1}$ \\[6pt]
        101111100 & $\frac{s(z_5+z_7)}{z_1z_3z_4z_5z_6z_7},\, \frac{(z_5+z_7)}{z_1z_3z_4z_5z_6z_7}\frac{E_1}{P_1},\, \frac{1}{z_1z_3z_4z_5z_6z_7}\frac{T E_2}{P_1}$ \\[6pt]
        111011100 & $\frac{s^2}{z_1z_2z_3z_5z_6z_7},\, \frac{s}{z_1z_2z_3z_5z_6z_7}\frac{E_1}{P_1},\, \frac{s(z_5+z_7)}{z_1z_2z_3z_5z_6z_7}\frac{E_2}{P_1}$ \\[6pt]
        111110100 & $\frac{s}{z_1z_2z_3z_4z_5z_7}\frac{sF_1}{P_1}, \quad \frac{s(z_5+z_7)}{z_1z_2z_3z_4z_5z_7}\frac{sF_2}{P_1}$ \\[6pt]
        \hline
    \end{tabular}
    \caption{Auxiliary $d\log$s needed for the outer-massive double box family.}
    \label{eq:omdbauxdlog}
\end{table}

The only non-trivial conversion from $d\log$-forms to Feynman integrals in the outer-massive double box family is that of $\hat{\varphi}^{\prime}_{3}$ in \eqref{eq:topsectordlogs}. We apply the same method as in the inner-massive double box family. We construct the auxiliary $d\log$s in the sub-sectors and list them in Table~\ref{eq:omdbauxdlog}. The polynomials appearing in the numerators are given by
\begin{align}
    E_1&=st-tz_5-tz_7-sz_6+(s+2t)z_8 \,, \nonumber \\
    E_2&=(z_5+z_7)E_1+2(s+2t)P_1-(s+2z_8)E_1 \,, \nonumber \\
    T&=\lambda(s,z_5,z_7)-4sm^2 \,, \nonumber \\
    F_1&=(z_8+2m^2)(z_5+z_7)+2P_1-(s+2z_8)(z_8+2m^2) \,, \nonumber \\
    F_2&=z_8+2m^2 \, ,
\end{align}
and $P_{1}$ is defined in Eq.~\eqref{eq:emdbtopsector}.

We subtract the top-sector components of
\begin{align}
\bra{\varphi'_3} = -2s^2 \bra{F_{1111111-10}} - 2s \bra{F_{1111111-1-1}} + \text{sub-sector integrals} \,,
\end{align}
to arrive at the remainder
\begin{equation}
\label{eq:emdbR3}
\mathcal{R}_{3}= \frac{\mathcal{N}}{2z_1z_2z_3z_4z_5z_6z_7P_1}-\frac{st(z_5+z_7)}{z_1z_2z_3z_4z_5z_6z_7} \, .
\end{equation}
The numerator $\mathcal{N}$ is given by
\begin{equation}
    \begin{aligned}
        \mathcal{N}&=z_1\mathcal{N}_1+z_2\mathcal{N}_2+z_3\mathcal{N}_3+z_4\mathcal{N}_4+z_6\mathcal{N}_6 \,, \\
        \mathcal{N}_1&=(z_5-z_8)E_1-(z_5-z_8)(z_5+z_7)E_2-2(st+sz_5-sz_8+2tz_5)P_1 \,, \\
        \mathcal{N}_2&=(z_5+z_7)E_1-TE_2+2s(s-z_5-z_7+2z_6)P_1 \,, \\
        \mathcal{N}_3&=(z_7-z_8)E_1-(z_7-z_8)(z_5+z_7)E_2-2(st+sz_7-sz_8+2tz_7)P_1 \,, \\
        \mathcal{N}_4&=-sE_1+s(z_5+z_7)E_2+2s(s+2t)P_1 \,, \\
        \mathcal{N}_6&=s^2(s+2z_8)(z_8+2m^2)=-s^2F_1+s^2(z_5+z_7)F_2+2s^2P_1 \,.
    \end{aligned}
\end{equation}
We can now use the auxiliary $d\log$s listed in Table~\ref{eq:omdbauxdlog} to cancel the terms with a $P_1$ denominator. We then have
\begin{align}
    \hat{\varphi}_{3} &=\frac{1}{z_1z_2z_3z_4z_5z_6z_7} \left[ -2s^2z_8-2sz_8z_9
     -(st+2tz_5)z_1-(st+2tz_7)z_3 \right. \nonumber
     \\
     &\hspace{10em} \left. +(s^2+2sz_6)z_2+2stz_4-st(z_5+z_7)+s^2z_6 \right] .
\end{align}
This is apparently a combination of Feynman integrals and we can arrive at the final result in Eq.~\eqref{eq:topsectorfeynint}.

\subsection{Some relations used in the construction for the top sector}

In this appendix, we discuss the relations among the polynomials $N_1$, $N_2$, $Q_1$, $R_2$ and $z_9R_2$ appearing in and below Eq.~\eqref{eq:imdb_R2}. We'd like to show that both $R_2$ and $z_9R_2$ can be expressed in the form
\begin{equation}
c_1 z_9N_1 + c_2 z_9 N_2 + c_3(z_9) Q_1 \, ,
\label{eq:finalform}
\end{equation}
where $c_1$ and $c_2$ are independent of $z_9$, and $c_3(z_9)$ is at most linear in $z_9$.

We first note that $R_2$ and $Q_1$ appear in the polynomial $\bar{P}_0(z_8,z_9)$ in Eq.~\eqref{eq:P0bar}:
\begin{align}
    \bar{P}_0(z_8,z_9) &= Q_1(z_9) z_8^2 + R_2(z_9) z_8 + C(z_9) \, ,
\end{align}
where we have suppressed the dependence on $z_5$ and $z_7$. In the construction for the variable $z_9$, we need to employ Sylvester's determinant identity Eq.~\eqref{eq:Sylvester}. In the current case it reads 
\begin{align}
    R_2^2-4Q_1C &= \bar{Q}_{2}(z_9) \, \bar{Q}_{3}(z_9) \, .
    \label{eq:Q2Q3}
\end{align}
The polynomials $N_1$ and $N_2$ comes from the square roots of $\bar{Q}_2$ (see Section~\ref{sec:square_roots}). Hence one can imagine that $R_2$, $N_1$ and $N_2$ are related through $Q_1$ and $\bar{Q}_2$. In fact, the relations are not restricted to the special case here, but are universally applicable to quadratic polynomials satisfying Sylvester's determinant identity. So hereafter we'll take $Q_1$, $Q_2$, $Q_3$, $R_2$ and $C$ to be generic quadratic polynomials of the variable $z$, where $Q_2$ and $Q_3$ do not share common factors. These polynomials satisfy
\begin{equation}
\left[ R_2(z) \right]^2 - 4 Q_1(z) C(z) = Q_2(z) Q_3(z) \, .
\label{eq:R2Q2Q3}
\end{equation}
Writing $Q_1$ as $(z-c_+)(z-c_-)$, we have
\begin{align}
    \frac{N_1(z)}{Q_1\sqrt{Q_2}} &=\frac{\sqrt{Q_2(z=c_{+})}}{(z-c_{+})\sqrt{Q_2}} + \frac{\sqrt{Q_2(z=c_{-})}}{(z-c_{-})\sqrt{Q_2}}\,,\nonumber
    \\
    \frac{(c_{+}-c_{-})N_2(z)}{Q_1\sqrt{Q_2}} &=\frac{\sqrt{Q_2(z=c_{+})}}{(z-c_{+})\sqrt{Q_2}} - \frac{\sqrt{Q_2(z=c_{-})}}{(z-c_{-})\sqrt{Q_2}}\,.
    \label{eq:N1N2_roots}
\end{align}
According to Eq.~\eqref{eq:R2Q2Q3}, we know that
\begin{align}
   \sqrt{Q_2(z=c_{\pm}) \, Q_{3}(z=c_{\pm})} = R_{2}(z=c_{\pm}) \, ,
    \label{eq:generalR2}
\end{align}
is a polynomial. Hence we can define the linear functions $S_2(z)$ and $S_3(z)$ (similar to those in Eqs.~\eqref{eq:generalsqrtQ} and \eqref{eq:generalR}), such that
\begin{align}
    \sqrt{Q_2(z=c_{\pm})} = S_2(z=c_{\pm}) \, , \quad
    \sqrt{Q_3(z=c_{\pm})} = S_3(z=c_{\pm}) \, .
    \label{eq:S2S3}
\end{align}
Plugging the above equations back to Eq.~\eqref{eq:N1N2_roots}, we find
\begin{align}
    \frac{N_1(z)}{Q_1\sqrt{Q_2}} &=\frac{(z-c_{-}) \, S_2(z=c_{+})+(z-c_{+}) \, S_2(z=c_{-})}{Q_1\sqrt{Q_2}}\, , \nonumber \\
    \frac{(c_{+}-c_{-})N_2(z)}{Q_1\sqrt{Q_2}} &=\frac{(z-c_{-}) \, S_2(z=c_{+})-(z-c_{+}) \, S_2(z=c_{-})}{Q_1\sqrt{Q_2}}\, .
    \label{eq:squareout}
\end{align}
Using that $S_2(z)$ is a linear function of $z$, we can rewrite the above in a more instructive form:
\begin{align}
    N_2(z) = S_2(z) \equiv A_2 z + B_2 \, , \quad
    2z\, N_2(z) = N_1(z) + (c_+ + c_-) N_2(z) + 2A_2 \, Q_1(z) \, .
    \label{eq:N1N2}
\end{align}

Now, using $N_2(z)=S_2(z)$ together with Eqs.~\eqref{eq:R2Q2Q3}, \eqref{eq:generalR2} and \eqref{eq:S2S3}, we can deduce that
\begin{equation}
R_2(z) = a \, Q_1(z) + A_3 \, zN_2(z) + B_3 N_2(z) \, ,
\end{equation}
where we have written $S_3(z) \equiv A_3 z + B_3$. We can get rid of the $B_3 N_2$ term using the following identity
\begin{align}
    2c_{+}c_{-}N_{2}(z)=-zN_{1}(z)+(c_{+}+c_{-})zN_{2}(z)+2B_2Q_{1}(z) \, ,
\end{align} 
which follows from Eq.~\eqref{eq:N1N2}. This shows that $R_2(z)$ can be written in the form of Eq.~\eqref{eq:finalform}, as presented in Eq.~\eqref{eq:imdb_R2}. To show that $z R_2(z)$ can also be written in this way, we just need to note that we can use the second equation in Eq.~\eqref{eq:N1N2} to reduce the power of $z$ in front of $N_2(z)$. Our purpose is then achieved.

We finally note that $Q_2$ and $Q_3$ are symmetric, and the above arguments also apply in case $1/\sqrt{Q_3}$ appears in the construction of $d\log$-forms.

\subsection{List of $d\log$-forms for all sectors}

Here we list all $d\log$-forms in the outer-massive double box family. The denominators $z_i$ are given by
\begin{align}
&\{k_1^2-m^2,\,(k_1-p_1)^2-m^2,\,(k_1-p_1-p_2)^2-m^2,\,(k_1-k_2)^2,\,(k_2-p_1-p_2)^2-m^2, \nonumber
\\
&(k_2-p_1-p_2-p_3)^2-m^2,\,k_2^2-m^2,\,(k_2-p_1)^2-m^2,\,(k_1-p_1-p_2-p_3)^2-m^2 \} \, .
\end{align}

\begin{itemize}

\item Sector \{111111100\}: $z_8$ as ISP.
\begin{align}
    \hat{\varphi}_{1}&=\frac{\sqrt{s(s-4m^2)}\sqrt{st(st-4m^2(s+t))}}{z_1z_2z_3z_4z_5z_6z_7} \, , \nonumber \\
    \hat{\varphi}_{2}&=\frac{\sqrt{s(s-4m^2)}(s z_8+t(z_5 + z_7))}{z_1z_2z_3z_4z_5z_6z_7} \, , \nonumber \\
    \hat{\varphi}_{3}&=\frac{1}{z_1z_2z_3z_4z_5z_6z_7} \left[ -2s^2z_8-2sz_8z_9
     -(st+2tz_5)z_1-(st+2tz_7)z_3 \right. \nonumber
     \\
     &\hspace{10em} \left. +(s^2+2sz_6)z_2+2stz_4-st(z_5+z_7)+s^2z_6 \right] .
\end{align}

\item Sector \{111111000\}: $z_9$ as ISP.
\begin{align}
    \hat{\varphi}_{4}&=\frac{\sqrt{st(st-4m^2(s+t))}}{z_1z_2z_3z_4z_5z_6} \, , \nonumber  \\
    \hat{\varphi}_{5}&=\frac{s(z_3-z_9)}{z_1z_2z_3z_4z_5z_6} \, .
\end{align}

\item Sector \{111011100\}: no ISP.
\begin{align}
\hat{\varphi_{6}} &= \frac{s^2}{z_1z_2z_3z_5z_6z_7} \, .
\end{align}

\item Sector \{111101000\}: $z_9$ as ISP.
\begin{align}
    \hat{\varphi}_{7} &= \frac{1-2\epsilon}{\epsilon}\frac{(z_9+m^2)\sqrt{st(st-4m^2(s+t))}}{4z_1z_2z_3z_4z_6 \, G(k_1-p_1-p_2-p_3,k_2-p_1-p_2-p_3)} \, ,\nonumber \\
    \hat{\varphi}_{8} &= \frac{1-2\epsilon}{\epsilon}\frac{(z_9+z_4-z_6)\sqrt{s(s(t-m^2)^2-4m^2t^2)}}{4z_1z_2z_3z_4z_6 \, G(k_1-p_1-p_2-p_3,k_2-p_1-p_2-p_3)} \, , \nonumber \\
    \hat{\varphi}_{9} &= \frac{1-2\epsilon}{\epsilon}\frac{sz_9(z_9+m^2)}{4z_1z_2z_3z_4z_6 \, G(k_1-p_1-p_2-p_3,k_2-p_1-p_2-p_3)} \, .
\end{align}

\item Sector \{111010100\}: no ISP.
\begin{align}
    \hat{\varphi}_{10}= \frac{1-2\epsilon}{\epsilon}\frac{s^2\sqrt{s(s-4m^2)}}{4z_1z_2z_3z_5z_7 \, G(k_2,p_1+p_2)} \,.
\end{align}

\item Sector \{110111000\}: $z_3$ and $z_9$ as ISPs.
\begin{align}
    &\hat{\varphi}_{11}= \frac{s+t}{z_1z_2z_4z_5z_6} \, , \nonumber \\
    &\hat{\varphi}_{12}= \frac{(z_5-z_6)(z_3-z_9)\sqrt{st(st-4m^2(s+t))}}{4z_1z_2z_4z_5z_6 \, G(k_2-p_1-p_2,k_1-p_1-p_2,p_3)} \, , \nonumber \\
    &\hat{\varphi}_{13}= \frac{(z_5-z_6)(z_3-z_9)(sz_9+tz_3)}{4z_1z_2z_4z_5z_6 \, G(k_2-p_1-p_2,k_1-p_1-p_2,p_3)} \, .
\end{align}

\item Sector \{101111000\}: $z_9$ as ISP (super-sector with $z_7$ for $\hat{\varphi}_{15}$).
\begin{align}
    \hat{\varphi}_{14} &= \frac{s}{z_1z_3z_4z_5z_6} \, , \nonumber \\
    \hat{\varphi}_{15} &= \frac{s\sqrt{s(s-4m^2)}}{z_1z_3z_4z_5z_6z_7} \, .
\end{align}

\item Sector \{101101000\}: $z_9$ as ISP.
\begin{align}
    \hat{\varphi}_{16} &= -\frac{1-2\epsilon}{\epsilon}\frac{4s(z_9+z_4-z_6)}{z_1z_3z_4z_6 \, G(\tilde{k}_1,\tilde{k}_2)} \, , \nonumber \\
    \hat{\varphi}_{17} &= \frac{1-2\epsilon}{\epsilon}\frac{s(z_9+m^2)}{4z_1z_3z_4z_6 \, G(\tilde{k}_1,\tilde{k}_2)} \, , \nonumber \\
    \hat{\varphi}_{18} &= -\frac{1-2\epsilon}{\epsilon}\frac{s^2z_9(z_9+m^2)\sqrt{s(s-4m^2)}}{16z_1z_3z_4z_6 \, G(\tilde{k}_1,\tilde{k}_2) \, G(\tilde{k}_1,p_1+p_2,p_3)} \, ,
\end{align}
where $\tilde{k}_1=k_1-p_1-p_2-p_3$ and $\tilde{k}_2=k_2-p_1-p_2-p_3$.

\item Sector \{111010000\}: no ISP.
\begin{align}
    \hat{\varphi}_{19}= \frac{1-\epsilon}{\epsilon}\frac{s}{z_1z_2z_3z_5 \, G(k_2-p_1-p_2)}  \, .
\end{align}

\item Sector \{011111000\}: $z_9$ as ISP.
\begin{align}
    \hat{\varphi}_{20}= \frac{t}{z_2z_3z_4z_5z_6} \, .
\end{align}

\item Sector \{110110000\}: $z_3$ as ISP.
\begin{align}
    \hat{\varphi}_{21}= \frac{1-2\epsilon}{\epsilon}\frac{s \, G(k_1-p_1-p_2)}{4z_1z_2z_4z_5 \, G(k_1-p_1-p_2,k_2-p_1-p_2)} \, .
\end{align}

\item Sector \{110101000\}: $z_9$ as ISP.
\begin{align}
\hat{\varphi}_{22}= \frac{1-2\epsilon}{\epsilon}\frac{t \, G(k_1-p_1-p_2)}{4z_1z_2z_4z_6 \, G(k_1-p_1-p_2,k_2-p_1-p_2)} \, .
\end{align}

\item Sector \{101010100\}: no ISP.
\begin{align}
\hat{\varphi}_{23}= \frac{(1-2\epsilon)^2}{\epsilon^2}\frac{s^3(s-4m^2)}{16z_1z_3z_5z_7 \, G(k_1,p_1+p_2) \, G(k_2,p_1+p_2)} \, .
\end{align}

\item Sector \{100110000\}: $z_3$ as ISP.
\begin{align}
    &\hat{\varphi}_{24}= \frac{(1-2\epsilon)^2}{\epsilon^2}\frac{s\sqrt{s(s-4m^2)} \, G(\tilde{k}_1)}{15z_1z_4z_5 \, G(\tilde{k}_1,\tilde{k}_2) \, G(\tilde{k}_1,p_1+p_2)} \, , \nonumber \\
    &\hat{\varphi}_{25}= \frac{(1-2\epsilon)^2}{\epsilon^2}\frac{sz_3 \, G(\tilde{k}_1)}{16z_1z_4z_5 \, G(\tilde{k}_1,\tilde{k}_2) \, G(\tilde{k}_1,p_1+p_2)} \, ,
\end{align}
where $\tilde{k}_1=k_1-p_1-p_2$ and $\tilde{k}_2=k_2-p_1-p_2$.

\item Sector \{010101000\}: $z_9$ as ISP.
\begin{align}
    &\hat{\varphi}_{26}=  \frac{(1-2\epsilon)^2}{\epsilon^2}\frac{t\sqrt{t(t-4m^2)} \, G(\tilde{k}_1)}{16z_2z_4z_6 \, G(\tilde{k}_1,\tilde{k}_2) \, G(\tilde{k}_1,p_2+p_3)} \, , \nonumber \\
    &\hat{\varphi}_{27}= \frac{(1-2\epsilon)^2}{\epsilon^2}\frac{tz_9 \, G(\tilde{k}_1)}{16z_2z_4z_6 \, G(\tilde{k}_1,\tilde{k}_2) \, G(\tilde{k}_1,p_2+p_3)} \, ,
\end{align}
where $\tilde{k}_1=k_1-p_1-p_2-p_3$ and $\tilde{k}_2=k_2-p_1-p_2-p_3$.

\item Sector \{101010000\}: no ISP.
\begin{align}
\hat{\varphi}_{28}= \frac{(1-2\epsilon)(1-\epsilon)}{\epsilon^2}\frac{s\sqrt{s(s-4m^2)}}{4z_1z_3z_5 \, G(k_2-p_1-p_2) \, G(k_1,p_1+p_2)} \, .
\end{align}

\item Sector \{100010000\}: no ISP.
\begin{align}
\hat{\varphi}_{29}= \frac{(1-\epsilon)^2}{\epsilon^2}\frac{1}{z_1z_5 \, G(k_1) \, G(k_2-p_1-p_2)} \, .
\end{align}

\end{itemize}

\section{Massless double box}\label{sec:masslessdb}

\begin{figure}[t!]
\centering
\includegraphics[width=0.6\textwidth]{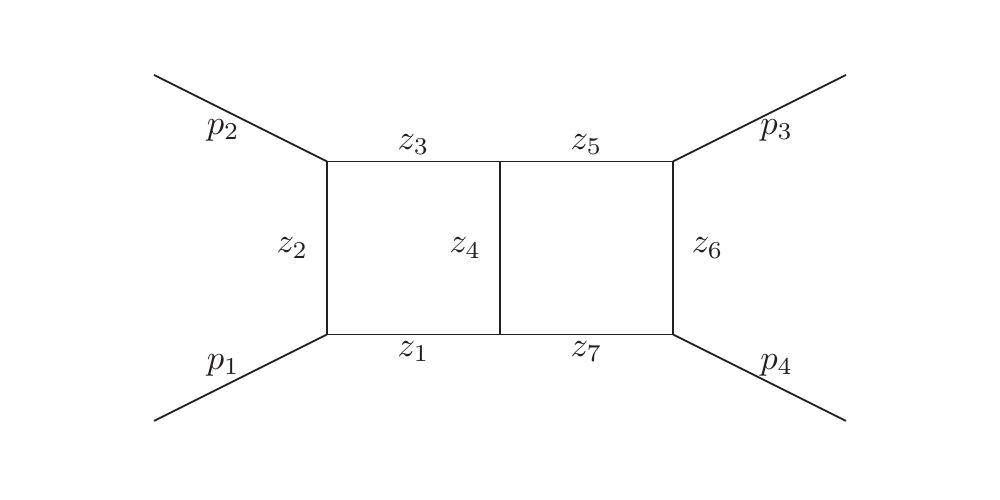}
\caption{\label{fig:masslessdb}Massless double box. All propagators and external legs are massless.}
\end{figure}

This is a simpler example since we have fewer mass scales in the problem. The diagram is depicted in Fig.~\ref{fig:masslessdb}, where all external momenta are outgoing. The propagator denominators $\{z_i\}$ ($i=1,\ldots,9$) are given by
\begin{multline}
\label{eq:plist}
\{ k_1^2, \, (k_1-p_1)^2, \, (k_1-p_1-p_2)^2, \, (k_1-k_2)^2,
\\
(k_2-p_1-p_2)^2, \, (k_2-p_1-p_2-p_3)^2, \, k_2^2, \, (k_2-p_1)^2, \, (k_1-p_1-p_2-p_3)^2 \} \, .
\end{multline}
The momentum invariants are $p_i^2 = 0$ for $i=1,\ldots,4$, and
\begin{equation}
(p_1+p_2)^2=s \, , \quad (p_2+p_3)^2=t \, , \quad (p_1+p_3)^2 = -s-t \, .
\end{equation}

We consider the top sector \{1,1,1,1,1,1,1,0,0\} and its sub-sectors. Using \kira we find 8 MIs in total, spanning the top sector and 6 sub-sectors. We list the $d\log$-forms in the following:
\begin{itemize}

\item Sector \{1,1,1,1,1,1,1,0,0\}: $z_9$ as ISP.
\begin{equation}
\hat{\varphi}_{1} = \frac{s^2 t}{z_1z_2z_3z_4z_5z_6z_7} \, , \quad
\hat{\varphi}_{2} = \frac{s^2 z_9}{z_1z_2z_3z_4z_5z_6z_7} \, .
\end{equation}

\item Sector \{1,1,1,1,0,1,0,0,0\}: $z_9$ as ISP.
\begin{equation}
\hat{\varphi}_{3} = -\frac{1-2\epsilon}{\epsilon} \frac{s t z_9}{4z_1z_2z_3z_4z_6 \, G(k_1-k_2, k_1-p_1-p_2-p_3) } \, .
\end{equation}

\item Sector \{1,1,0,1,1,1,0,0,0\}: $z_7$ and $z_8$ as ISPs.
\begin{equation}
    \hat{\varphi}_{4}=\frac{s+t}{z_1z_2z_4z_5z_6} \, .
\end{equation}

\item Sector \{101101000\}: $z_9$ as ISP.
\begin{equation}
    \hat{\varphi}_{5}=-\frac{1-2\epsilon}{\epsilon}\frac{sz_9}{4z_1z_3z_4z_6 \, G(k_2-p_1-p_2-p_3,k_1-p_1-p_2-p_3)} \, .
\end{equation}

\item Sector \{101010100\}: no ISP.
\begin{equation}
    \hat{\varphi}_{6}= \frac{(1-2\epsilon)^2}{\epsilon^2}\frac{s^4}{16z_1z_3z_5z_7 \, G(k_1,p_1+p_2) \, G(k_2,p_1+p_2)} \, .
\end{equation}

\item Sector \{010101000\}: $z_8$ as ISP.
\begin{equation}
    \hat{\varphi}_{7}=\frac{(1-2\epsilon)^2}{\epsilon^2}\frac{t^2z_8}{16z_2z_4z_6 \, G(k_2-p_1,k_1-p_1) \, G(k_2-p_1,p_2+p_3)} \, .
\end{equation}

\item Sector \{001100100\}: $z_5$ as ISP.
\begin{equation}
    \hat{\varphi}_{8}=\frac{(1-2\epsilon)^2}{\epsilon^2}\frac{s^2z_5}{16z_3z_4z_7 \, G(k_2-p_1-p_2,k_1-p_1-p_2) \, G(k_2-p_1-p_2,p_1+p_2)} \, .
\end{equation}
Note that this sector is symmetric with respect to the previous one under the replacements
\begin{equation}
z_3\leftrightarrow z_2 \, , \quad z_5\leftrightarrow z_8 \, , \quad z_7\leftrightarrow z_6 \, , \quad s \leftrightarrow t \, .
\end{equation}
\end{itemize}

Some of the above $d\log$-forms already appear to be Feynman integrals, and the others can be converted using dimensional recurrence relations. Hence we don't bother to invoke intersection theory here. The results are given by
\begin{align}
\bra{\varphi_1}&= s^2 t \bra{F_{1111111}} , \nonumber \\
\bra{\varphi_2}&= s^2 \bra{F_{11111110-1}} , \nonumber \\
\bra{\varphi_3}&=\frac{3 (2 \epsilon -1) (3 \epsilon -1) }{2 \epsilon^2}\bra{F_{1011010}}+\frac{3 s (2 \epsilon -1)}{\epsilon }\bra{F_{1111010}} , \nonumber \\
\bra{\varphi_4}&=(s+t)\bra{F_{1101110}}  , \nonumber \\
\bra{\varphi_5}&=-\frac{(2 \epsilon -1) (3 \epsilon -1) }{2 \epsilon^2}\bra{F_{1011010}} , \nonumber \\
\bra{\varphi_6}&=\frac{(2 \epsilon -1)^2}{\epsilon ^2}\bra{F_{1010101}} , \nonumber \\
\bra{\varphi_{7}}&=\frac{3 (2 \epsilon -1) (3 \epsilon -2) (3 \epsilon -1)}{2 t \epsilon ^3}\bra{F_{0101010}} , \nonumber \\
\bra{\varphi_{8}}&=\frac{3 (2 \epsilon -1) (3 \epsilon -2) (3 \epsilon -1)}{2 s \epsilon ^3}\bra{F_{0011001}} .
\end{align}

It is straightforward to derive the differential equations of the above basis with respect to $s$ and $t$. We may multiply the basis by a factor of $(-s)^{2\epsilon}$ to make it dimensionless, and introduce the dimensionless variable $x=t/s$. Denoting the basis as $\vec{\phi}$, we can write the differential equations as
\begin{equation}
\partial_{x}\vec{\phi} = \epsilon \left( \frac{A_1}{x} + \frac{A_2}{x+1}\right) \vec{\phi} \, ,
\end{equation}
where the two matrices are given by    
\begin{align}
A_1&=
\begin{pmatrix}
-2 & 0 & -4 & 12 & 0 & 0 & -4 & -4 \\
-1 & 1 & -4 & 18 & 3 & -1 & -6 & -4 \\
0 & 0 & -2 & 0 & 0 & 0 & -2 & 0 \\
0 & 0 & 0 & -2 & 0 & 0 & -2/3 & 2/3 \\
0 & 0 & 0 & 0 & 0 & 0 & 0 & 0 \\
0 & 0 & 0 & 0 & 0 & 0 & 0 & 0 \\
0 & 0 & 0 & 0 & 0 & 0 & -2 & 0 \\
0 & 0 & 0 & 0 & 0 & 0 & 0 & 0
\end{pmatrix}
, \nonumber
\\
A_2&=
\begin{pmatrix}
2 & -2 & 4 & -12 & 6 & -2 & 4 & 8 \\
1 & -1 & 4 & -18 & -3 & -1 & 6 & 4 \\
0 & 0 & 1 & 0 & -3 & 0 & 2 & 0 \\
0 & 0 & 0 & 2 & 0 & 0 & 0 & 0 \\
0 & 0 & 0 & 0 & 0 & 0 & 0 & 0 \\
0 & 0 & 0 & 0 & 0 & 0 & 0 & 0 \\
0 & 0 & 0 & 0 & 0 & 0 & 0 & 0 \\
0 & 0 & 0 & 0 & 0 & 0 & 0 & 0 \\
\end{pmatrix}
.
\end{align}
We see that the equations are of the $\epsilon$-form, and the solutions can be easily expressed as HPLs.

\section{The two-loop triangle family relevant to the $HW^+W^-$ vertex}
\label{sec:HWW}

\begin{figure}[t!]
    \centering
    \includegraphics[width=0.6\textwidth]{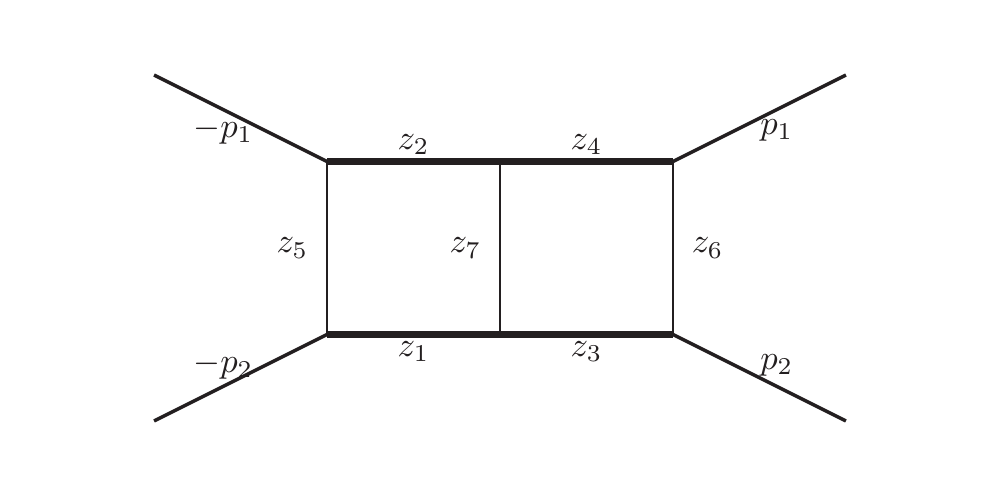}
    \caption{\label{fig:hww}The integral family relevant to the $HW^+W^-$ vertex. Note that we only consider its various sub-sectors obtained by pinching some propagators. Internal thick lines represent propagators with mass $m_t$, while internal thin lines represent massless propagators. The external momenta are outgoing with $p_1^{2}=m_{1}^{2}$ and $p_{2}^{2}=m_{2}^{2}$.}
\end{figure}

This two-loop triangle family is defined by four massive and three massless propagators. The diagram is depicted in Fig.~\ref{fig:hww}, where all external momenta are outgoing. The propagator denominators are given by
\begin{align}
&\{k_1^2 - m_t^2 , (k_1-p_1-p_2)^2 - m_t^2, k_2^2 - m_t^2, (k_2-p_1-p_2)^2 - m_t^2, (k_1-p_2)^2, \nonumber \\
&\;(k_2-p_2)^2, (k_1-k_2)^2  \} \, ,
\end{align}
where the external momenta $p_1$ and $p_2$ satisfy 
\begin{align}
p_1^2 = m_1^2 \, , \quad p_2^2 = m_2^2 \, , \quad (p_1+p_2)^2 = s \, .
\end{align}
For convenience we define the abbreviation $\lambda \equiv \lambda(s, m_1^2, m_2^2)$ where the K\"all\'en function is defined as
\begin{align}
\lambda(x, y, z) = x^2 + y^2 +z^2 -2xy-2yz-2xz \, .
\end{align}
This family is relevant to the $HW^+W^-$ vertex in the standard model \cite{DiVita:2017xlr, Ma:2021cxg}. There are 38 MIs in 24 unique sectors. We perform the construction as follows:
\begin{itemize}
\item Sector \{1, 1, 1, 1, 1, 1, 0\}, $z_7$ as ISP.
\begin{align}
\hat{\varphi}_1 &= \frac{\lambda}{z_1z_2z_3z_4z_5z_6} \,.
\end{align}

\item Sector \{1, 1, 1, 1, 0, 1, 0\}: no ISP.
\begin{align}
\hat{\varphi}_2 &= -\frac{1-2\epsilon}{\epsilon}\frac{s\sqrt{\lambda}\sqrt{s(s-4m_t^2)}}{4z_1z_2z_3z_4z_6 \, G(k_1, p_1+p_2)} \, .
\end{align}

\item Sector \{1, 1, 1, 1, 0, 0, 0\}: no ISP.
\begin{align}
\hat{\varphi}_3 &= \frac{(1-2\epsilon)^2}{\epsilon^2} \frac{s^3(s-4m_t^2)}{16z_1z_2z_3z_4 \, G(k_1, p_1+p_2) \, G(k_2, p_1+p_2)}.
\end{align}

\item Sector \{1, 0, 1, 1, 1, 0, 1\}: $z_6$ as ISP.
\begin{align}
P_{2}(z_1, z_3, z_5, z_6, z_7) &=-4G(k_1, k_2, p_2) \,,  \nonumber  \\
P_{3}(z_3, z_4, z_6) &= -4G(k_2, p_1, p_2) \,, \nonumber \\
\hat{\varphi}_4 &= \frac{\sqrt{\lambda}}{z_1z_3z_4z_5z_7} \,,  \nonumber\\
\hat{\varphi}_5 &= \frac{\sqrt{\lambda}\sqrt{s(s-4m_t^2)}}{z_1z_3z_4z_5z_7}\frac{1}{P_{3}}\frac{\partial P_{3}}{\partial z_4} \,,  \nonumber\\
\hat{\varphi}_6 &= \frac{1}{z_1z_3z_4z_5z_7}\left[ \frac{1}{P_{3}} \frac{\partial P_{3}}{\partial z_4} \frac{\partial P_{3}}{\partial z_6} - 2 \frac{\partial^2 P_{3}}{\partial z_4 \partial z_6} \right] \,,  \nonumber\\
\hat{\varphi}_7 &= \frac{\sqrt{\lambda}(m_2^2 -m_t^2)}{z_1z_3z_4z_5z_7} \frac{1}{P_{2}}\frac{\partial P_{2}}{\partial z_5} \,.
\end{align}

\item Sector \{1, 0, 1, 1, 1, 1, 0\}: no ISP.
\begin{align}
\hat{\varphi}_8 &= -\frac{1-2\epsilon}{\epsilon} \frac{\sqrt{\lambda} m_2^2(m_2^2 -m_t^2)}{4z_1z_3z_4z_5z_6 G(k_1, p_2)} \,.
\end{align}

\item Sector \{1, 0, 1, 1, 1, 0, 0\}: no ISP.
\begin{align}
\hat{\varphi}_9 &= \frac{(1-2\epsilon)^2}{\epsilon^2}\frac{s m_2^2(m_2^2 - m_t^2)\sqrt{s(s-4m_t^2)}}{16z_1z_3z_4z_5 \, G(k_2, p_1+p_2) \, G(k_1, p_2)} \,.
\end{align}

\item Sector \{1, 0, 1, 0, 1, 1, 0\}: no ISP.
\begin{align}
\hat{\varphi}_{10} &=  \frac{(1-2\epsilon)^2}{\epsilon^2} \frac{m_2^4(m_2^2-m_t^2)^2}{16z_1z_3z_5z_6 \, G(k_1, p_2) \, G(k_2, p_2)} \,.
\end{align}

\item Sector \{0, 1, 1, 1, 1, 0, 1\}: $z_6$ as ISP.

The $d\log$-forms $\hat{\varphi}_{11}$, $\hat{\varphi}_{12}$, $\hat{\varphi}_{13}$ and $\hat{\varphi}_{14}$ in this sector can be obtained from sector \{1, 0, 1, 1, 1, 0, 1\} by the replacements:
\begin{align}
z_3 \leftrightarrow z_4 \,, \quad m_1 \leftrightarrow m_2 \,, \quad z_1 \leftrightarrow z_2 \,. \nonumber 
\end{align}

\item Sector \{0, 1, 1, 1, 1, 1, 0\}: no ISP.
\begin{align}
\hat{\varphi}_{15} &= -\frac{1-2\epsilon}{\epsilon} \frac{\sqrt{\lambda}m_1^2(m_1^2-m_t^2)}{4z_2z_3z_4z_5z_6 \, G(k_1, p_1)} \,.
\end{align}

\item Sector \{0, 1, 1, 1, 1, 0, 0\}: no ISP.
\begin{align}
\hat{\varphi}_{16} &= \frac{(1-2\epsilon)^2}{\epsilon^2} \frac{s m_1^2(m_1^2 -m_t^2)\sqrt{s(s-4m_t^2)}}{16z_2z_3z_4z_5 \, G(k_2, p_1+p_2) \, G(k_1, p_1)} \,.
\end{align}

\item Sector \{0, 1, 1, 1, 0, 1, 0\}: no ISP.
\begin{align}
\hat{\varphi}_{17} &=  \frac{1-\epsilon}{\epsilon} \frac{\sqrt{\lambda}}{z_2z_3z_4z_6 \, G(k_1-p_1-p_2)} \,.
\end{align}

\item Sector \{0, 1, 1, 1, 0, 0, 0\}: no ISP.
\begin{align}
\hat{\varphi}_{18} &= -\frac{(1-\epsilon)(1-2\epsilon)}{\epsilon^2} \frac{s\sqrt{s(s-4m_t^2)}}{4z_2z_3z_4 \, G(k_1-p_1-p_2) \, G(k_2, p_1+p_2)} \,.
\end{align}

\item Sector \{0, 1, 1, 0, 1, 1, 1\}: $z_1$ as ISP.
\begin{align}
\hat{\varphi}_{19} &= \frac{\sqrt{\lambda}}{z_2z_3z_5z_6z_7} .
\end{align}

\item Sector \{0, 1, 1, 0, 1, 0, 1\}: $z_1$ as ISP.
\begin{align}
P_{1}(z_1) &=  G(k_1) \,, \nonumber \\
P_{2}(z_1, z_2, z_5) &= -4G(k_1, p_1, p_2) \,, \nonumber \\
P_{3}(z_1, z_3, z_7) &= -4G(k_1, k_2) \,, \nonumber \\
\hat{\varphi}_{20} &= \frac{1-2\epsilon}{\epsilon} \frac{\sqrt{\lambda} P_{1}}{z_2z_3z_5z_7 P_{3}} \,, \nonumber \\
\hat{\varphi}_{21} &= \frac{1-2\epsilon}{\epsilon} \frac{\lambda(m_1^2-m_t^2) z_1P_{1}}{z_2z_3z_5z_7 P_{2}P_{3}} \,, \nonumber \\
\hat{\varphi}_{22} &= \frac{1-2\epsilon}{\epsilon} \frac{\sqrt{\lambda} z_1 P_{1}}{z_2z_3z_5z_7 P_{2}P_{3}} \frac{\partial P_{2}}{\partial z_1} \,.
\end{align}

\item Sector \{0, 1, 1, 0, 1, 1, 0\}: no ISP.
\begin{align}
\hat{\varphi}_{23} &= \frac{(1-2\epsilon)^2}{\epsilon^2} \frac{m_1^2 m_2^2(m_1^2 -m_t^2)(m_2^2 - m_t^2)}{16z_2z_3z_5z_6 \, G(k_1, p_1) \, G(k_2, p_2)}.
\end{align}

\item Sector \{0, 1, 1, 0, 0, 1, 1\}: $z_4$ ISP.

The $d\log$-forms $\hat{\varphi}_{24}$, $\hat{\varphi}_{25}$ and $\hat{\varphi}_{26}$ in this sector can be obtained from sector \{0, 1, 1, 0, 1, 0, 1\} by the replacements:
\begin{align}
z_1 \leftrightarrow z_4 \,, \quad z_2 \leftrightarrow z_3 \,, \quad z_5 \leftrightarrow z_7 \,, \quad m_1 \leftrightarrow m_2 \,.  \nonumber 
\end{align}

\item Sector \{0, 1, 1, 0, 0, 0, 1\}: $z_1$ as ISP.
\begin{align}
\hat{\varphi}_{27} &= \frac{(1-2\epsilon)^2}{16\epsilon^2}\frac{s\sqrt{s(s-4m_t^2)}}{z_2z_3z_7}\frac{G(k_1)}{G(k_1, p_1+ p_2) \, G(k_1, k_2)} , \nonumber \\
\hat{\varphi}_{28} &= \frac{(1-2\epsilon)^2}{16\epsilon^2}\frac{s z_1 \, G(k_1)}{z_2z_3z_7 \, G(k_1, p_1+ p_2) \, G(k_1, k_2)}.
\end{align}

\item Sector \{0, 1, 1, 0, 0, 1, 0\}: no ISP.
\begin{align}
\hat{\varphi}_{29} &= -\frac{(1-\epsilon)(1-2\epsilon)}{\epsilon^2} \frac{m_2^2(m_2^2-m_t^2)}{4z_2z_3z_6 \, G(k_1-p_1-p_2) \, G(k_2,p_2)} \,.
\end{align}

\item Sector \{0, 1, 0, 1, 1, 1, 0\}: no ISP.
\begin{align}
\hat{\varphi}_{30} &= \frac{(1-2\epsilon)^2}{16\epsilon^2} \frac{m_1^4(m_1^2 - m_t^2)^2}{z_2z_4z_5z_6 \, G(k_1, p_1) \, G(k_2, p_1)}.
\end{align}

\item Sector \{0, 1, 0, 1, 0, 1, 0\}: $z_5$ as ISP.
\begin{align}
\hat{\varphi}_{31} &= \frac{(1-2\epsilon)^2}{16\epsilon^2} \frac{m_1^4(m_1^2 - m_t^2)}{z_2z_4z_6 \, G(k_1, p_1) \, G(k_2, p_1)} \,.
\end{align}

\item Sector \{0, 1, 0, 1, 0, 0, 0\}: no ISP.
\begin{align}
\hat{\varphi}_{32} &= \frac{(1-\epsilon)^2}{\epsilon^2} \frac{1}{z_2z_4 \, G(k_1-p_1-p_2) \, G(k_2-p_1-p_2)} \,.
\end{align}

\item Sector \{0, 0, 1, 1, 1, 0, 1\}: $z_6$ as ISP.
\begin{align}
\hat{\varphi}_{33} &= -\frac{1-2\epsilon}{4\epsilon} \frac{\sqrt{\lambda} \, z_6}{z_3z_4z_5z_7 \, G(k_1, k_2)} \,, \nonumber \\
\hat{\varphi}_{34} &= \frac{1-2\epsilon}{16\epsilon} \frac{\lambda\sqrt{s(s-4m_t^2)} \, z_6^2}{z_3z_4z_5z_7 \, G(k_1, k_2) \, G(k_2, p_1, p_2)} \,.
\end{align}

\item Sector \{0, 0, 1, 0, 1, 0, 1\}: $z_1$ as ISP.
\begin{align}
\hat{\varphi}_{35} &= \frac{(1-2\epsilon)^2}{16\epsilon^2} \frac{m_2^2(m_2^2 -m_t^2)}{z_3z_5z_7} \frac{G(k_1)}{G(k_1, p_2) \, G(k_1, k_2)} \,, \nonumber \\
\hat{\varphi}_{36} &= \frac{(1-2\epsilon)^2}{16\epsilon^2} \frac{m_2^2 z_1}{z_3z_5z_7} \frac{G(k_1)}{G(k_1, p_2) \, G(k_1, k_2)} \,.
\end{align}

\item Sector \{0, 0, 0, 1, 1, 0, 1\}: $z_6$ as ISP.
\begin{align}
\hat{\varphi}_{37} &= \frac{(1-2\epsilon)^2}{16\epsilon^2} \frac{m_1^2 z_6^2}{z_4z_5z_7 \, G(k_2, p_1) \, G(k_1, k_2)} \,, \nonumber \\
\hat{\varphi}_{38} &= \frac{(1-2\epsilon)^2}{16\epsilon^2} \frac{m_1^2(m_1^2 -m_t^2) z_6}{z_4z_5z_7 \, G(k_2, p_1) \, G(k_1, k_2)} \,.
\end{align}
\end{itemize}

We now list the canonical basis in terms of Feynman integrals. For convenience, we introduce the following dimensionless variables:
\begin{gather}
u = -\frac{s}{4m_t^2} \,, \quad v= -\frac{m_1^2}{4m_t^2} \,, \quad w = -\frac{m_2^2}{4m_t^2} \,,  \nonumber \\
R_1 = \sqrt{u(u+1)} \,, \quad R_2 = \sqrt{\lambda(u, v, w)} \,.
\end{gather}
\begin{align}
\bra{\varphi_1} &= 16 R_2^2 m_t^4 \bra{F_{1111110}},  \nonumber \\
\bra{\varphi_2} &= -\frac{16 R_1 R_2 m_t^4}{\epsilon } \bra{F_{2111010} }, \nonumber \\
\bra{\varphi_3} &= \frac{16 R_1^2  m_t^4}{\epsilon ^2} \bra{F_{2121000}}, \nonumber \\
\bra{\varphi_4} &= 4 R_2  m_t^2 \bra{F_{1011101}}, \nonumber \\ 
\bra{\varphi_5} &= -\frac{16 R_1 R_2  m_t^4}{\epsilon } \bra{F_{1012101}} , \nonumber \\
\bra{\varphi_6} &= -\frac{4  m_t^2 (u+v-w)}{\epsilon } \bra{F_{0210011}} -8  m_t^2 (u-v+w) \bra{F_{1011101}} \nonumber \nonumber \\
& -\frac{8 u  m_t^4 (2 u-2 v-2 w+1)}{\epsilon } \bra{F_{1012101}} +\frac{8 u  m_t^2}{\epsilon } \bra{F_{10121-11}}, \nonumber \\
\bra{\varphi_7} &= \frac{4 R_2 (4 w+1)  m_t^4}{\epsilon } \bra{F_{1011201}}, \nonumber \\
\bra{\varphi_8} &= -\frac{16 R_2 w  m_t^4}{\epsilon } \bra{F_{2011110}}-\frac{2 R_2  m_t^2}{\epsilon } \bra{F_{0211010}}, \nonumber \\
\bra{\varphi_9} &= \frac{16 R_1 w m_t^4}{\epsilon ^2} \bra{F_{2021100}}  +\frac{2 R_1  m_t^2}{\epsilon ^2}\bra{ F_{0212000}} , \nonumber \\
\bra{\varphi_{10}} &= \frac{16 w^2  m_t^4}{\epsilon ^2} \bra{F_{2020110}} +\frac{4 w  m_t^2}{\epsilon ^2}\bra{ F_{0220010}} +\frac{1}{4\epsilon ^2}\bra{ F_{0202000}} , \nonumber \\
\bra{\varphi_{11}} &=  4 R_2  m_t^2 \bra{ F_{0111101}}, \nonumber \\
\bra{\varphi_{12}} &=  -\frac{16 R_1 R_2  m_t^4}{\epsilon } \bra{ F_{0121101}}, \nonumber \\
\bra{\varphi_{13}} &=  -\frac{4  m_t^2 (u-v+w)}{\epsilon } \bra{F_{0120101}} -8  m_t^2 (u+v-w) \bra{F_{0111101}}  \nonumber \\
& -\frac{8 u  m_t^4 (2 u-2 v-2 w+1)}{\epsilon } \bra{F_{0121101}} +\frac{8 u  m_t^2}{\epsilon } \bra{F_{01211-11}} , \nonumber \\
\bra{\varphi_{14}} &=  \frac{4 R_2 (4 v+1)  m_t^4}{\epsilon }  \bra{F_{0111201}}, \nonumber \\
\bra{\varphi_{15}} &=  -\frac{16 R_2 v  m_t^4}{\epsilon } \bra{F_{0211110}} -\frac{2 R_2 m_t^2}{\epsilon } \bra{F_{0211010} } , \nonumber \\
\bra{\varphi_{16}} &=  \frac{16 R_1 v  m_t^4}{\epsilon ^2} \bra{F_{0221100}} +\frac{2 R_1  m_t^2}{\epsilon ^2} \bra{ F_{0212000}} , \nonumber \\
\bra{\varphi_{17}} &= \frac{4 R_2  m_t^2}{\epsilon } \bra{F_{0211010}} , \nonumber \\
\bra{\varphi_{18}} &= \frac{4 R_1 m_t^2}{\epsilon ^2}\bra{ F_{0212000}} , \nonumber \\
\bra{\varphi_{19}} &= 4 R_2 m_t^2 \bra{F_{0110111}}  , \nonumber \\
\bra{\varphi_{20}} &= \frac{2 R_2  m_t^2}{\epsilon } \bra{F_{0110102}} +\frac{2 R_2 m_t^2}{\epsilon }\bra{F_{0120101} }  , \nonumber \\
\bra{\varphi_{21}} &= -\frac{2  m_t^2 f(u,v,w)}{\epsilon }\bra{F_{0110102}}-\frac{(4 w+1)  m_t^2}{2 \epsilon ^2}\bra{F_{0010202}} +\frac{(8 w-1) m_t^2}{\epsilon ^2}\bra{F_{0020201}}  \nonumber \\
& -\frac{4  m_t^2 f(u,v,w)}{\epsilon }\bra{F_{0120101}}-\frac{4 v  m_t^2}{\epsilon^2}\bra{F_{0202010}}-\frac{1}{\epsilon ^2}\bra{ F_{0202000}} \nonumber \\
& -\frac{2 m_t^4 h(u,v,w)}{\epsilon ^2}  \bra{F_{0220101}}-\frac{6 u  m_t^2}{\epsilon ^2}\bra{ F_{0220001}}  , \nonumber \\
\bra{\varphi_{22}} &=  -\frac{4 R_2  m_t^2}{\epsilon }\bra{ F_{0110102}} , \nonumber \\
\bra{\varphi_{23}} &= \frac{16 v w  m_t^4}{\epsilon ^2} \bra{F_{0220110}}+\frac{2 v  m_t^2}{\epsilon ^2}\bra{F_{0202010}} +\frac{2 w
   m_t^2}{\epsilon ^2} \bra{ F_{0220010}}+\frac{1}{4 \epsilon ^2} \bra{F_{0202000} } , \nonumber \\
\bra{\varphi_{24}} &= \frac{2 R_2  m_t^2}{\epsilon } \bra{F_{0110012}} +\frac{2 R_2  m_t^2}{\epsilon }\bra{ F_{0210011}} , \nonumber \\
\bra{\varphi_{25}} &= -\frac{2  m_t^2 f(u,w,v)}{\epsilon }\bra{F_{0110012}}-\frac{(4 v+1)  m_t^2}{2 \epsilon ^2}\bra{F_{0001202}} +\frac{(8 v-1) m_t^2}{\epsilon ^2}\bra{F_{0002201} } \nonumber \\
& -\frac{4  m_t^2 f(u,w,v)}{\epsilon }\bra{F_{0210011}}-\frac{4 w  m_t^2}{\epsilon^2}\bra{F_{0220010}}-\frac{1}{\epsilon ^2} \bra{F_{0202000}} \nonumber \\
& -\frac{2 m_t^4 h(u,w,v)}{\epsilon ^2}  \bra{F_{0220011}}-\frac{6 u  m_t^2}{\epsilon ^2} \bra{F_{0220001}}  , \nonumber \\
\bra{\varphi_{26}} &=  -\frac{4 R_2  m_t^2}{\epsilon } \bra{F_{0110012}}, \nonumber \\
\bra{\varphi_{27}} &= -\frac{2 R_1  m_t^2}{\epsilon ^2} \bra{F_{0210002}}-\frac{R_1 m_t^2}{\epsilon ^2}  \bra{F_{0220001}} , \nonumber \\
\bra{\varphi_{28}} &=  \frac{u m_t^2}{\epsilon ^2} \bra{F_{0220001}} , \nonumber \\
\bra{\varphi_{29}} &=  \frac{4 w m_t^2}{\epsilon ^2} \bra{F_{0220010}}  +\frac{1}{2 \epsilon ^2}\bra{ F_{0202000}}, \nonumber \\
\bra{\varphi_{30}} &= \frac{16 v^2  m_t^4}{\epsilon ^2}\bra{ F_{0202110}} +\frac{4 v  m_t^2}{\epsilon ^2} \bra{F_{0202010}}+\frac{1}{4\epsilon ^2}\bra{F_{0202000}}  , \nonumber \\
\bra{\varphi_{31}} &= \frac{2 v  m_t^2}{\epsilon ^2} \bra{F_{0202010}} +\frac{1}{4 \epsilon ^2} \bra{F_{0202000}} , \nonumber \\
\bra{\varphi_{32}} &= \frac{1}{\epsilon ^2} \bra{F_{0202000}} , \nonumber \\
\bra{\varphi_{33}} &=  \frac{4 R_2  m_t^2}{\epsilon } \bra{F_{0011201}}, \nonumber \\
\bra{\varphi_{34}} &=  \frac{4 R_1 (2 \epsilon -1)  m_t^2}{\epsilon ^2} \bra{F_{0012101}} +\frac{4 R_1 (2 \epsilon -1)  m_t^2}{\epsilon ^2} \bra{ F_{0021101}}+\frac{4 R_1  m_t^2}{\epsilon }\bra{F_{0011201}} , \nonumber \\
\bra{\varphi_{35}} &=  -\frac{(4 w+1)  m_t^2}{4 \epsilon ^2} \bra{F_{0010202}} -\frac{(4 w+1)  m_t^2}{2 \epsilon ^2}\bra{F_{0020201}} , \nonumber \\
\bra{\varphi_{36}} &=  \frac{(4 w+1)  m_t^2}{4 \epsilon ^2} \bra{F_{0010202}} +\frac{m_t^2}{2 \epsilon ^2}\bra{F_{0020201}}  , \nonumber \\
\bra{\varphi_{37}} &= \frac{(4 v-1)  m_t^2}{2 \epsilon ^2} \bra{F_{0002201}}-\frac{(4 v+1)  m_t^2}{4 \epsilon ^2}\bra{F_{0001202}} , \nonumber \\
\bra{\varphi_{38}} &= -\frac{(4 v+1) m_t^2}{4 \epsilon ^2} \bra{F_{0001202}} -\frac{(4 v+1)  m_t^2}{2 \epsilon ^2}\bra{F_{0002201}} .
\end{align}
where
\begin{align}
f(u,v,w) = 1+ 2u +2v -2w \,, \quad h(u,v,w) = 1 + 4v +4u(1+4v) -4w \,.
\end{align}

\bibliographystyle{JHEP}
\bibliography{references_inspire.bib,references_local.bib}

\end{document}